\documentclass[reprint, 10pt, amsmath, amssymb, floatfix, aps, prd,  superscriptaddress, nofootinbib, nobibnotes,twocolumn]{revtex4-1}

\pdfoutput=1
\usepackage{graphicx}
\graphicspath{{figures/}}
\usepackage[utf8]{inputenc}
\usepackage{amsmath}
\usepackage{amsfonts}
\usepackage{amssymb}
\usepackage{multirow}
\usepackage{tcolorbox}
\usepackage{CJKutf8}
\usepackage{color}
\usepackage[colorlinks=true, linkcolor=red, citecolor=blue]{hyperref}
\usepackage{orcidlink}
\usepackage[capitalise]{cleveref}
\usepackage{acro}
\usepackage{adjustbox}
\usepackage{array}
\usepackage{natbib}
\usepackage{dblfnote}
\DFNalwaysdouble
\usepackage{slashed}
\usepackage{dcolumn}
\usepackage{hyperref}
\usepackage{geometry}
\usepackage{subfig} 
\usepackage{overpic}
\usepackage{inputenc}
\usepackage{caption}
\def\({\left(}
\def\){\right)}
\def\[{\left[}
\def\]{\right]}
\def\be{\begin{eqnarray}}
\def\ee{\end{eqnarray}}

\def\lag{\mathcal{L}}

\DeclareAcronym{GW}{
  short = GW ,
  long = gravitational wave ,
  short-plural = s 
}
\DeclareAcronym{LIGO}{
  short = LIGO ,
  long = Laser Interferometer Gravitational-wave Observatory ,
  short-plural = 
}
\DeclareAcronym{LISA}{
  short = LISA ,
  long = Laser Interferometer Space Antenna ,
  short-plural =  
}
\DeclareAcronym{SKA}{
  short = SKA ,
  long = Square Kilometre Array ,
  short-plural =  
}  

\DeclareAcronym{PGW}{
  short = PGW ,
  long = primordial gravitational wave ,
  short-plural = s 
}
\DeclareAcronym{TSIGW}{
	short = TSIGW ,
	long = tensor-scalar induced gravitational wave ,
	short-plural =  s
}
  
\DeclareAcronym{SNR}{
	short = SNR ,
	long = signal-to-noise ratio ,
	short-plural = 
}

\DeclareAcronym{PTA}{
	short = PTA ,
	long = pulsar timing array ,
	short-plural = 
}

\DeclareAcronym{KDE}{
  short = KDE ,
  long = kernel density estimator ,
  short-plural = s
}

\DeclareAcronym{FLRW}{
  short = FLRW ,
  long = Friedmann-Lemaitre-Robertson-Walker ,
  short-plural =  
}

\DeclareAcronym{SIGW}{
	short = SIGW ,
	long = scalar induced gravitational wave ,
	short-plural =  s
}

\DeclareAcronym{TIGW}{
	short = TIGW ,
	long = tensor induced gravitational wave ,
	short-plural =  s
}

\DeclareAcronym{PBH}{
	short = PBH ,
	long = primordial black hole ,
	short-plural =  s
}

\DeclareAcronym{SMBHB}{
  short = SMBHB ,
  long = supermassive black hole binary ,
  short-plural = s
}

\DeclareAcronym{CMB}{
	short = CMB ,
	long = cosmic microwave background ,
	short-plural =  
}
\DeclareAcronym{DM}{
	short = DM ,
	long = dark matter ,
	short-plural =  
}

\DeclareAcronym{BBN}{
	short = BBN ,
	long = Big-Bang nucleosynthesis ,
	short-plural =  
}

\DeclareAcronym{SGWB}{
	short = SGWB ,
	long = stochastic gravitational	wave background ,
	short-plural =  s
}

\DeclareAcronym{LSS}{
	short = LSS ,
	long = large scale structure ,
	short-plural =  
}

\DeclareAcronym{RD}{
	short = RD ,
	long = radiation-dominated ,
	short-plural =  
}

\DeclareAcronym{BAO}{
	short = BAO ,
	long = baryon acoustic oscillations ,
	short-plural = 
}
\DeclareAcronym{nymtg}{
        short = NYMTG,
        long = Nieh-Yan modified teleparallel gravity ,
        short-plural =
}

\begin{document}
%\begin{CJK}{UTF8}{}

\title{Tensor induced gravitational waves}

\author{Fei-Yu Chen\orcidlink{0000-0002-8674-9316}}
\affiliation{Center for Joint Quantum Studies and Department of Physics,
School of Science, Tianjin University, Tianjin 300350, China}

\author{Jing-Zhi Zhou\orcidlink{0000-0003-2792-3182}} 
\email{zhoujingzhi@tju.edu.cn}
\affiliation{Center for Joint Quantum Studies and Department of Physics,
School of Science, Tianjin University, Tianjin 300350, China}

\author{Di Wu\orcidlink{0000-0001-7309-574X}}
\affiliation{School of Fundamental Physics and Mathematical Sciences, Hangzhou Institute for Advanced Study, University of Chinese Academy of Sciences, Hangzhou 310024, China}

\author{Zhi-Chao Li\orcidlink{0009-0005-7984-2626}}
\affiliation{Center for Joint Quantum Studies and Department of Physics,
School of Science, Tianjin University, Tianjin 300350, China}

\author{Peng-Yu Wu\orcidlink{0009-0001-5797-3829}}
\affiliation{Center for Joint Quantum Studies and Department of Physics,
School of Science, Tianjin University, Tianjin 300350, China}

\begin{abstract}
Primordial gravitational waves on small scales are not tightly constrained by current cosmological observations, which allows for the possibility of large amplitudes at small scales. We investigate second-order \acp{TIGW} sourced by primordial gravitational waves and present the corresponding corrections to the total energy density spectrum of gravitational wave. We analyze primordial gravitational waves with large amplitudes generated by various models at small scales. Our results indicate that when primordial gravitational waves on small scales sufficiently dominate the current \acp{PTA} observations, corrections to the total energy density spectrum from second-order \acp{TIGW} may become pronounced in certain frequency bands.

\end{abstract}

\maketitle
%%%%%%%%%%%%%%%%%%%%%%%%%%%%%%%%%%%%%%%%
%%%%%%%%%%%%%%%%%%%%%%%%%%%%%%%%%%%%%%%
\acresetall

\section{Introduction}\label{sec:1.0}
In cosmic evolution, the background spacetime is described by the \ac{FLRW} metric \cite{Baumann:2022mni}. Due to the symmetry of \ac{FLRW} spacetime, the corresponding metric perturbations can be decomposed into scalar, vector, and tensor perturbations with tensor perturbations representing \acp{GW} in the \ac{FLRW} background \cite{Bardeen:1980kt,Kodama:1984ziu,Malik:2008im}. For first-order cosmological perturbations, these three types of perturbations do not couple and evolve independently. However, when considering second-order or higher-order cosmological perturbations, different types of perturbations will couple with each other. Specifically, the equations of motion for second-order cosmological perturbations include terms from all three types of first-order cosmological perturbations. Different types of cosmological perturbations couple through higher-order cosmological perturbation equations, leading to the evolution of higher-order cosmological perturbations being influenced by all kinds of lower-order cosmological perturbations.

Primordial perturbations generated during inflation serve as initial conditions that affect the evolution of first-order cosmological perturbations \cite{Lyth:2005fi,Weinberg:2005vy,Bassett:2005xm,Arkani-Hamed:2003juy,Kachru:2003sx,Armendariz-Picon:1999hyi,Peebles:1998qn,Kobayashi:2010cm,Bezrukov:2010jz,Arkani-Hamed:2018kmz}. Through the coupling between first-order and second-order cosmological perturbations, primordial perturbations will induce second-order cosmological perturbations. In recent years, higher-order cosmological perturbations induced by lower-order perturbations have received widespread attention. Numerous physical processes associated with higher-order cosmological perturbations have been systematically investigated. These include second-order and third-order \acp{SIGW} \cite{Domenech:2021ztg,Ananda:2006af,Baumann:2007zm,Chang:2023vjk}, the effects of second-order density perturbations induced by primordial tensor perturbations on \ac{LSS} \cite{Bari:2021xvf,Bari:2022grh,Abdelaziz:2025qpn,Chang:2022aqk}, the influence of second-order density perturbations induced by primordial curvature perturbations on the thresholds and probability distribution functions of \acp{PBH} \cite{Zhou:2023itl,Nakama:2016enz,Nakama:2015nea,DeLuca:2023tun}, and the impact of second-order induced tensor and vector perturbations on \ac{CMB} \cite{Mollerach:2003nq,Cyr:2023pgw,Gurian:2021rfv,DAmico:2007ngk,Yamauchi:2012bc,Saga:2015apa}. These studies cover various physical processes across different scales during cosmic evolution. With current cosmological observations, such as \ac{CMB}, \ac{LSS}, \ac{BAO}, \ac{PTA}, and \ac{LISA}, we can investigate the impact of these higher-order induced cosmological perturbations on observations at different scales to determine the physical properties of the corresponding primordial perturbations.

On large scales ($\gtrsim$1 Mpc), the power spectrum of primordial curvature perturbations is determined to be approximately scale-invariant by cosmological observation experiments such as \ac{CMB} and \ac{LSS}, with an amplitude $A_{\zeta} \approx 2\times 10^{-9}$, and the tensor-to-scalar ratio $r$ is limited to $r<0.064$ at a $95\%$ confidence level \cite{Planck:2018vyg}. This excludes the possibility of significant primordial curvature perturbations and \acp{PGW} on large scales. Due to the small amplitude of primordial perturbations on large scales, higher-order induced cosmological perturbations do not have a particularly significant impact on large-scale cosmological observations. However, on small scales ($\lesssim$1 Mpc), the primordial curvature perturbations have not been well constrained by small-scale observations, with the upper limit of $A_{\zeta}$ reaching around $0.1$ \cite{NANOGrav:2023hvm}. Potential large-amplitude primordial perturbations on small scales make the observational effects of higher-order cosmological perturbations very important.

In 2023, the \ac{PTA} collaborations NANOGrav \cite{NANOGrav:2023gor,NANOGrav:2023hvm}, EPTA \cite{EPTA:2023fyk}, PPTA \cite{Reardon:2023gzh}, and CPTA \cite{Xu:2023wog} have reported positive evidence for an isotropic \ac{SGWB} within the nHz frequency range. As reported by the NANOGrav collaboration, second-order \acp{SIGW} generated from large-amplitude primordial curvature perturbations at small scales exhibit the highest Bayes factor compared to other \ac{SGWB} models \cite{NANOGrav:2023hvm}. Therefore, \acp{SIGW} are among the most likely dominant sources of the \ac{SGWB} in the nHz frequency band, sparking intense research interest \cite{Assadullahi:2009nf,Alabidi:2013lya,Domenech:2020ssp,Domenech:2020kqm,Inomata:2019zqy,Wang:2019kaf,Liu:2023pau,Domenech:2024rks,Harigaya:2023pmw,Zhu:2023gmx,Pearce:2023kxp,Inomata:2019ivs,Zhang:2022dgx,Mangilli:2008bw,Yu:2024xmz,Loverde:2022wih,Saga:2014jca,Sui:2024nip,Madge:2023dxc,Domenech:2020ers}. Similar to primordial curvature perturbations, \acp{PGW} on small scales have not been tightly constrained by observations. In this study, we investigate the potential existence of large-amplitude \acp{PGW} on small scales and the second-order \acp{TIGW} that arise from \acp{PGW}. Our analysis focuses on the following five questions:
\tcbset{colback=gray!20, colframe=gray!20, boxrule=0.5mm, arc=0mm, auto outer arc, width=\linewidth} 
\begin{tcolorbox} 
\noindent
\textbf{ 1, What kinds of models can generate large-amplitude \acp{PGW} on small scales?}

\

\noindent
\textbf{ 2, Given a small-scale power spectrum of \acp{PGW}, how can we calculate the energy density spectrum of second-order \acp{TIGW}?}

\

\noindent
\textbf{ 3, Under what conditions on the \acp{PGW} do the corrections from second-order \acp{TIGW} become significant? }

\

\noindent
\textbf{ 4, If the small-scale \acp{PGW} dominate current \ac{PTA} observations, will the second-order \acp{TIGW} affect the \ac{PTA} observations?}

\

\noindent
\textbf{ 5, Are there any other methods to constrain \acp{PGW} on small scales, apart from the observation of the \ac{SGWB}?}
\end{tcolorbox}
\noindent
In this paper, we address these questions in detail and place constraints on small-scale \acp{PGW} using current cosmological observations.

This paper is organized as follows. In Sec.~\ref{sec:2.0}, we investigate the equation of motion and kernel function of second-order \acp{TIGW} and provide the corresponding formula for calculating their energy density spectrum. In Sec.~\ref{sec:3.0}, we analyze various models capable of generating large-amplitude \acp{PGW} on small scales, along with the associated corrections from \acp{TIGW} to the total energy density spectrum. In Sec.~\ref{sec:4.0}, we discuss constraints on small-scale \acp{PGW} based on current \ac{PTA} observations and other cosmological data. Finally, we summarize our results and provide further discussions in Sec.~\ref{sec:5.0}.

\section{Tensor induced gravitational waves}\label{sec:2.0}
In this section, we provide a detailed review of the derivation of the equation of motion for second-order \acp{TIGW}, and examine the properties of the corresponding second-order kernel functions. Finally, we present a comprehensive derivation of the energy density spectrum associated with the second-order \acp{TIGW}. The line element of perturbed spacetime can be expressed as
\cite{Malik:2008im}
\begin{eqnarray}\label{eq:dS}
	\mathrm{d} s^{2}=a^{2}\left[- \mathrm{d} \eta^{2}+\left( \delta_{i j} +h^{(1)}_{ij}+\frac{1}{2}h^{(2)}_{ij}\right)\mathrm{d} x^{i} \mathrm{d} x^{j}\right]  ,
\end{eqnarray}
where $h^{(n)}_{ij}$$\left( n=1,2 \right)$ are $n$th-order tensor perturbations. The Fourier components $ h^{\lambda,(1)}_{\mathbf{k}}(\eta)$$(\lambda=+,\times)$ of $h_{ij}^{(n)}(\mathbf{x},\eta)$ in Eq.~(\ref{eq:dS}), expressed in terms of the polarization tensors $\varepsilon_{i j}^{\lambda}(\mathbf{k})$$(\lambda=+,\times)$, are defined as
\begin{equation}\label{eq:h}
\begin{aligned}
h_{i j}^{(n)}(\mathbf{x}, \eta)=\int \frac{\mathrm{d}^3 k}{(2 \pi)^{3 / 2}} e^{i \mathbf{k} \cdot \mathbf{x}}\sum_{\lambda}h_{\mathbf{k}}^{\lambda,(n)}(\eta) \varepsilon_{i j}^{\lambda}(\mathbf{k}) \  .
\end{aligned}
\end{equation}
In Eq.~(\ref{eq:h}), the polarization tensors with momentum $\mathbf{k}$ are defined as
\begin{equation}\label{eq:pt1}
	\begin{aligned}
		\varepsilon^{\times}_{ij}\left(\mathbf{k}  \right)=\frac{1}{\sqrt{2}}\left( e_i\left( \mathbf{k} \right)\bar{e}_j\left( \mathbf{k} \right)+\bar{e}_i\left( \mathbf{k} \right)e_j\left( \mathbf{k} \right)  \right) \ ,
	\end{aligned} 
\end{equation}
\begin{equation}
	\begin{aligned}
		\varepsilon^{+}_{ij}\left(\mathbf{k}  \right)=\frac{1}{\sqrt{2}}\left( e_i\left( \mathbf{k} \right)e_j\left( \mathbf{k} \right)-\bar{e}_i\left( \mathbf{k} \right)\bar{e}_j\left( \mathbf{k} \right)  \right) \ ,
	\end{aligned} 
\end{equation}
where $e_i\left( \mathbf{k} \right)$ and $\bar{e}_i\left( \mathbf{k} \right)$ are transverse polarization vectors in three dimensional momentum space. Since we are primarily concerned with \acp{PGW} possessing large amplitudes on small scales, first-order scalar and vector perturbations are neglected in Eq.~(\ref{eq:dS}). In contrast, studies of \acp{SIGW} typically retain first-order scalar perturbations while discarding both the first-order tensor and vector modes. It is worth noting that \acp{SIGW} are generally computed in Newtonian gauge. When alternative gauge choices are adopted, the second-order kernel functions and the corresponding energy density spectrum of \acp{SIGW} may exhibit potential gauge dependence. The behavior of \acp{SIGW} under different gauges and their approximate gauge invariance have been systematically investigated in Refs.~\cite{Hwang:2017oxa,DeLuca:2019ufz,Lu:2020diy,Tomikawa:2019tvi,Domenech:2020xin,Inomata:2019yww,Yuan:2019fwv,Yuan:2024qfz,Kugarajh:2025pjl}. In contrast, \acp{TIGW} exhibit no gauge dependence, as metric perturbations take the same form in all common gauges, rendering Eq.~(\ref{eq:dS}) gauge invariant.

\subsection{Equation of motion}\label{sec:2.1}
To derive the equation of motion for second-order tensor-induced gravitational waves, we substitute Eq.~(\ref{eq:dS}) into the Einstein field equations and solve the cosmological perturbation equations iteratively at each perturbative order. The transverse-traceless part of the spatial–spatial component of the first-order perturbation of the Einstein field equation in \ac{FLRW} background is given by
\begin{eqnarray}\label{eq:1oh}
    h_{ij}^{(1)''}(\eta,\mathbf{x})+2 \mathcal{H} h_{ij}^{(1)'}(\eta,\mathbf{x})-\Delta h_{ij}^{(1)}(\eta,\mathbf{x})=0 \ ,
\end{eqnarray}
where $\mathcal{H}=a'(\eta)/a(\eta)$ is the conformal Hubble parameter.  For a constant equation of state
\begin{eqnarray}\label{eq:wH}
    w=\frac{p^{(1)}}{\rho^{(1)}}=\text{constant}  \ , \   \mathcal{H}=\frac{2}{(1+3w)\eta} \ .
\end{eqnarray}
As shown in Eq.~(\ref{eq:1oh}) and  Eq.~(\ref{eq:wH}), the equation of motion for first-order tensor perturbations is influenced by the parameter $w$. In momentum space, the solutions for the first-order tensor modes $h^{\lambda,(1)}_{\mathbf{k}}(\eta)$ under different values of $w$ can be expressed as 
\begin{eqnarray}\label{eq:Th1}
    h^{\lambda,(1)}_{\mathbf{k}}(\eta)= \mathbf{h}^{\lambda}_{\mathbf{k}} T_h(x) \ ,
\end{eqnarray}
where $\mathbf{h}^{\lambda}_{\mathbf{k}}$ is the \acp{PGW} and $x=|\mathbf{k}|\eta=k\eta$. The first-order transfer function $T_h(x)$ in Eq.~(\ref{eq:Th1}) is given by \cite{Domenech:2019quo,Gong:2019mui}
\begin{equation}\label{eq:Tw}
    T_h(x)=\left(\frac{2}{x}  \right)^{\beta}\Gamma[1+\beta] J_\beta(x) \ , 
\end{equation}
where $\beta=(3-3w)/(2+6w)$. $J_\beta(x)$ is the Bessel function of the first kind, and $\Gamma[1+\beta]$ denotes the Gamma function, which satisfies $\lim_{x \rightarrow 0} x^{-\nu} J_\nu(x)=2^{-\nu} / \Gamma[1+\nu]$. During the \ac{RD} era, characterized by $w=1/3$ and $\beta=1/2$, the first-order tensor modes $h^{\lambda,(1)}_{\mathbf{k}}(\eta)$ can be expressed as
\begin{eqnarray}\label{eq:Th1}
    h^{\lambda,(1)}_{\mathbf{k}}(\eta)= \mathbf{h}^{\lambda}_{\mathbf{k}} T_h(x)=\mathbf{h}^{\lambda}_{\mathbf{k}}\frac{\sin x}{x}  \ .
\end{eqnarray}

By extracting the transverse traceless component of the spatial-spatial part of the second-order perturbation in the Einstein field equations, we obtain the following equation of motion of second-order \acp{TIGW}
\begin{equation}\label{eq:h2o}
	\begin{aligned}
		h_{ij}^{(2)''}(\eta,\mathbf{x})&+2 \mathcal{H} h_{ij}^{(2)'}(\eta,\mathbf{x})-\Delta h_{ij}^{(2)}(\eta,\mathbf{x}) \\
        &=-4 \Lambda_{ij}^{lm}\mathcal{S}^{(2)}_{lm}(\eta,\mathbf{x}) \ .
	\end{aligned}
\end{equation}
In Eq.~(\ref{eq:h2o}), the decomposed operator to extract the transverse and traceless terms is given by
\begin{eqnarray}
    \Lambda_{ij}^{lm}=\left(\mathcal{T}_{i}^{l} \mathcal{T}_{j}^{m}-\frac{1}{2} \mathcal{T}_{i j} \mathcal{T}^{lm}\right) \ ,
\end{eqnarray}
where
\begin{equation}
	\mathcal{T}_{j}^{i} \equiv \delta_{i}^{i}-\partial^{i} \Delta^{-1} \partial_{j} \ .
\end{equation} 
The source term $\mathcal{S}^{(2)}_{lm}(\eta,\mathbf{x}) $ in Eq.~(\ref{eq:h2o}) consists of the product of two first-order tensor perturbations, and its explicit expression is given by
\begin{equation}\label{eq:s2}
	\begin{aligned}
		\mathcal{S}^{(2)}_{lm}(\eta,\mathbf{x})&=\frac{1}{2}\left( -h^{b,(1)'}_{l} h^{(1)'}_{mb} +\partial_c h^{(1)}_{mb}\partial^c h^{b,(1)}_{l} \right.\\
        &\left.-h^{bc,(1)}\partial_c \partial_m h^{(1)}_{lb}  -\partial_b h^{(1)}_{mc}\partial^c h^{b,(1)}_{l}  \right. \\
	&\left. +\frac{1}{2} h^{bc,(1)}\partial_l \partial_m h^{(1)}_{bc}+ h^{bc,(1)}\partial_c \partial_b h^{(1)}_{lm}\right.\\
        &\left.-h^{bc,(1)}\partial_c \partial_l h^{(1)}_{mb} \right) \ .
	\end{aligned}
\end{equation}
Eq.~(\ref{eq:h2o}) and Eq.~(\ref{eq:s2}) describe the equation of motion of second-order \acp{TIGW}. It is important to note that, similar to the equation of motion of first-order tensor perturbations given in Eq.~(\ref{eq:1oh}), Eq.~(\ref{eq:h2o}) and Eq.~(\ref{eq:s2}) are valid for an arbitrary dominant era. The influence of parameter $w$ on second-order \acp{TIGW} manifests only through the conformal Hubble parameter $\mathcal{H}$ in Eq.~(\ref{eq:1oh}) and Eq.~(\ref{eq:h2o}); parameter $w$ does not explicitly appear in Equations Eq.~(\ref{eq:1oh}), Eq.~(\ref{eq:h2o}), and Eq.~(\ref{eq:s2}). Furthermore, second-order \acp{TIGW} are unaffected by the sound speed $c_s=\sqrt{p^{(1)}/\rho^{(1)}}$; the corresponding energy density spectrum remains unchanged regardless of variations in $c_s$.  This behavior is in sharp contrast to that of second-order \acp{SIGW}, whose energy density spectrum is significantly sensitive to changes in $c_s$ \cite{Balaji:2023ehk,Jin:2023wri,Liu:2023hpw,Li:2023uhu,Chen:2024fir,Wang:2025aon}. More precisely, the influence of the sound speed $c_s$ on induced gravitational waves can be evaluated through the analysis of cosmological perturbation equations. In the Newtonian gauge, when all first-order perturbations are retained, the first-order density perturbation can be expressed as \cite{Zhou:2024doz}
\begin{equation}\label{eq:rhocs}
	\begin{aligned}
		\rho^{(1)}=\frac{1}{\kappa a^2}\left(-6 \mathcal{H}\left(\mathcal{H} \phi^{(1)}+\psi^{(1)^{\prime}}\right)+2 \Delta \psi^{(1)}\right) \ , 
	\end{aligned}
\end{equation}
where $\psi^{(1)}$ and $\phi^{(1)}$ are first-order scalar perturbations. As shown in Eq.~(\ref{eq:rhocs}), first-order density perturbations are independent of first-order tensor and vector perturbations. Consequently, when small-scale first-order scalar perturbations are neglected and only large-amplitude \acp{PGW} are considered on small scales, the first-order density perturbation $\rho^{(1)}\approx 0$. Under these conditions, the contribution of second-order \acp{SIGW} to the total energy density spectrum of gravitational waves is significantly smaller than that of second-order \acp{TIGW}, and the sound speed $c_s$ does not affect the equation of motion for second-order \acp{TIGW}. The source term of second-order \acp{TIGW} $\mathcal{S}^{(2)}_{lm}$ in Eq.~(\ref{eq:s2}) arises entirely from the perturbative expansion of the Einstein tensor $G_{\mu\nu}$. Here, it is important to note that this property does not hold for third-order and higher-order \acp{TIGW}. For instance, in the case of third-order \acp{TIGW}, first-order tensor perturbations can induce a large second-order density perturbation $\rho^{(2)}$, through which the sound speed $c_s^2=p^{(2)}/\rho^{(2)}$ influences the second-order scalar perturbations and subsequently affects the third-order \acp{TIGW} sourced by the coupling of second-order scalar and first-order tensor perturbations.

\subsection{Second-order kernel functions}\label{sec:2.2}
In this subsection, we investigate the solution to the equation of motion for second-order \acp{TIGW}, along with the corresponding second-order kernel functions $I^{(2)}$. As shown in Eq.~(\ref{eq:s2}), the source term $\mathcal{S}^{(2)}_{lm}(\eta,\mathbf{x})$ of the second-order \acp{TIGW} consists of the product of two first-order tensor perturbations. Based on the form of the source term $\mathcal{S}^{(2)}_{lm}(\eta,\mathbf{x})$, the solution is decomposed into the following five components:
\begin{equation}\label{eq:h10}
	\begin{aligned}
		h^{\lambda,(2)}_{\mathbf{k}}(\eta)=\sum^{5}_{i=1} h_{\mathbf{k},i}^{\lambda,(2)}(\eta) \ ,
	\end{aligned}
\end{equation}
where $h_{\mathbf{k},i}^{\lambda,(2)}(\eta)$$(i=1,2,3,4,5)$ are given by
\begin{eqnarray}
	h^{\lambda,(2)}_{\mathbf{k},i}(\eta)&=&-\int\frac{\mathrm{d}^3p}{(2\pi)^{3/2}}\varepsilon^{\lambda,lm} \left( \mathbf{k}\right)I^{(2)}_{i}\left(u,v,x\right)\nonumber\\
&\times&\mathbb{P}^{\lambda_1\lambda_2}_{lm,i}~\mathbf{h}^{\lambda_1}_{\mathbf{k}-\mathbf{p}}\mathbf{h}^{\lambda_2}_{\mathbf{p}} \ ,\label{eq:hhall}
\end{eqnarray}
where $|\mathbf{k}-\mathbf{p}|=u|\mathbf{k}|$ and $|\mathbf{p}|=v|\mathbf{k}|$. The momentum polynomials $\mathbb{P}^{\lambda_1\lambda_2}_{lm,i}$ are given by
\begin{eqnarray}
\mathbb{P}^{\lambda_1\lambda_2}_{lm,1}&=&\varepsilon^{\lambda_1,b}_{l}\left(\mathbf{k}-\mathbf{p}\right)\varepsilon^{\lambda_2}_{bm}\left(\mathbf{p}\right)k^2 \ ,\label{eq:Pp1}\\
\mathbb{P}^{\lambda_1\lambda_2}_{lm,2}&=&2\varepsilon^{\lambda_2}_{mb}\left( \mathbf{p} \right)\varepsilon^{\lambda_1,bc}\left( \mathbf{k}-\mathbf{p} \right)p_cp_l \ ,\label{eq:Pp2}\\
     \mathbb{P}^{\lambda_1\lambda_2}_{lm,3}&=&\varepsilon^{\lambda_1}_{mc}\left(\mathbf{k}-\mathbf{p}\right)\varepsilon^{\lambda_2}_{lb}\left(\mathbf{p}\right)\left( k-p \right)^bp^c \ ,\label{eq:Pp3}\\
     \mathbb{P}^{\lambda_1\lambda_2}_{lm,4}&=&-\varepsilon^{\lambda_1}_{bc}\left(\mathbf{k}-\mathbf{p}\right)\varepsilon^{\lambda_2}_{lm}\left(\mathbf{p}\right)p^bp^c \ ,\label{eq:Pp4} \\
     \mathbb{P}^{\lambda_1\lambda_2}_{lm,5}&=&-\varepsilon^{\lambda_1,bc}\left(\mathbf{k}-\mathbf{p}\right)\varepsilon^{\lambda_2}_{bc}\left(\mathbf{p}\right)\frac{p_lp_m}{2} \ .	\label{eq:Pp5}
\end{eqnarray}
The second-order kernel functions $I^{(2)}_{i}\left(u,v,x\right)$ $(i=1,2,3,4,5)$ in Eq.~(\ref{eq:hhall}) can be calculated using the Green's function method, and the corresponding explicit expressions are given by
\begin{equation}\label{eq:I}
	\begin{aligned}
		I^{(2)}_i\left( u,v,x \right)=\frac{4}{k^2} \int_{0}^{x} \mathrm{d} \bar{x} \left( G_h\left(x, \bar{x}  \right) f_i\left( u,v,\bar{x} \right)  \right) \ , 
	\end{aligned}
\end{equation}
where the Green’s function $ G_h(x, \bar{x})$ can be expressed in terms of the Bessel functions of the first kind $J_{\beta}$ and the second kind $Y_{\beta}$ \cite{Domenech:2019quo}
\begin{equation}\label{eq:Gw}
    G_h(x, \bar{x})=\frac{\pi}{2} \frac{\bar{x}^{1+\beta}}{x^\beta}\left(J_\beta(\bar{x}) Y_\beta(x)-Y_\beta(\bar{x}) J_\beta(x)\right) \ ,
\end{equation}
and
\begin{eqnarray}
	f_{1}^{(2)}\left(u,v,x\right)&=&-\frac{1-u^2-v^2}{4}T_{h}\left( ux \right)T_{h}\left( vx \right)-\frac{uv}{2} \nonumber\\
    &\times&\frac{\mathrm{d}}{\mathrm{d}(ux)}T_{h}\left( ux \right)\frac{\mathrm{d}}{\mathrm{d}(vx)}T_{h}\left( vx \right) \ , 
	\label{eq:f1} \\
	f_i^{(2)}\left( u,v,x \right)&=&\frac{1}{2}T_{h}\left( ux \right)T_{h}\left( vx \right) \ ,  \nonumber\\
    &&(i=2,3,4,5)  \label{eq:fi} \ .
\end{eqnarray}
In Eq.~(\ref{eq:I}), both the Green's function $G_h\left(x ,\bar{x} \right)$ and the source function $f_i(u,v,x)$ depend on the parameter of the equation of state: $w$. By substituting Eq.~(\ref{eq:Tw}) into Eq.~(\ref{eq:f1}) and Eq.~(\ref{eq:fi}), the explicit form of the source function $f_i(u,v,x)$ can be obtained. In combination with the Green's function provided in Eq.~(\ref{eq:Gw}), the specific expression for the second-order kernel functions $I^{(2)}_i$ under arbitrary dominant epoch can be derived by evaluating the time integral in Eq.~(\ref{eq:I}). In this paper, we focus on the second-order \acp{TIGW} during the \ac{RD} era. Under this condition, we have $w=1/3$ and $\beta=1/2$. The corresponding Green's function $G_h(x,\bar{x})$ in Eq.~(\ref{eq:I}) can be expressed as
\begin{eqnarray}
     G_h(x, \bar{x})=\frac{\bar{x}}{x}\sin\left(x-\bar{x}  \right) \ .
\end{eqnarray}
The first-order transfer function during \ac{RD} era is provided in Eq.~(\ref{eq:Th1}). By evaluating the time integral in Eq.~(\ref{eq:I}), we can obtain the analytical form of the second-order kernel function $I_i^{(2)}(u,v,x)$ in the \ac{RD} era
\begin{eqnarray}
        I^{(2)}_1(u,v,x)=\frac{1}{k^2}\left(
        \frac{\sin x}{4 x} 
        - \frac{\sin(u x) \sin(v x)}{4 u v x^2}\right)\ , \label{eq:033I}
\end{eqnarray}
\begin{eqnarray}
    &&I^{(2)}_i(u,v,x)=-\frac{\cos x}{8 u v x}\Bigg[ \mathrm{Si}\left((1+v-u )x \right)  \nonumber\\
    &&~~ -\mathrm{Si}\left((1-u-v)x  \right)- \mathrm{Si}\left(( 1+u+v )x\right) \nonumber\\
    &&~~ +\mathrm{Si}\left((1+u-v )x \right)\Bigg]+\frac{\sin x}{8 u v x}\Bigg[\mathrm{Ci}\left((1+v-u )x \right) \nonumber\\
    &&~~ +\mathrm{Ci}\left((1+u-v )x \right)-\mathrm{Ci}\left(|1-u-v|x \right) \nonumber\\
    &&~~ -\mathrm{Ci}\left((1+u+v )x \right)+\ln \left|\frac{1-(u+v)^2}{1-(u-v)^2}\right| \Bigg] \ , \nonumber\\
    &&~~~~~    (i=2,3,4,5) \ , \label{eq:0ijI}
\end{eqnarray}
where $\mathrm{Si}(x)$ and $\mathrm{Ci}(x)$ are defined as follows
\begin{eqnarray}
    \mathrm{Si}(x)=\int_0^x \mathrm{~d} \bar{x} \frac{\sin \bar{x}}{\bar{x}}  , \  \mathrm{Ci}(x)=-\int_x^{\infty} \mathrm{d} \bar{x} \frac{\cos \bar{x}}{\bar{x}} \ .
\end{eqnarray}
In Eq.~(\ref{eq:033I}) and Eq.~(\ref{eq:0ijI}), we present the results showing that two types of second-order kernel functions for \acp{TIGW} exhibit damped oscillations as $x=|\mathbf{k}|\eta$ increases. In order to evaluate second-order \acp{TIGW} detectable at present, we analyze the asymptotic properties of the second-order kernel function for large values of $x$. Consequently, we make use of the approximations $\lim_{x\to\pm \infty} \mathrm{Si}(x)=\pm \pi/2$ and $\lim_{x\to \infty} \mathrm{Ci}(x)=0$. In the asymptotic limit as $x$ approaches infinity, the second-order kernel function can be approximated as follows
\begin{eqnarray}
        I^{(2)}_3&=&\frac{1}{k^2}
        \frac{\sin x}{4 x} \ , \label{eq:33I}\\
        I^{(2)}_i&=&\frac{1}{8k^2uvx}\Bigg(\ln\left|\frac{1 - (u + v)^2}{1 - (u - v)^2}\right| \sin x -\pi \cos x \nonumber \\
        &\times& \Theta\left(u-|1-v|\right)
        \Bigg) \  , \ (i=2,3,4,5) \ ,\nonumber\\ 
        \label{eq:ijI}
\end{eqnarray}
where $\Theta(x)$ represents the Heaviside theta function. Eq.~(\ref{eq:33I}) and Eq.~(\ref{eq:ijI}) characterize the asymptotic behavior of the kernel function for second-order \acp{TIGW} as $x$ approaches infinity. In the following subsection, we will utilize this result to derive the expression for the energy density spectrum of second-order \acp{TIGW}.

\subsection{Energy density spectrum}\label{sec:2.3}

To derive the energy density spectrum of second-order \acp{TIGW}, it is necessary to calculate the two-point correlation function of the second-order \acp{TIGW} and the corresponding power spectrum. The power spectrum of the $n$-th order gravitational wave $\mathcal{P}_{h}^{(n)}(k)$ is defined as
\begin{equation}\label{eq:Ph}
  \left\langle h^{\lambda,(n)}_{\mathbf{k}}(\eta) h^{\lambda^{\prime},(n)}_{\mathbf{k}^{\prime}}\left(\eta\right)\right\rangle=\delta^{\lambda \lambda^{\prime}} \delta\left(\mathbf{k}+\mathbf{k}^{\prime}\right) \frac{2 \pi^2}{k^3} \mathcal{P}_{h}^{(n)}(k) \ .  
\end{equation}
The total energy density fraction of \acp{GW} up to second order can be written as \cite{Maggiore:1999vm}
\begin{equation}\label{eq:Omega}
	\begin{aligned}
		\Omega_{\mathrm{GW}}^{\mathrm{tot}}(\eta, k)&=\frac{\rho^{(1)}_{\mathrm{GW}}(\eta, k)+\frac{1}{4}\rho^{(2)}_{\mathrm{GW}}(\eta, k)}{\rho_{\mathrm{tot}}(\eta)} \\
        &=\Omega_{\mathrm{GW}}^{(1)}(\eta, k)+\Omega_{\mathrm{GW}}^{(2)}(\eta, k)\\
		&=\frac{1}{6}\left(\frac{k}{\mathcal{H}}\right)^{2} \left( \mathcal{P}^{(1)}_{h}(k)+\frac{1}{4}\mathcal{P}^{(2)}_{h} (k)\right) \ ,
	\end{aligned}
\end{equation}
where $\mathcal{P}^{(1)}_{h}(k)$ and $\mathcal{P}^{(2)}_{h}(k)$ represent the power spectra of first-order tensor perturbation and second-order \acp{TIGW}, respectively. As illustrated in Eq.~(\ref{eq:Ph}) and Eq.~(\ref{eq:Omega}), the total energy density spectra of \acp{GW} can be calculated based on the corresponding two-point correlation functions. By substituting Eq.~(\ref{eq:hhall}) into Eq.~(\ref{eq:Ph}), we obtain
\begin{equation}\label{eq:sumhh}
	\begin{aligned}
    &\delta^{\lambda \lambda^{\prime}} \delta\left(\mathbf{k}+\mathbf{k}^{\prime}\right) \frac{2 \pi^2}{k^3} \mathcal{P}_{h}^{(2)}\\
		&=\sum_{i,j=1}^{5}\int\frac{\mathrm{d}^3p}{(2\pi)^{3/2}}\int\frac{\mathrm{d}^3p'}{(2\pi)^{3/2}}\varepsilon^{\lambda,lm} \left( \mathbf{k}\right) \varepsilon^{\lambda,rs} \left( \mathbf{k}'\right)  \\
&\times ~\mathbb{P}^{\lambda_1\lambda_2}_{lm,i}\mathbb{P}^{\lambda^{'}_1\lambda^{'}_2}_{rs,j}~I^{(2)}_{i}\left(u,v,x\right)I^{(2)}_{j}\left(u',v',x'\right)\\
&\times~ \langle  \mathbf{h}^{\lambda_1}_{\mathbf{k}-\mathbf{p}}\mathbf{h}^{\lambda_2}_{\mathbf{p}}  \mathbf{h}^{\lambda_1^{'}}_{\mathbf{k}'-\mathbf{p}'}\mathbf{h}^{\lambda_2^{'}}_{\mathbf{p}'} \rangle  \ ,
	\end{aligned}
\end{equation}
where $|\mathbf{k}'-\mathbf{p}'|=k'u'$, $|\mathbf{p}'|=k'v'$, and $x'=k'\eta$. The four-point correlation function of \acp{PGW} in Eq.~(\ref{eq:sumhh}) can be expressed, via Wick’s theorem, as a product of two primordial gravitational wave power spectra
\begin{equation}\label{eq:Wick}
	\begin{aligned}
		&\langle \mathbf{h}_{\mathbf{k}-\mathbf{p}}^{\lambda_1}\mathbf{h}_{\mathbf{p}}^{\lambda_2} \mathbf{h}_{\mathbf{k}'-\mathbf{p}'}^{\lambda_1'}\mathbf{h}_{\mathbf{p}'}^{\lambda_2'} \rangle \\
        &=\langle \mathbf{h}_{\mathbf{k}-\mathbf{p}}^{\lambda_1} \mathbf{h}_{\mathbf{k}'-\mathbf{p}'}^{\lambda_1'}  \rangle\langle \mathbf{h}_{\mathbf{p}}^{\lambda_2}\mathbf{h}_{\mathbf{p}'}^{\lambda_2'} \rangle +\langle \mathbf{h}_{\mathbf{k}-\mathbf{p}}^{\lambda_1}\mathbf{h}_{\mathbf{p}'}^{\lambda_2'} \rangle \langle \mathbf{h}_{\mathbf{p}}^{\lambda_2} \mathbf{h}_{\mathbf{k}'-\mathbf{p}'}^{\lambda_1'} \rangle \\
		&=\delta\left(\mathbf{k}+\mathbf{k}'\right) \frac{(2\pi^2)^2}{p^3|\mathbf{k}-\mathbf{p}|^3} \left(\delta^{\lambda_1\lambda'_1}\delta^{\lambda_2\lambda'_2}\delta\left(\mathbf{p}+\mathbf{p}'\right) \right.\\
&~~\left.+~\delta^{\lambda_1\lambda'_2}\delta^{\lambda_2\lambda'_1}\delta\left(\mathbf{k}'-\mathbf{p}'+\mathbf{p}\right) \right)\mathcal{P}_{h}(uk)\mathcal{P}_{h}(vk) \ ,
	\end{aligned}
\end{equation}
where $\mathcal{P}_h(k)$ is the power spectrum of \acp{PGW}. By substituting Eq.~(\ref{eq:Pp1})--Eq.~(\ref{eq:Pp5}) and Eq.~(\ref{eq:Wick}) into Eq.~(\ref{eq:Ph}), we can derive the expression for the power spectrum of second-order \acp{TIGW}. Substituting the resulting power spectrum into Eq.~(\ref{eq:Omega}) and simplifying yields the expression for the energy density spectrum of second-order \acp{TIGW} during the \ac{RD} era
\begin{equation}
    \begin{aligned}
        &\Omega_{\mathrm{GW}}^{(2)}(k)= \int_{0}^{\infty} \mathrm{d}v \int_{|1-v|}^{1+v} \mathrm{d}u\, 
        \mathcal{P}_h(uk) \mathcal{P}_h(vk) \\
        &~~\times \frac{1}{3145728\, u^8 v^8} 
        \Big((u - v)^2 - 1\Big)^2 \Big((u + v)^2 - 1\Big)^2 \\
        &~~\times \Bigg[
        64 u^2 v^2 \Big(1 + u^4 + v^4 + 6 (u^2 + v^2) + 6 u^2 v^2\Big) \\
        &~~ + 16 u v \Big(1 + u^6 + v^6 + 15 (u^4 + v^4) + 15 \left(u^2 \right. \\
        &~~\left. + v^2\right)+ 15 u^2 v^2 (u^2 + v^2 + 6) \Big)
        \ln \left| \frac{1 - (u + v)^2}{1 - (u - v)^2} \right| \\
        &~~\quad + \Big(
            1 - 7 u^8 + 4 v^2 + 126 v^4 + 116 v^6 + 9 v^8 \\
        &~~\quad - 12 u^6 (5 + 7 v^2)
            + 2 u^4 (7 + 118 v^2 + 35 v^4) \\
        &~~\quad + 4 u^2 (13 + 105 v^2 + 151 v^4 + 35 v^6)
        \Big) \\
        &~~\quad \times \left( \pi^2 + \ln^2 \left| \frac{1 - (u + v)^2}{1 - (u - v)^2} \right| \right)
        \Bigg] \ ,
    \end{aligned}
    \label{eq:1Ott}
\end{equation}
where $\mathcal{P}_h(k)$ is the power spectrum of \acp{PGW}. In Eq.~(\ref{eq:1Ott}), we have performed an oscillatory average of the squared kernel function using the relations $\sin^2 x\sim 1/2$ and $\cos^2 x\sim 1/2$ \cite{Yuan:2021qgz}. Similarly, by substituting Eq.~(\ref{eq:Th1}) into Eq.~(\ref{eq:Ph}) and Eq.~(\ref{eq:Omega}), and applying an oscillatory average, we obtain the energy density spectrum of first-order gravitational waves
\begin{eqnarray}\label{eq:1pk}
    \Omega_{\mathrm{GW}}^{(1)}(k)=\frac{x^2}{6}\mathcal{P}_h^{(1)}\left( k,\eta \right)=\frac{1}{12} \mathcal{P}_h(k) \ .
\end{eqnarray}
Given a specific form of the power spectrum of \acp{PGW} $\mathcal{P}_h(k)$, we can use Eq.~(\ref{eq:1Ott}), Eq.~(\ref{eq:1pk}) and Eq.~(\ref{eq:Omega}) to calculate the total energy density spectrum of gravitational waves. Furthermore, Eq.~(\ref{eq:Omega}) provides the energy density spectrum of \acp{PGW}$+$\acp{TIGW} during the \ac{RD} era. Taking into account the thermal history of the universe, the current total energy density spectrum $\bar{\Omega}^{(2)}_{\mathrm{GW,0}}(k)$ can be expressed as
\begin{equation}
    \Omega^{\mathrm{tot}}_{\mathrm{GW,0}}(k) = \Omega_{\mathrm{rad},0}\left(\frac{g_{*,\rho,e}}{g_{*,\rho,0}}\right)\left(\frac{g_{*,s,0}}{g_{*,s,e}}\right)^{4/3}\Omega^{\mathrm{tot}}_{\mathrm{GW}}(k) \ ,
\end{equation}
where $\Omega_{\mathrm{rad},0}$ ($ =4.2\times 10^{-5}h^{-2}$) is the energy density fraction of radiation today.

\section{Large \acp{PGW} on small scales}\label{sec:3.0}
In this section, we investigate how to construct \acp{PGW} with large amplitudes on small scales. To address this question, we begin with the equation of motion of gravitational waves. In general relativity, the standard form of the equation of motion of \acp{PGW} can be expressed as
\begin{eqnarray}\label{eq:Pgw0}
    h_{ij}'' +2 \mathcal{H} h_{ij}'+k^2h_{ij}=0 \ .
\end{eqnarray}
To generate large-amplitude \acp{PGW} on small scales, it is necessary to modify Eq.~(\ref{eq:Pgw0}). There are two possible approaches: (1) artificially introduce additional source terms, such as spectator fields \cite{Gorji:2023ziy,Gorji:2023sil,Biagetti:2013kwa,Fu:2023aab}; and (2) modify the dynamical terms in the equation of motion, for example by modified gravity or invoking sound speed resonance \cite{Cai:2021uup,Jiang:2024woi,Cai:2020ovp,Addazi:2024gew,Guzzetti:2016mkm}. 

In this section, we focus on the models of small-scale \acp{PGW} presented in Refs.~\cite{Fu:2023aab,Cai:2021uup,Jiang:2024woi}. We will calculate the energy density spectrum of \acp{PGW} on small scales under different models, and analyze the modifications to the total energy density spectrum caused by second-order \acp{TIGW} under various model parameters. Specifically, we consider \ac{nymtg}, where the Lagrangian takes the form \cite{Li:2020xjt,Li:2021wij}
\begin{align}
    \lag & = \lag_{\rm Einstein} + \lag_{\rm NY} + \lag_{\rm other}\nonumber\\
    & = \frac{R}{2\kappa} + \frac{\alpha}{4}\phi\mathcal T^{A\mu\nu}(\star\mathcal T)_{A\mu\nu} + \lag_{\rm other} \ , \label{eq:Lamg}
\end{align}
where $\mathcal T^A{}_{\mu\nu}$ is the torsion two-form, $\star$ represents the Hodge dual, and $(\star\mathcal T)_{A\mu\nu} = 1/2\epsilon_{\mu\nu\rho\sigma}\mathcal T_A{}^{\rho\sigma}$. The parameter $\alpha$ is the coupling constant, $\phi$ is the field emerging as a dynamical scalar field. Other fields, including $\phi$, are minimally coupled to gravity, and their Lagrangians are given by $\lag_{\rm other}$. In Eq.~(\ref{eq:Lamg}), the Nieh–Yan term does not modify the equations of motion for the cosmological background spacetime, but only alters the form of the cosmological perturbation equations. In \ac{nymtg} model, the equation of motion of \acp{PGW} can be expressed as \cite{Fu:2023aab}
\begin{equation}
    h_{ij}'' + 2 \mathcal{H} h_{ij}' + (k^2 + k\alpha A_\lambda\phi')h_{ij}= 0 \ ,
\end{equation}
where the polarization index $\lambda = L$ and $\lambda =R$ denote left and right polarization, respectively, with $A_L = -1$ and $A_R = 1$. During inflation, if the effective mass $\omega^2 = k^2 + k\alpha A\phi'$ becomes negative, the corresponding mode would undergo a tachyonic instability, thus amplifying the \acp{PGW}. In the following subsections, we shall consider the specific forms of $\lag_{\rm other}$ and calculate the corresponding small-scale power spectra of \acp{PGW}.

\subsection{Model 1}\label{sec:M1}
\begin{figure}[htbp]
\centering
\includegraphics[width=\linewidth]{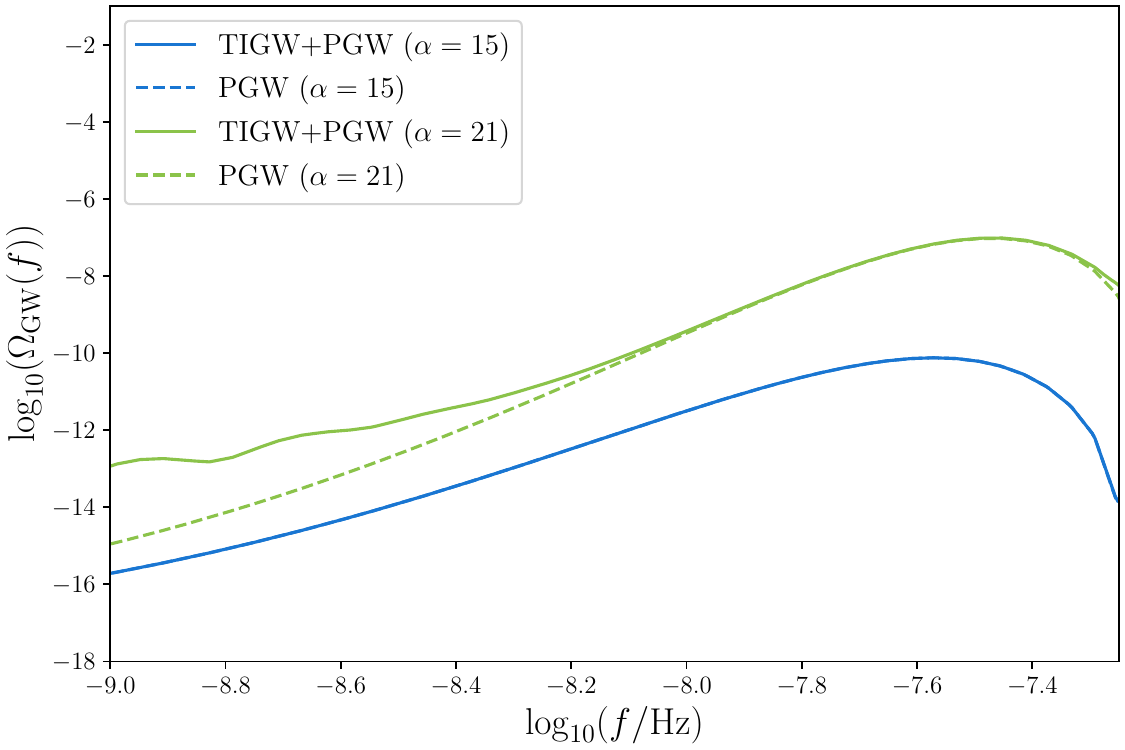}
\caption{Energy density spectra in Model 1. Dashed curves represent \acp{PGW}, and solid curves represent the total contribution from \acp{PGW} and second-order \acp{TIGW}. The variation in spectra corresponds to different values of the coupling constant $\alpha$, as indicated in the legend. } \label{fig:spectrum_model1}
\end{figure}

We consider the following simple scalar Lagrangian
\begin{equation}\label{eq:other1}
    \lag_{\rm other} = -\nabla_\mu\phi\nabla^\mu\phi - V(\phi)  \ ,
\end{equation}
where \cite{Cai:2021uup}
\begin{equation}\label{eq:other2}
    V(\phi) = \frac12 m^2\phi^2 + \Lambda^4 \frac{\phi}{f}\sin\left(\frac{\phi}{f}\right) \ ,
\end{equation}
with
\begin{equation}
    m = 10^{-6} \ ,\Lambda = 5.512\times 10^{-4} \ , f = 0.3044 \ . 
\end{equation}
Fig.~\ref{fig:spectrum_model1} presents the energy density spectra of \acp{PGW} and \acp{PGW}$+$\acp{TIGW} for different values of the parameter $\alpha$ in Eq.~(\ref{eq:Lamg}). Since the energy density spectrum of \acp{PGW} $\Omega_{\mathrm{GW}}^{(1)}$ is proportional to $A_h$, while the second order energy density spectrum of second-order \acp{TIGW} $\Omega_{\mathrm{GW}}^{(2)}$ is proportional to $A_h^{2}$, noticeable modifications from \acp{TIGW} to the total energy density spectrum only emerge when the amplitude of the power spectrum of \acp{PGW} $A_h$ is sufficiently large. Moreover, for Model 1, the correction to the total energy density spectrum from second-order \acp{TIGW} is primarily concentrated in the low-frequency regime.

\subsection{Model 2}
\begin{figure}[htbp]
\centering
\includegraphics[width=\linewidth]{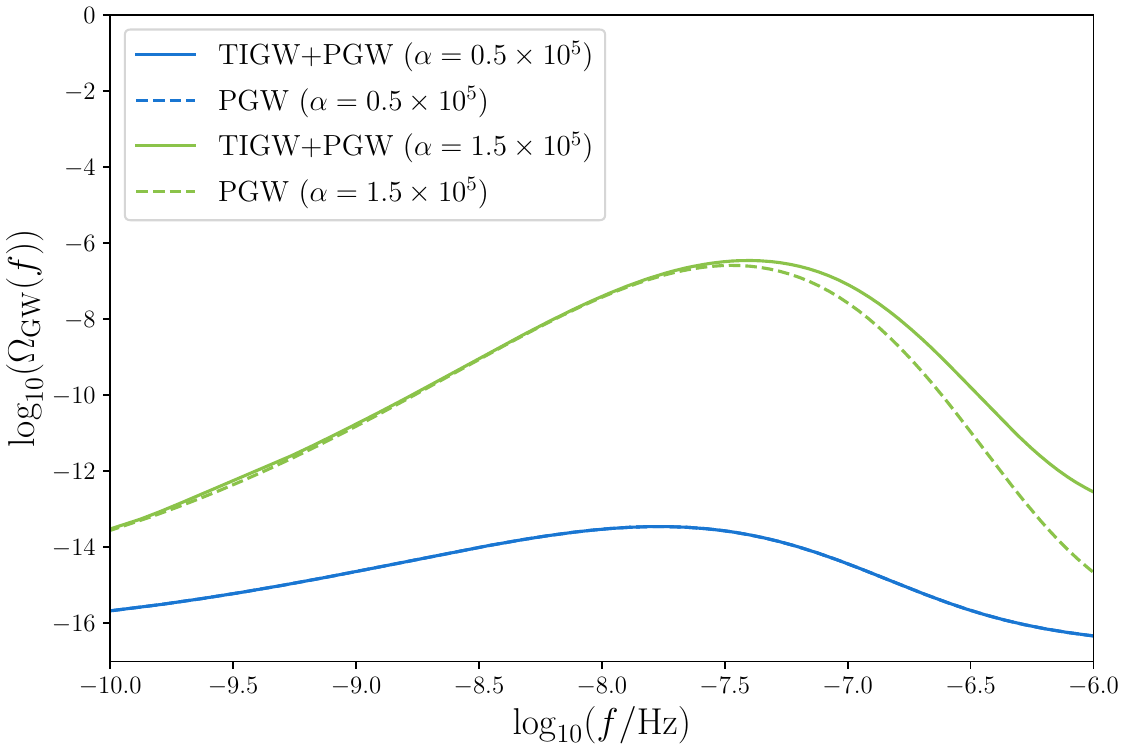}
\caption{Energy density spectra in Model 2. Dashed curves represent \acp{PGW}, and solid curves represent the total contribution from \acp{PGW} and second-order \acp{TIGW}. The variation in spectra corresponds to different values of the coupling constant $\alpha$, as indicated in the legend. } \label{fig:spectrum_model2}
\end{figure}

In Model 1, we consider the Lagrangian $\lag_{\rm other}$ as given in Eq.~(\ref{eq:other1}). In addition to the scalar field $\phi$, an extra scalar field $\varphi$ can also be incorporated into $\lag_{\rm other}$, resulting in the following Lagrangian \cite{Fu:2023aab}
\begin{equation}\label{eq:other3}
    \lag_{\rm other} = -\nabla_\mu\phi\nabla^\mu\phi - \nabla_\mu\varphi\nabla^\mu\varphi - V_\phi(\phi) - V_\varphi(\varphi) \ ,
\end{equation}
where the potentials are given by
\begin{align}
    V_\phi(\phi) & = \frac12m^2\phi^2 + \frac{\Lambda^4}{\sigma}\sin\left(\frac{\phi}{\sigma}\right),\\
    V_\varphi(\varphi) & = V_0\left[1 - \exp(-\sqrt{2/3}\varphi)\right]^2,
\end{align}
with
\begin{gather}
    V_0 = 9.75 \times 10^{-11}, \, \sigma = 0.0002, \,\nonumber\\
    m = \sqrt{0.16 V_0}, \, \Lambda = 2.76 \times 10^{-5}.
\end{gather}
As shown in Fig.~\ref{fig:spectrum_model2}, Model 2 differs from Model 1 in that the correction to the total energy density spectrum from second-order \acp{TIGW} is mainly concentrated in the high-frequency region

\subsection{Model 3}
\begin{figure}[htbp]
\centering
\includegraphics[width=\linewidth]{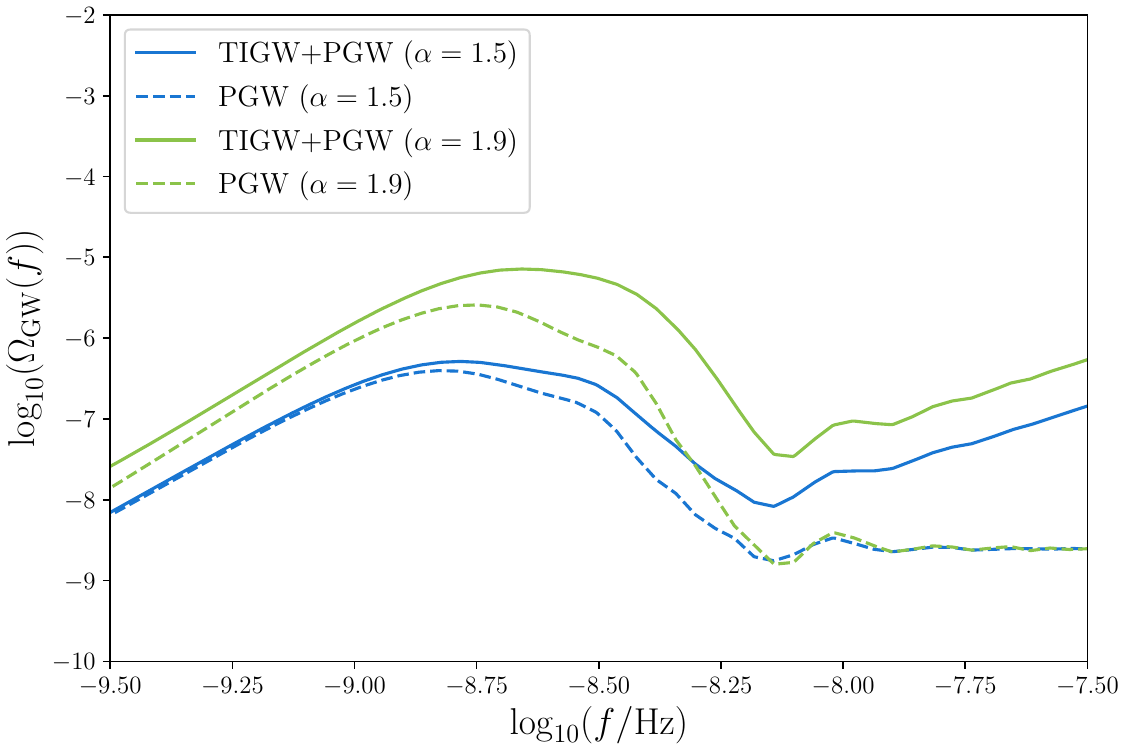}
\caption{Energy density spectra in Model 3. Dashed curves represent \acp{PGW}, and solid curves represent the total contribution from \acp{PGW} and second-order \acp{TIGW}. The variation in spectra corresponds to different values of the coupling constant $\alpha$, as indicated in the legend. } \label{fig:spectrum_model3}
\end{figure}

We consider two scalar fields $\phi$ and $\varphi$, with the Lagrangian given by \cite{Jiang:2024woi}
\begin{align}
    \lag_{\rm other} & = -\frac{g_1(\phi)}{2}\nabla_\mu\phi\nabla^\mu\phi + \frac{g_2(\phi)}{4}(\nabla_\mu\phi\nabla^\mu\phi)^2\nonumber\\
    & \qquad - V_\phi(\phi) - \frac12\nabla_\mu\varphi\nabla^\mu\varphi - V_\varphi(\varphi) \ ,
\end{align}
where
\begin{align}
    g_1(\phi) & = \frac{2}{1 + e^{-q_1(\phi - \phi_0)}} + \frac{1}{1 + e^{q_2(\phi - \phi_3)}}\nonumber\\
    & \qquad - \frac{f_1 e^{2\phi}}{1 + f_1e^{2\phi}}\ , \\
    g_2(\phi) & = \frac{f_2}{1 + e^{-q_2(\phi - \phi_3  )}}\frac{1}{1 + e^{q_3(\phi - \phi_0)}}\ , \\
    V_\phi(\phi) & = \frac12 m^2\phi^2\frac{1}{1 + e^{q_2(\phi - \phi_2)}}\nonumber\\
    &  + \lambda\left[1 - \frac{(\phi - \phi_2)^2}{\sigma^2}\right]^2\frac{1}{1 + e^{-q_4(\phi - \phi_1)}} \ ,
\end{align}
with the parameters chosen as
\begin{gather}
    q_1 = 10, \, q_2 = 6, \, q_3 = 10, \, q_4 = 4, \, f_1 = 1, \nonumber\\
    f_2 = 40, \,\lambda = 0.01, \, m = 4.5\times 10^{-6}.
\end{gather}
The potential $V_\varphi(\varphi)$ has the same form as \eqref{eq:other2}, with the parameters given by
\begin{equation}
    m = 3.32 \times 10^{-6}, \, \Lambda = 5.761 \times 10^{-6}, \, f = 10^{-5}.
\end{equation}
The corresponding energy density spectra of PGW and PGW+TIGW are shown in Fig.~\ref{fig:spectrum_model3}. For Model 3, the amplitude of the \acp{PGW} is significantly larger than in the previous two models, resulting in a more substantial correction from second-order \acp{TIGW}.

\subsection{Model 4}
\begin{figure}[htbp]
\centering
\includegraphics[width=\linewidth]{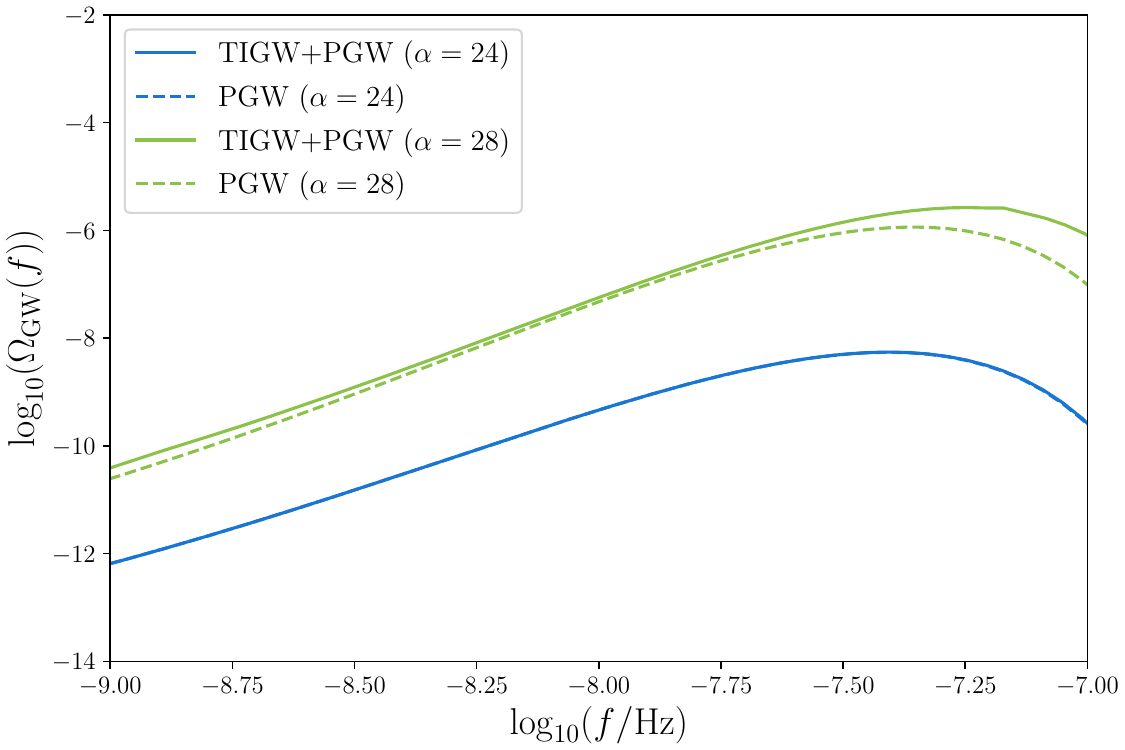}
\caption{Energy density spectra in Model 4. Dashed curves represent \acp{PGW}, and solid curves represent the total contribution from \acp{PGW} and second-order \acp{TIGW}. The variation in spectra corresponds to different values of the coupling constant $\alpha$, as indicated in the legend. } \label{fig:spectrum_model4}
\end{figure}

We consider the same form of the Lagrangian as in Sec.~\ref{sec:M1}, with the potential replaced by \cite{Fu:2023aab}
\begin{equation}
    V(\phi) = \begin{cases}
        V_0 + C_+(\phi - \phi_0) \ , & \qquad \phi > \phi_0 \ , \\
        V_0 + C_-(\phi - \phi_0) \ , & \qquad \phi \leqslant\phi_0  \ ,
    \end{cases}.
\end{equation}
and we fix the parameters as
\begin{equation}
    V_0 = 10^{-14}\ , C_+ = 10^{-14} \ , C_- = 10^{-15} \ , \phi_0 = 6\ .
\end{equation}
As shown in Fig.~\ref{fig:spectrum_model4}, in the model considered here, the contribution from second-order \acp{TIGW} is primarily concentrated in the high-frequency region, with negligible effects in the low-frequency region. Furthermore, compared with the other three models, the second-order \acp{TIGW} introduces relatively minor corrections to the energy density spectrum in this model.

\section{Detection of \acp{TIGW}}\label{sec:4.0}
In Sec.~\ref{sec:2.0} and Sec.~\ref{sec:3.0}, we systematically investigated various models that generate large-amplitude \acp{PGW} on small scales, as well as the corrections to the total energy density spectrum of second-order \acp{TIGW}. In this section, we examine the constraints imposed on the parameter space of the power spectrum of \acp{PGW} by current observations. Current cosmological observations impose three types of constraints on the small-scale \ac{PGW}:

\noindent
(1) Upper limits on the energy density inferred from large-scale observations including \ac{CMB}, \ac{BAO}, and \ac{BBN} \cite{Zhou:2024yke,Cang:2022jyc,Ben-Dayan:2019gll,Wang:2023sij}.

\noindent
(2) Direct detections of \ac{SGWB} across different frequency bands from experiments such as \ac{PTA}, \ac{LISA}, and Taiji \cite{NANOGrav:2023hvm,LISACosmologyWorkingGroup:2022jok,Iacconi:2024hmg,Flauger:2020qyi,TaijiScientific:2021qgx}.

\noindent
(3) \acp{PBH} formed by second-order induced density perturbations impose constraints on small-scale \acp{PGW} \cite{Nakama:2016enz,Zhou:2023itl,Nakama:2015nea}.  

\noindent
Furthermore, for \acp{SIGW}, the power spectrum of primordial curvature perturbations $\mathcal{P}_{\zeta}(k)$ can be further constrained through observations such as $\mu$-distortions, which in turn impose limits on the energy density spectrum of \acp{SIGW} \cite{Chluba:2013dna,Bringmann:2025cht,Zhou:2025djn,Tagliazucchi:2023dai,Chluba:2012we,Chluba:2012gq,Sharma:2024img,Byrnes:2024vjt}.

\subsection{Constraints from large-scale cosmological observations}\label{sec:4.1}
There are two methods by which large-scale cosmological observations constrain \ac{SGWB}. One approach treats the \ac{SGWB} as an additional radiation component, thereby affecting the effective number of relativistic species $N_{\mathrm{eff}}$. Since the impact of the \ac{SGWB} on $N_{\mathrm{eff}}$ must not exceed the current observational uncertainty in $N_{\mathrm{eff}}$, the total energy density spectrum of the background must satisfy the following condition \cite{Ben-Dayan:2019gll,Cang:2022jyc,Wang:2023sij,Zhou:2024yke}
\begin{eqnarray}\label{eq:rhup1}
  \int_{f_{\mathrm{min}}}^{\infty} h^2\Omega_{\mathrm{GW},0}(k) \mathrm{d}\left(\ln k\right) < 1.3 \times 10^{-6}\frac{\Delta N_{\text{eff}}}{0.234} \   ,
\end{eqnarray}
 where $\Delta N_{\text{eff}}= N_{\text{eff}}-3.046$. Here, we use the $N_{\text{eff}}$ limits provided by \cite{Planck:2018vyg}, which report $N_{\text{eff}}=3.04 \pm 0.22$ at a $95\%$ confidence level for the \texttt{ Planck} + \ac{BAO} + \ac{BBN} data. By applying Gaussian statistics, this translates to a $95\%$ confidence level upper limit of $\Delta N_{\text{eff}}<0.175$ \cite{Cang:2022jyc}. Another method involves using observational data from the \ac{CMB} and \ac{BAO}, which requires the energy density spectrum of the \ac{SGWB} to fulfill the condition 
\begin{equation}\label{eq:rhup2}
\int_{f_{\mathrm{min}}}^{\infty} h^2\Omega_{\mathrm{GW},0}(k) \mathrm{d}\left(\ln k\right) < 2.9\times 10^{-7}  \ ,
\end{equation}
at $95\%$ confidence level for \ac{CMB}$+$\ac{BAO} data \cite{Clarke:2020bil}. It should be noted that the large-scale cosmological observation constraints in Eq.~(\ref{eq:rhup2}) are stronger than the constraints obtained from the relativistic degrees of freedom $N_{\text{eff}}$ in Eq.~(\ref{eq:rhup1}). In this paper, we will jointly consider the constraints imposed by the two aforementioned methods on the parameter space of the energy density spectrum of gravitational wave. The constraints from large-scale cosmological observations on the parameter space of different models will be presented in the next subsection.

\subsection{PTA observations}\label{sec:4.2}
In this section, we focus on the observations of \acp{PGW} and second-order \acp{TIGW} in the \ac{PTA} frequency band, as well as the constraints on the energy density of gravitational wave from large-scale cosmological observations. More precisely, for \ac{PTA} observation, we employ the \ac{KDE} representations of the free spectra to construct the likelihood function \cite{Lamb:2023jls,Moore:2021ibq,Mitridate:2023oar}
\begin{equation} \label{eq:likelihood}
    \ln \mathcal{L}(d|\theta) = \sum_{i=1}^{N_f} p(\Phi_i,\theta)\ .
\end{equation}
In Eq.~(\ref{eq:likelihood}), $p(\Phi_i,\theta)$ represents the probability of $\Phi_i$ given the parameter $\theta$, and $\Phi_i = \Phi(f_i)$ denotes the time delay
\begin{equation} \label{eq:timedelay}
    \Phi(f) = \sqrt{\frac{H_0^2 \Omega_{\mathrm{GW}}(f)}{8\pi^2 f^5 T_{\mathrm{obs}}}} \ ,
\end{equation}
where $H_0=h\times 100 \mathrm{km/s/Mpc}$ is the present-day value of the Hubble constant. In this study, we employ the kernel density estimate (KDE) representation of the HD-correlated free spectrum extracted from the first 14 frequency components of the NANOGrav 15-year dataset \cite{Nanograv:KDE}. And the Bayesian analysis is performed via \textsc{bilby} \cite{bilby_paper} using its integrated \textsc{dynesty} nested sampler \cite{Speagle:2019ivv,dynesty_software}. To rigorously evaluate the viability of competing explanations for the current PTA signal, we investigate a hybrid model scenario where both \acp{SMBHB} and \acp{TIGW} contribute jointly. The energy density spectrum of SMBHBs is characterized by \cite{NANOGrav:2023hvm}
\begin{equation} \label{eq:SMBHB}
    \Omega_{\mathrm{GW}}^{\mathrm{BH}}(f) = \frac{2\pi^2 A_{\mathrm{BHB}}^2}{3H_0^2 h^2} (\frac{f}{\mathrm{year}^{-1}})^{5-\gamma_{\mathrm{BHB}}}\mathrm{year}^{-2} \ ,
\end{equation}
with the prior distribution for $\log_{10}A_{\mathrm{BHB}}$ assumed to follow a multivariate Gaussian distribution \cite{NANOGrav:2023hvm}
\begin{equation} \label{eq:prior_SMBHB}
\begin{aligned}
    \boldsymbol{\mu}_{\mathrm{BHB}}&=\begin{pmatrix} -15.6
 \\ 4.7 \end{pmatrix} \ , \\
\boldsymbol{\sigma}_{\mathrm{BHB}}&=0.1\times \begin{pmatrix}
2.8  & -0.026\\
-0.026  & 2.8
\end{pmatrix} \ .
\end{aligned}
\end{equation} 
As shown in Fig.~\ref{fig:corner_PGW_model1} $\sim$ Fig.~\ref{fig:corner_TIGW_model4}, we present the posterior distributions of parameters for four different models as determined by current \ac{PTA} observations. The blue curves represent the posterior distributions that take into account the influence of \ac{SMBHB}. The constraints on the parameter space of the \ac{SGWB} from large-scale cosmological observations discussed in Sec.~\ref{sec:4.1} are represented by the black solid and grey dashed lines. Furthermore, to better assess the plausibility of different models in explaining current \ac{PTA} observations, we perform a detailed analysis of Bayes factors between competing models. The Bayes factor is defined as $B_{i,j} = \frac{Z_i}{Z_j}$, where $Z_i$ represents the evidence of model $H_i$. Fig.~\ref{fig:bayes} presents the Bayes factors corresponding to different models. In the following, we analyze the results of the four models introduced in Sec.~\ref{sec:3.0} individually.

\subsubsection{Model 1}
\begin{figure}[htbp]
\centering
\includegraphics[width=\linewidth]{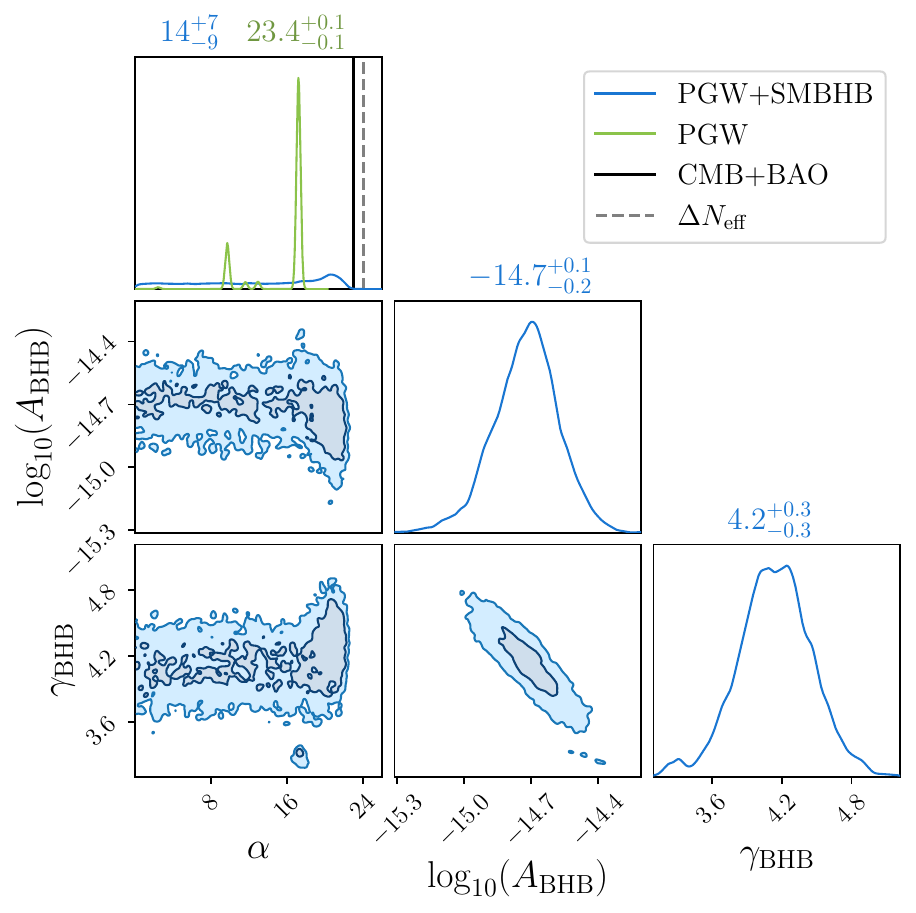}
\caption{The corner plot of the posterior distributions for model 1. The contours in the off-diagonal panels denote the $68\% $ and $95 \%$ credible intervals of the 2D posteriors. The numbers above the figures represent the median values and $1$-$\sigma$ ranges of the parameters. The blue and green solid curves correspond to the \acp{PGW} energy spectrum with or without \ac{SMBHB}. The black solid line and grey dashed line denote the upper bounds from \ac{CMB} and \ac{BAO} observations in Eq.~(\ref{eq:rhup2}) and $\Delta N_{\mathrm{eff}}$ in Eq.~(\ref{eq:rhup1}).} \label{fig:corner_PGW_model1}
\end{figure}

\begin{figure}[htbp]
\centering
\includegraphics[width=\linewidth]{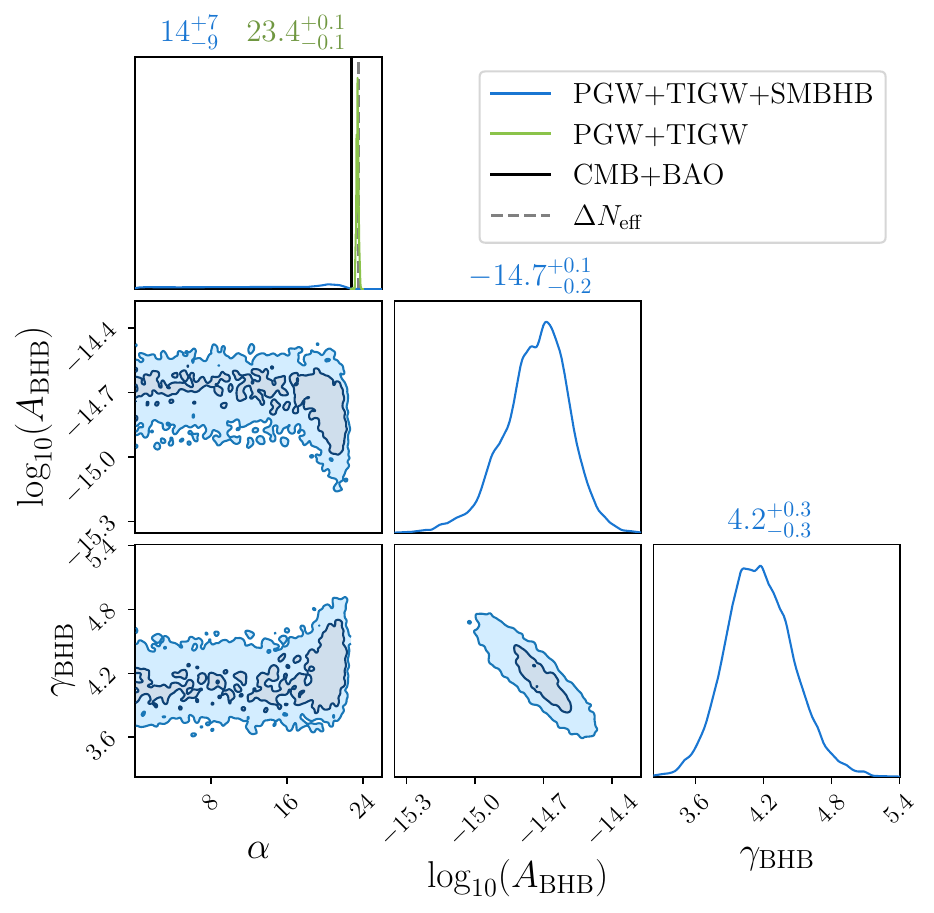}
\caption{The corner plot of the posterior distributions for model 1. The contours in the off-diagonal panels denote the $68\% $ and $95 \%$ credible intervals of the 2D posteriors. The numbers above the figures represent the median values and $1$-$\sigma$ ranges of the parameters. The blue and green solid curves correspond to the \acp{PGW}$+$\acp{TIGW} energy spectrum with or without \ac{SMBHB}. The black solid line and grey dashed line denote the upper bounds from \ac{CMB} and \ac{BAO} observations in Eq.~(\ref{eq:rhup2}) and $\Delta N_{\mathrm{eff}}$ in Eq.~(\ref{eq:rhup1}).} \label{fig:corner_TIGW_model1}
\end{figure}

Fig.~\ref{fig:corner_PGW_model1} presents the posterior distributions of \ac{PGW} and \ac{PGW}+\ac{SMBHB}, along with constraints from large-scale cosmological observations. The prior distribution of $\alpha$ is set as a uniform distribution over the interval $[0, 26]$. When only \acp{PGW} are considered, \ac{PTA} data yield the median value of the
posterior distribution of the parameter $\alpha=23.4$. The large-scale cosmological constraints require that $\alpha$ lies to the left of the black solid and grey dashed lines in the figure. As shown in Fig.~\ref{fig:corner_PGW_model1}, when the effects of second-order \acp{TIGW} are neglected, Model 1 can consistent with large-scale cosmological  constraints. However, as illustrated in Fig.~\ref{fig:corner_TIGW_model1}, the peak of the posterior distribution of $\alpha$ inferred from \ac{PTA} data now lies to the right of the black line, indicating that \acp{PGW}+\acp{TIGW} in Model 1 cannot simultaneously satisfy both \ac{PTA} observations and large-scale cosmological constraints. Additionally, blue curves in Fig.~\ref{fig:corner_PGW_model1} and Fig.~\ref{fig:corner_TIGW_model1} show the posterior distribution under the influence of \ac{SMBHB}. Only when the influence of \ac{SMBHB} is taken into account can Model 1 simultaneously comply with large-scale cosmological constraints and jointly dominate the current \ac{PTA} observations with \ac{SMBHB}.

Fig.~\ref{fig:bayes} shows the Bayes factors between different models. The magnitude of the Bayes factor serves as an important criterion for assessing whether a \ac{SGWB} model fits current \ac{PTA} observations. Without considering \ac{SMBHB}, Model 1 yields a Bayes factor of $1.29\times 10^{-5}$ for \acp{PGW}+\acp{TIGW}. Even when \acp{PGW}+\acp{TIGW} and \ac{SMBHB} are assumed to jointly dominate the \ac{PTA} signal, the corresponding Bayes factor is only $1.11$. Thus, Model 1 does not provide a satisfactory fit to current \ac{PTA} data.

\subsubsection{Model 2}
\begin{figure}[htbp]
\centering
\includegraphics[width=\linewidth]{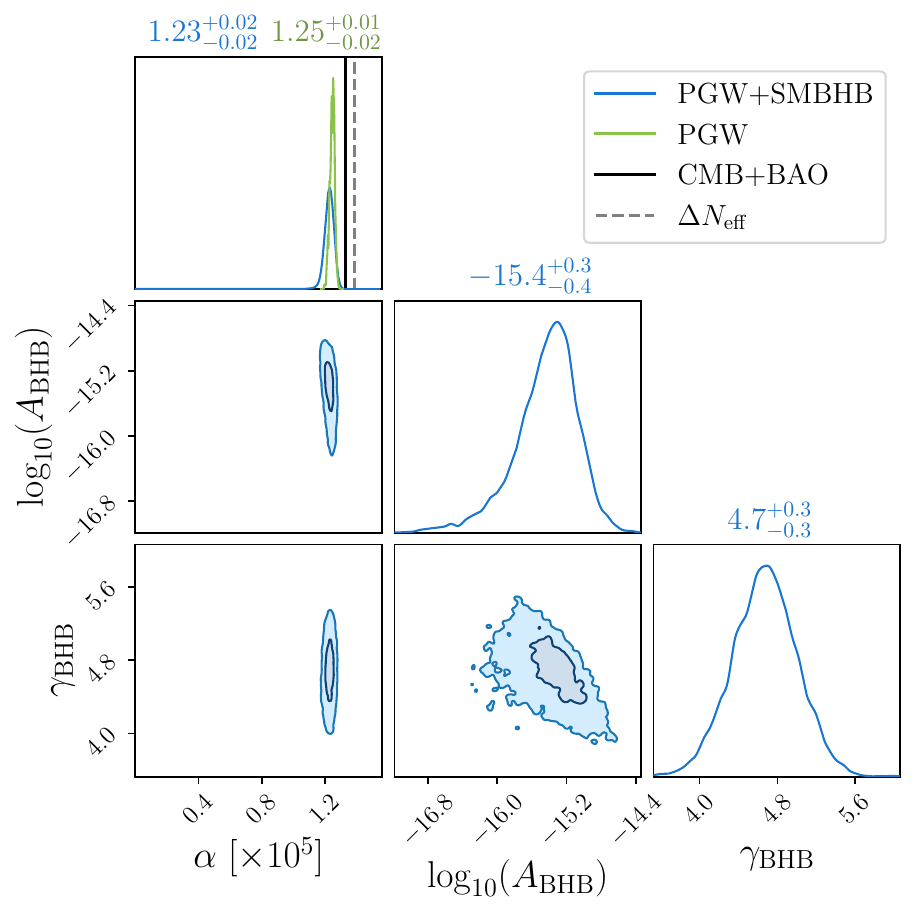}
\caption{The corner plot of the posterior distributions for model 2. The contours in the off-diagonal panels denote the $68\% $ and $95 \%$ credible intervals of the 2D posteriors. The numbers above the figures represent the median values and $1$-$\sigma$ ranges of the parameters. The blue and green solid curves correspond to the \acp{PGW} energy spectrum with or without \ac{SMBHB}. The black solid line and grey dashed line denote the upper bounds from \ac{CMB} and \ac{BAO} observations in Eq.~(\ref{eq:rhup2}) and $\Delta N_{\mathrm{eff}}$ in Eq.~(\ref{eq:rhup1}).} \label{fig:corner_PGW_model2}
\end{figure}

\begin{figure}[htbp]
\centering
\includegraphics[width=\linewidth]{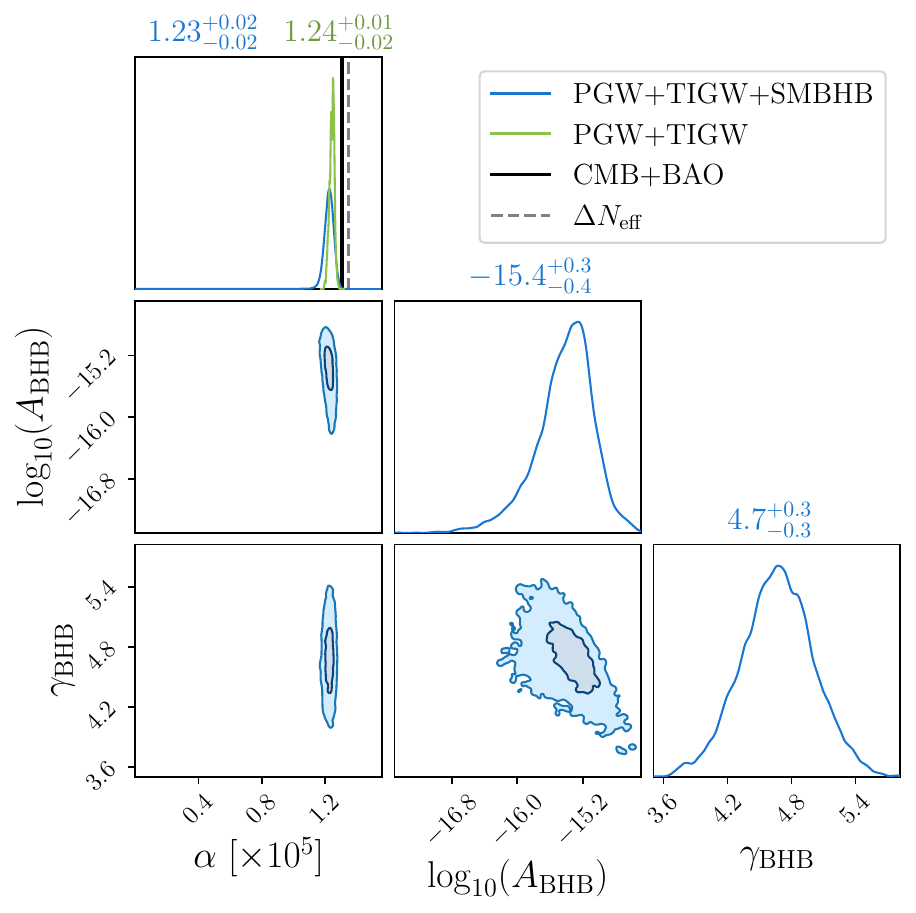}
\caption{The corner plot of the posterior distributions for model 2. The contours in the off-diagonal panels denote the $68\% $ and $95 \%$ credible intervals of the 2D posteriors. The numbers above the figures represent the median values and $1$-$\sigma$ ranges of the parameters. The blue and green solid curves correspond to the \acp{PGW}$+$\acp{TIGW} energy spectrum with or without \ac{SMBHB}. The black solid line and grey dashed line denote the upper bounds from \ac{CMB} and \ac{BAO} observations in Eq.~(\ref{eq:rhup2}) and $\Delta N_{\mathrm{eff}}$ in Eq.~(\ref{eq:rhup1}).} \label{fig:corner_TIGW_model2}
\end{figure}

For Model 2, the posterior distributions derived from current \ac{PTA} observations are shown in Fig.~\ref{fig:corner_PGW_model2} and Fig.~\ref{fig:corner_TIGW_model2}. The prior distribution of $\alpha$ is set as a uniform distribution over the interval $[0, 1.56\times 10^{5}]$. Unlike Model 1, the posterior distribution of parameter $\alpha$ in Model 2 lies entirely to the left of the black solid line, indicating that Model 2 can dominate the current \ac{PTA} signal while remaining consistent with large-scale cosmological observations, regardless of whether \ac{SMBHB} are considered. In addition, the green curves in Fig.~\ref{fig:corner_PGW_model2} and Fig.~\ref{fig:corner_TIGW_model2} identify the values of parameter $\alpha$ as $1.25$ and $1.24$, respectively, highlighting the influence of second-order \acp{TIGW} on the \ac{PTA} observations. As the energy density spectrum of \acp{PGW} increases with larger values of $\alpha$, the inclusion of \acp{TIGW} allows for sufficient \ac{SGWB} to match \ac{PTA} observations with relatively smaller $\alpha$ values.


\begin{figure}[htbp]
\centering
\includegraphics[width=\linewidth]{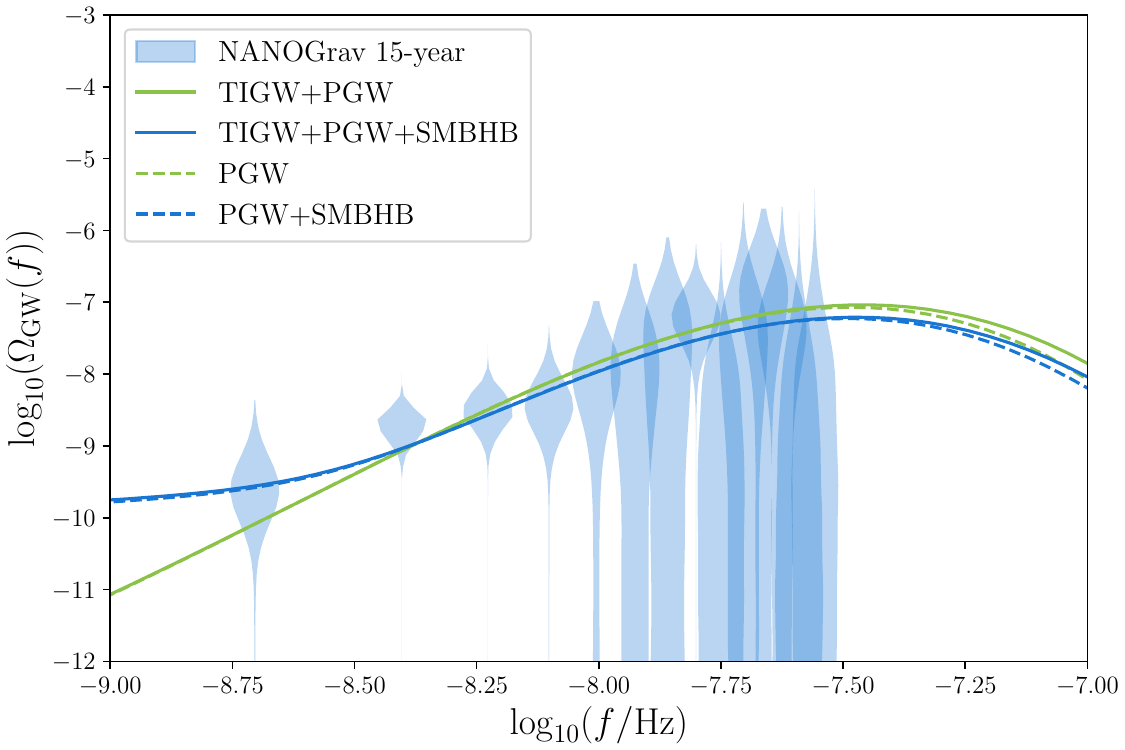}
\caption{The energy density spectra of \acp{PGW} and \acp{PGW}$+$\acp{TIGW} with or without \acp{SMBHB} for model 2. The energy density spectra derived from the free spectrum of the NANOGrav 15-year are shown in blue. The blue and green curves represent the energy density spectra of \acp{GW} with different line styles labeled in the figure. These parameters are selected based on the median values of the posterior distributions.} \label{fig:violinplot_model2}
\end{figure}

Furthermore, as shown in Fig.~\ref{fig:bayes}, the Bayes factors for Model 2 are $63.94$ and $16.83$ with and without considering \ac{SMBHB}, respectively. This indicates that the \ac{SGWB} generated by \ac{PGW}+\ac{TIGW} in Model 2 is more likely to dominate the current \ac{PTA} observations than that produced by \ac{SMBHB}. Since Model 2 provides a good fit to the current \ac{PTA} observational data, the corresponding energy density spectrum is presented in Fig.~\ref{fig:violinplot_model2}. As shown in Fig.~\ref{fig:violinplot_model2}, when the \acp{PGW} in Model 2 dominate the current \ac{PTA} observations, the second-order \acp{TIGW} introduce a noticeable correction in the high-frequency region of the total energy density spectrum.

\subsubsection{Model 3}
\begin{figure}[htbp]
\centering
\includegraphics[width=\linewidth]{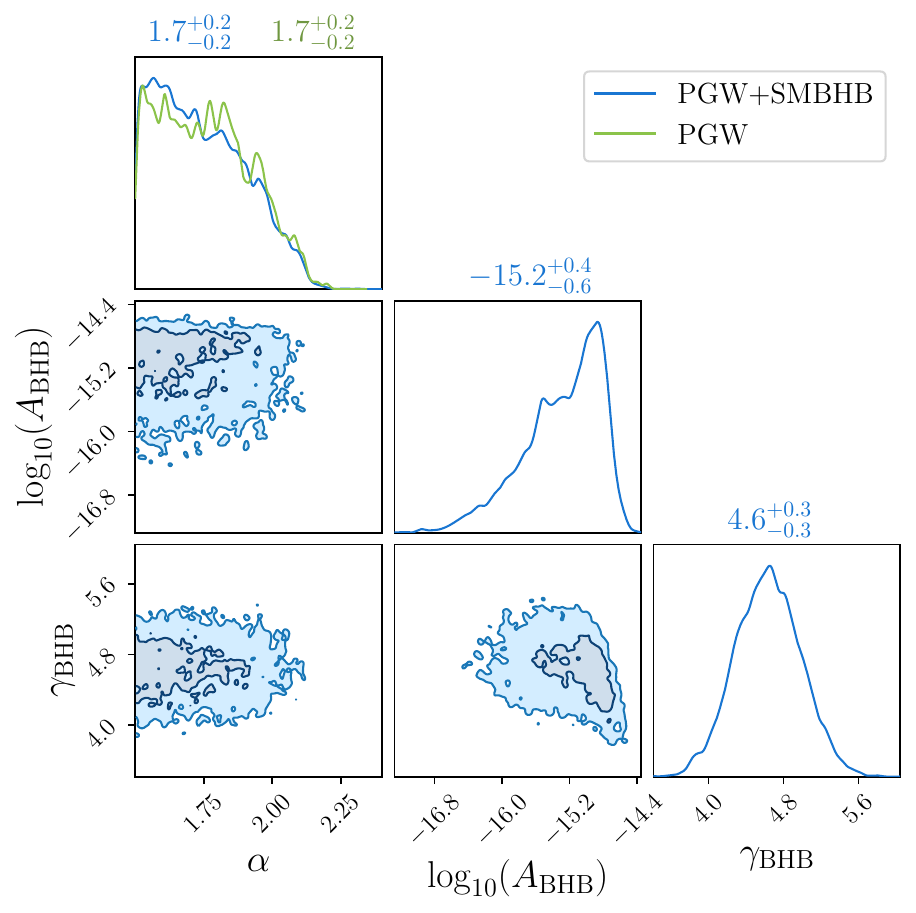}
\caption{The corner plot of the posterior distributions for model 3. The contours in the off-diagonal panels denote the $68\% $ and $95 \%$ credible intervals of the 2D posteriors. The numbers above the figures represent the median values and $1$-$\sigma$ ranges of the parameters. The blue and green solid curves correspond to the \acp{PGW} energy spectrum with or without \ac{SMBHB}.} \label{fig:corner_PGW_model3}
\end{figure}

\begin{figure}[htbp]
\centering
\includegraphics[width=\linewidth]{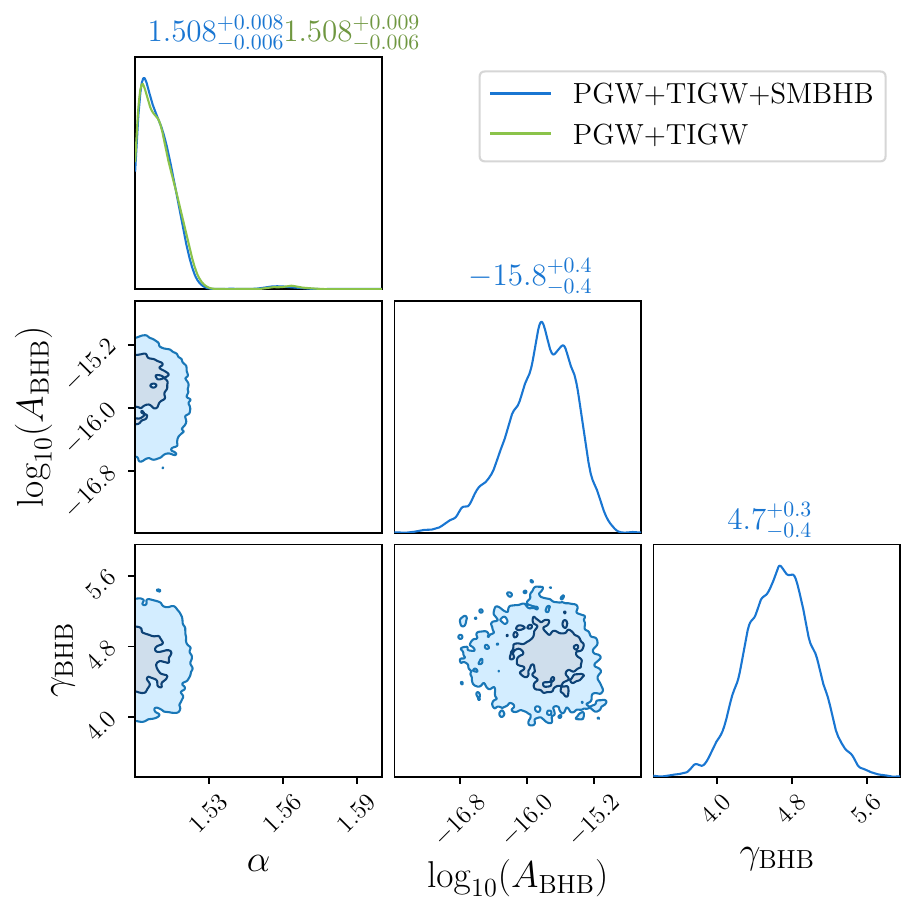}
\caption{The corner plot of the posterior distributions for model 3. The contours in the off-diagonal panels denote the $68\% $ and $95 \%$ credible intervals of the 2D posteriors. The numbers above the figures represent the median values and $1$-$\sigma$ ranges of the parameters. The blue and green solid curves correspond to the \acp{PGW}$+$\acp{TIGW} energy spectrum with or without \ac{SMBHB}.} \label{fig:corner_TIGW_model3}
\end{figure}

Based on current \ac{PTA} observations, Fig.~\ref{fig:corner_PGW_model3} and Fig.~\ref{fig:corner_TIGW_model3} present the posterior distribution of the parameter space for Model 3. The prior distribution of $\alpha$ is set as a uniform distribution over the interval $[1.5, 2.4]$. In contrast to Models 1 and Model 2, the constraints from large-scale cosmological observations—represented by the black solid and grey dashed curves—are not included in  Fig.~\ref{fig:corner_PGW_model3} and Fig.~\ref{fig:corner_TIGW_model3}  for Model 3. This is because, within the considered parameter range of Model 3, the total energy density spectrum of gravitational wave does not satisfy current large-scale cosmological constraints for any value of the parameter $\alpha$. This implies that Model 3 cannot simultaneously satisfy large-scale cosmological constraints while dominating current \ac{PTA} observations. Additionally, as shown in Fig.~\ref{fig:bayes}, the Bayes factor for Model 3 remains around $10^{-30}$ regardless of whether \ac{SMBHB} are considered. Therefore, even without incorporating large-scale cosmological limits, Model 3 performs poorly in fitting current \ac{PTA} data.

\subsubsection{Model 4}
\begin{figure}[htbp]
\centering
\includegraphics[width=\linewidth]{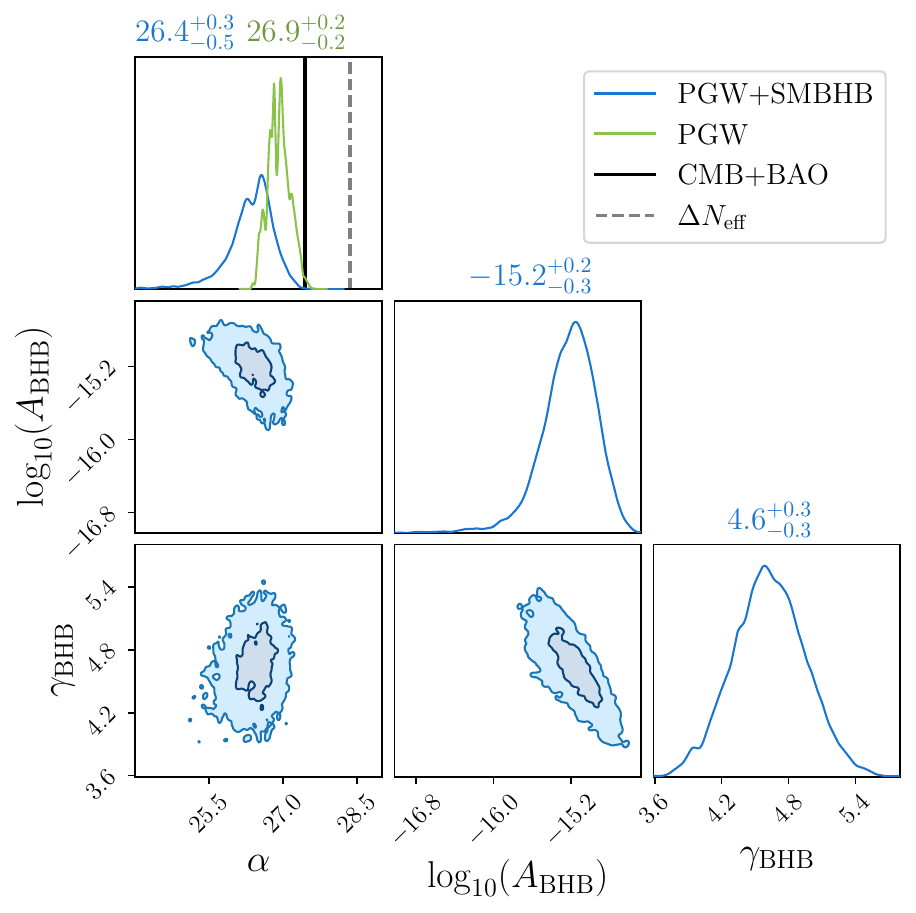}
\caption{The corner plot of the posterior distributions for model 4. The contours in the off-diagonal panels denote the $68\% $ and $95 \%$ credible intervals of the 2D posteriors. The numbers above the figures represent the median values and $1$-$\sigma$ ranges of the parameters. The blue and green solid curves correspond to the \acp{PGW} energy spectrum with or without \ac{SMBHB}. The black solid line and grey dashed line denote the upper bounds from \ac{CMB} and \ac{BAO} observations in Eq.~(\ref{eq:rhup2}) and $\Delta N_{\mathrm{eff}}$ in Eq.~(\ref{eq:rhup1}).} \label{fig:corner_PGW_model4}
\end{figure}

\begin{figure}[htbp]
\centering
\includegraphics[width=\linewidth]{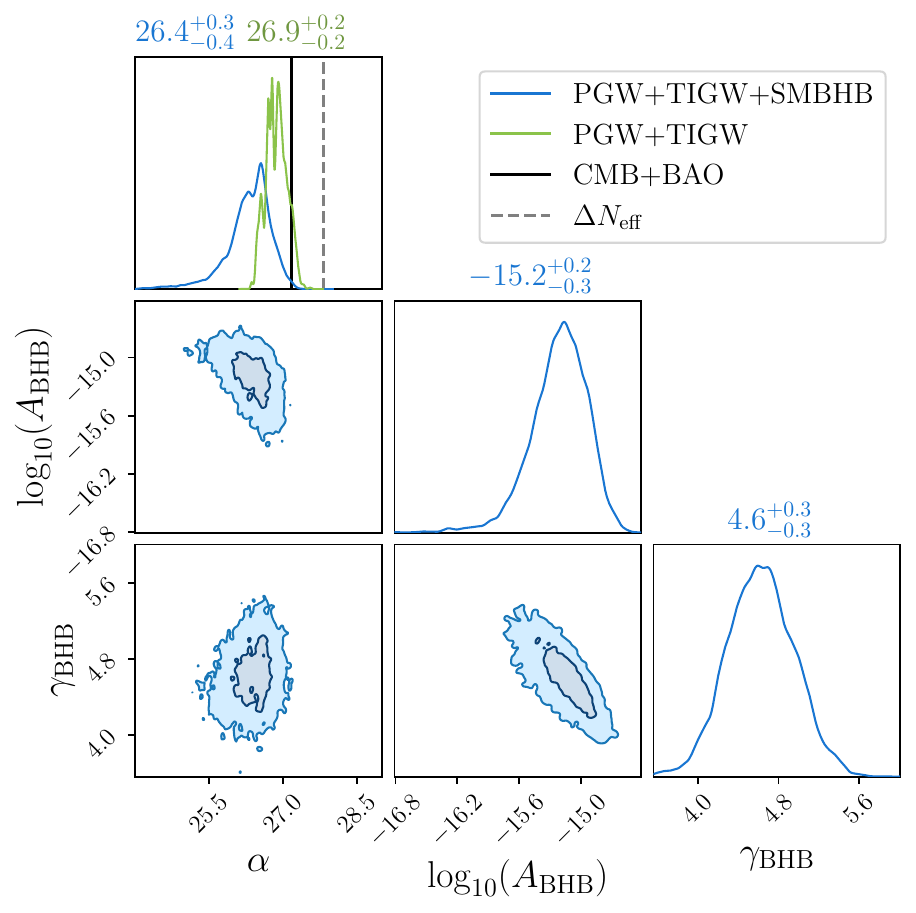}
\caption{The corner plot of the posterior distributions for model 4. The contours in the off-diagonal panels denote the $68\% $ and $95 \%$ credible intervals of the 2D posteriors. The numbers above the figures represent the median values and $1$-$\sigma$ ranges of the parameters. The blue and green solid curves correspond to the \acp{PGW}$+$\acp{TIGW} energy spectrum with or without \ac{SMBHB}. The black solid line and grey dashed line denote the upper bounds from \ac{CMB} and \ac{BAO} observations in Eq.~(\ref{eq:rhup2}) and $\Delta N_{\mathrm{eff}}$ in Eq.~(\ref{eq:rhup1}).} \label{fig:corner_TIGW_model4}
\end{figure}

\begin{figure}[htbp]
\centering
\includegraphics[width=\linewidth]{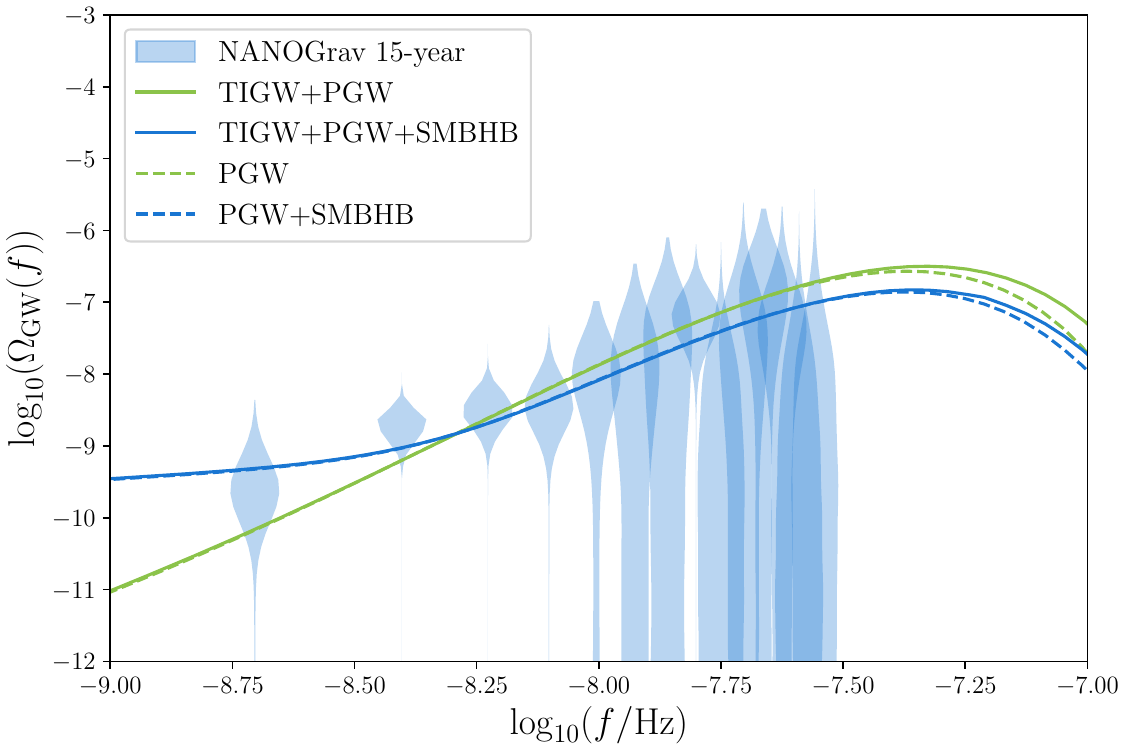}
\caption{The energy density spectra of \acp{PGW} and \acp{PGW}$+$\acp{TIGW} with or without \acp{SMBHB} for model 4. The energy density spectra derived from the free spectrum of the NANOGrav 15-year are shown in blue. The blue and green curves represent the energy density spectra of \acp{GW} with different line styles labeled in the figure. These parameters are selected based on the median values of the posterior distributions.} \label{fig:violinplot_model4}
\end{figure}

\begin{figure*}[htbp]
    \centering
    \includegraphics[width=1.8\columnwidth]{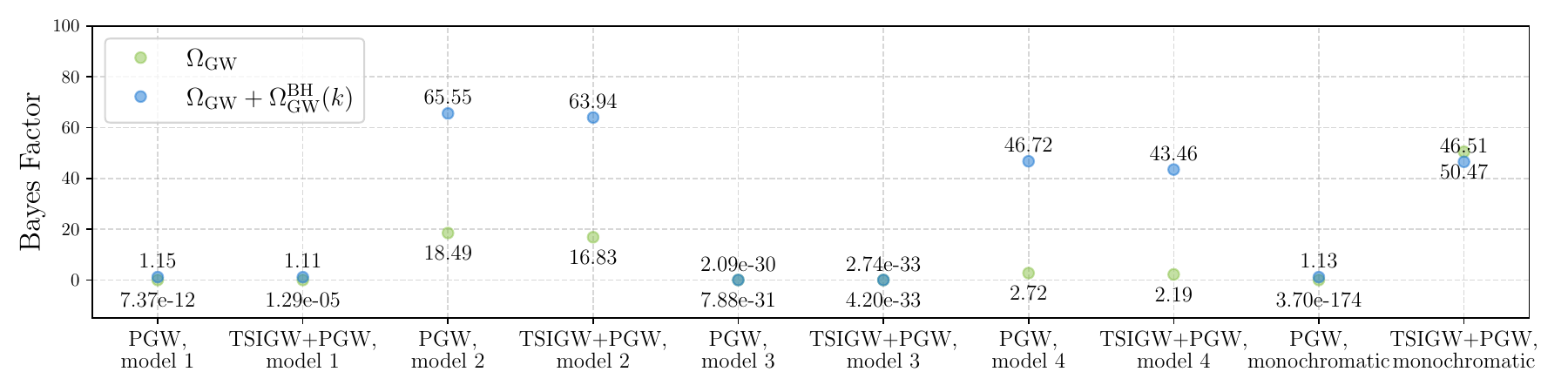}
\caption{\label{fig:bayes} The Bayes factors between different models. The vertical axis represents the Bayes factor of different models relative to \ac{SMBHB}, and the horizontal axis represents the different models. The green dots are for models without \ac{SMBHB} and the blue dots are for models in combination with the \ac{SMBHB} signal. }
\end{figure*}

As shown in Fig.~\ref{fig:corner_PGW_model4} and Fig.~\ref{fig:corner_TIGW_model4}, similar to Model 2, Model 4 is capable of dominating the current \ac{PTA} observations while still satisfying large-scale cosmological constraints. And the prior distribution of $\alpha$ is set as a uniform distribution over the interval $[24, 30]$. The corresponding energy density spectrum for Model 4 is presented in Fig.~\ref{fig:violinplot_model4}. As shown in Fig.~\ref{fig:violinplot_model4}, when the \acp{PGW} predicted by Model 4 dominate the current \ac{PTA} observations, the second-order \acp{TIGW} introduce noticeable corrections only in the region of the total energy density spectrum above $10^{-7.5}$ Hz, without affecting the low-frequency spectrum. Therefore, for Model 4, the second-order \acp{TIGW} do not exert a significant impact on the current \ac{PTA} observations. Moreover, as indicated in Fig.~\ref{fig:bayes}, unlike Model 2, Model 4 yields a Bayes factor below 5 when the influence of \ac{SMBHB} is not considered. Therefore, compared to Model 4, Model 2 is more likely to account for the current \ac{PTA} observations.

\subsection{Primordial black holes from primordial gravitational waves}\label{sec:4.3}
It is well known that if sufficiently large primordial curvature perturbations $\zeta_{\mathbf{k}}$ are generated at small scales during inflation, they will re-enter into the horizon after inflation and induce significant density perturbations $\delta^{(1)}= \rho^{(1)}/\rho^{(0)}$, thereby leading to the formation of \acp{PBH} \cite{Carr:1975qj,Carr:2020gox,Carr:2020xqk,Carr:2016drx,Carr:2009jm}. Given a power spectrum of the primordial curvature perturbation $\mathcal{P}_{\zeta}(k)$, the corresponding \ac{PBH} abundance $f_{\mathrm{PBH}}$ can be calculated using either threshold statistics or peak theory \cite{Musco:2018rwt,Germani:2018jgr,Ferrante:2022mui,Shimada:2024eec,Iovino:2024tyg,Musco:2020jjb,Young:2014ana,Green:2004wb,DeLuca:2019qsy}. For specific inflation models, the power spectrum $\mathcal{P}_{\zeta}(k)$ depends on the parameters of the model, which in turn causes $f_{\mathrm{PBH}}$ to vary with these parameters \cite{Ballesteros:2017fsr,Kannike:2017bxn,Ezquiaga:2017fvi,Garcia-Bellido:2017mdw,Firouzjahi:2023ahg,Ferraz:2024bvd}. By leveraging current observational constraints on \ac{PBH} abundance, we can place limits on the small-scale primordial power spectrum $\mathcal{P}_{\zeta}(k)$, thereby constraining the parameter space of inflation models at small scales \cite{Kristiano:2022maq,Franciolini:2023pbf,Wang:2023ost,Wu:2025gwt,Choudhury:2023jlt,Barker:2024mpz,Su:2025mam,Wang:2024nmd}. 

In this study, we focus solely on \acp{PGW} with large amplitudes on small scales while neglecting potentially significant primordial curvature perturbations. Under such conditions, the formation of \acp{PBH} appears to be unlikely. However, as previously noted, small-scale \acp{PGW} can induce higher-order density perturbations, which may still lead to \ac{PBH} formation. Specifically, when sufficiently large \acp{PGW} exist on small scales, they couple to second-order cosmological perturbations via the second-order perturbation equations, thereby generating second-order density perturbation \cite{Zhou:2023itl,Nakama:2016enz}
\begin{equation}\label{eq:rpbh2}
    \begin{aligned}
       \delta^{(2)}&=\frac{\rho^{(2)}}{\rho^{(0)}}=-2\phi^{(2)}+\frac{2}{3\mathcal{H}^2}\Delta \psi^{(2)}-\frac{2}{3\mathcal{H}}\Delta B^{(2)} \\
    &-\frac{2}{\mathcal{H}}\psi^{(2)'}+S^{(2)}_{\rho} \ ,
    \end{aligned}
\end{equation}
where $\phi^{(2)}$,$\psi^{(2)}$, and $B^{(2)}$ denote second-order scalar perturbations in the comoving gauge, and $S^{(2)}_{\rho}$ represents a source term composed of products of the first-order tensor perturbation. The existence of the second-order density perturbation $\delta^{(2)}$ can lead to the formation of \acp{PBH}, thereby allowing current observational constraints on \ac{PBH} abundance to place limits on small-scale \acp{PGW}.

\begin{figure}[htbp]
\centering
\includegraphics[width=\linewidth]{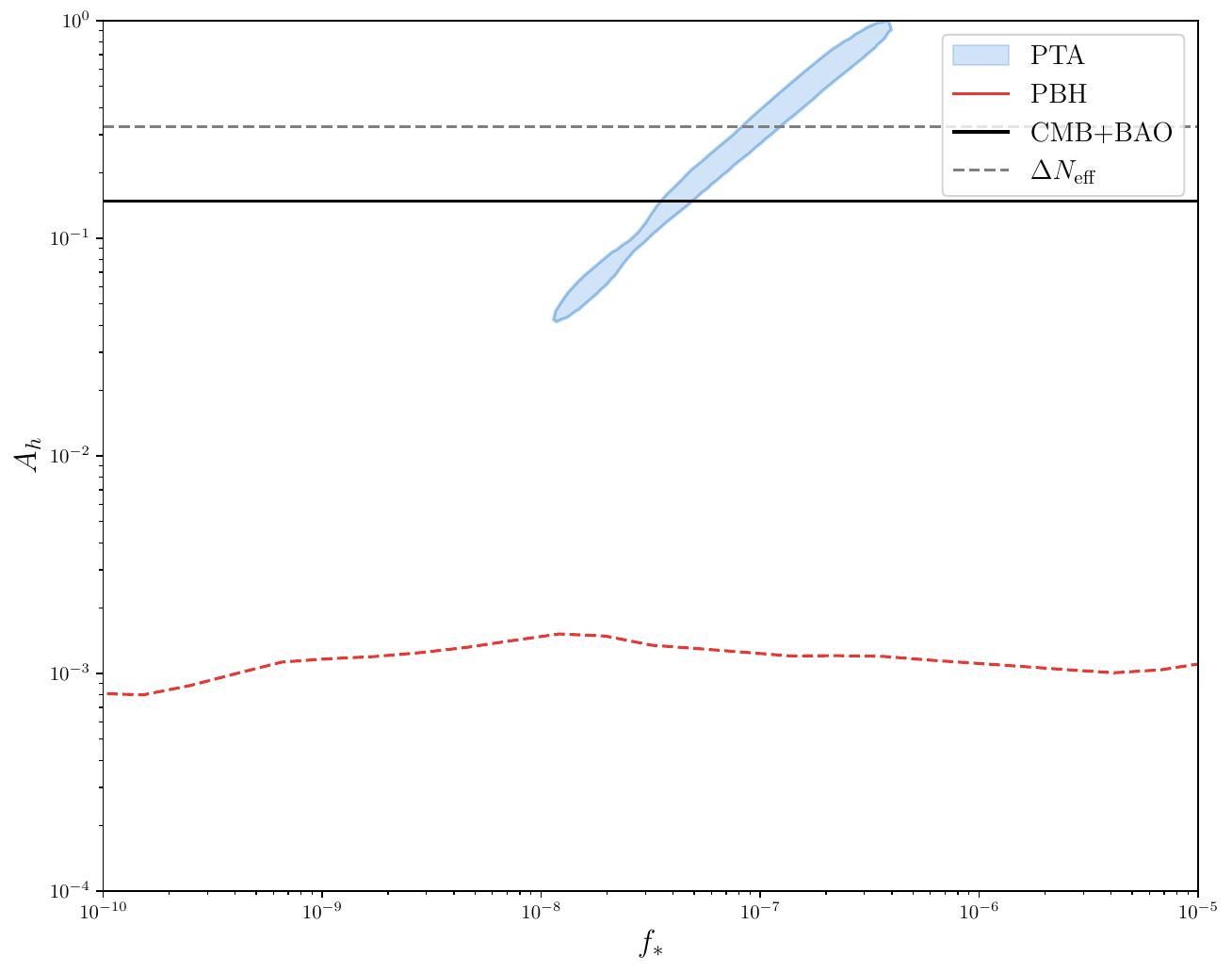}
\caption{The blue shaded region represents the $68\%$ credible intervals corresponding to the two-dimensional posterior distribution. And the prior distributions of $\log_{10}(A_{h})$ and $\log_{10}(f_*/\mathrm{Hz})$ are set as uniform distributions over the intervals $[-4,0]$ and $[-10,-5]$, respectively. The black solid line and grey dashed line denote the upper bounds from \ac{CMB} and \ac{BAO} observations in Eq.~(\ref{eq:rhup2}) and $\Delta N_{\mathrm{eff}}$ in Eq.~(\ref{eq:rhup1}). The red curve represents the observational constraints from \acp{PBH} on the parameter space of the primordial power spectrum.} \label{fig:constrain_mono}
\end{figure}

Utilizing results from Ref.~\cite{Nakama:2015nea}, we investigate the constraints on small-scale \acp{PGW} imposed by \ac{PBH} abundance, assuming a monochromatic \ac{PGW} power spectrum: $\mathcal{P}_h=A_{h}k_*\delta\left( k-k_* \right)$. As illustrated in Fig.~\ref{fig:constrain_mono}, we show the current constraints on the amplitude of a monochromatic primordial power spectrum from \ac{PBH}$+$\ac{CMB}$+$\ac{BAO}$+$\ac{PTA} observations. The blue shaded region in Fig.~\ref{fig:constrain_mono} represents the posterior distribution derived from \ac{PTA} observations, while the red and black (grey) curves indicate upper bounds on the amplitude $A_{h}$ from \ac{PBH} abundance and large-scale cosmological data, respectively. The allowed parameter space from current cosmological observations corresponds to the region below the red and black curves. If \acp{PGW}$+$\acp{TIGW} are assumed to dominate the \ac{PTA} observations, the parameters of the primordial power spectrum must also lie within the blue region. Therefore, under the assumption of a monochromatic primordial power spectrum, \ac{PGW}+\ac{TIGW} cannot account for the dominant contribution in current \ac{PTA} observations. Furthermore, even if \acp{PGW}$+$\acp{TIGW} are not the dominant source of the current \ac{PTA} signal, the \ac{SGWB} observations still provide an upper bound on the energy density spectrum of \acp{PGW}$+$\acp{TIGW}. 

It should be noted that the specific form of the second-order induced density perturbation $\delta^{(2)}$ in Eq.~(\ref{eq:rpbh2}) depends on the gauge choice in cosmological perturbation theory. In the comoving gauge, Refs.~\cite{Bari:2021xvf,Bari:2022grh,Abdelaziz:2025qpn} and Refs.~\cite{Nakama:2016enz,Zhou:2023itl,Nakama:2015nea} investigated the influence of second-order induced density perturbations on \ac{LSS} and the probability distribution function of \ac{PBH} formation, respectively. Additionally, Refs.~\cite{DeLuca:2023tun,Inomata:2020cck} studied analytical solutions of second-order scalar-induced density perturbations in the Newtonian gauge and examined their impact on the threshold of \ac{PBH}. Whether second-order induced density perturbations exhibit intrinsic gauge dependence, whether an approximate gauge invariance akin to that of \acp{SIGW} exists, and how gauge choice affects the calculation of \ac{LSS} and \ac{PBH}, remain open questions that have not yet been systematically addressed.

In this section, we only analyze the constraints on the parameter space of a monochromatic primordial power spectrum arising from \ac{PBH} abundance. For the models presented in Sec.~\ref{sec:3.0}, it is necessary to compute the power spectrum of the corresponding second-order induced density perturbations and analyze their impact on the probability distribution function and threshold of \ac{PBH} formation.  These related issues may be explored more systematically in future research.

\section{Conclusion and discussion}\label{sec:5.0}
In this paper, we investigated the large-amplitude \acp{PGW} on small scales, as well as the kernel functions and energy density spectra of the corresponding second-order \acp{TIGW}. We systematically studied the five questions raised in Sec.~\ref{sec:1.0} and summarize the conclusions for each of them below:

\noindent
\textbf{Large \acp{PGW} on small scales:} To generate large-amplitude \acp{PGW} on small scales, one must modify the equation of motion of \acp{PGW} during inflation. Two commonly adopted approaches are: (1) coupling the equation of motion of \acp{PGW} to additional fields; and (2) altering the dynamical terms on the left-hand side of the equation of motion. Both methods introduce new physical processes or mechanisms during inflation, making the detection of enhanced small-scale \acp{PGW} a potential probe of new  physics in the early universe. In Sec.~\ref{sec:3.0} and Sec.~\ref{sec:4.0}, we investigate primordial gravitational waves generated by four distinct models and examine current cosmological constraints on each of them.

\noindent
\textbf{Second-order \acp{TIGW}:} To evaluate the correction to the total energy density spectrum of gravitational waves arising from second-order \acp{TIGW}, we derive and solve the second-order cosmological perturbation equations and calculate the two-point correlation function of the second-order \acp{TIGW}. In Sec.~\ref{sec:2.0}, these complex computations are encapsulated in Eq.~(\ref{eq:1Ott}). Given a primordial power spectrum $\mathcal{P}_h(k)$, Eq.~(\ref{eq:1Ott}) allows us to directly calculate the energy density spectrum of second-order \acp{TIGW} during the \ac{RD} era.

\noindent
\textbf{\acp{TIGW} and \ac{PTA} observations}: Sec.~\ref{sec:3.0} and Sec.~\ref{sec:4.0} address the third and fourth questions outlined in Sec.~\ref{sec:1.0}. As discussed in Sec.~\ref{sec:3.0}, corrections to the total gravitational wave energy density spectrum caused by second-order \acp{TIGW} become significant only when the amplitude of \acp{PGW} is sufficiently large. Based on our analysis in Sec.~\ref{sec:4.0}, Model 1 and Model 3 are excluded by current cosmological observations and therefore cannot dominate the current \ac{PTA} observations. Both Models 2 and 4 are capable of driving the current PTA signals, with Model 2 being more favorable due to its higher Bayes factor. For Model 2, the presence of second-order \acp{TIGW} leads to a reduction in the value of parameter $\alpha$ inferred from current \ac{PTA} data. Table.~\ref{ta:1} summarizes the constraints imposed by current cosmological observations on the four models discussed in Sec.~\ref{sec:3.0}.
\begin{table}[h!]
\centering
\begin{tabular}{lccc}
\hline \hline & \ac{PTA}  & $\alpha$ & Bayes factors \\
\hline \rule{0pt}{12pt}
Model 1  &  $\times$ & $\alpha<22.76$ & $1.29\times 10^{-5}$ \\
\rule{0pt}{12pt}
Model 2 & $\checkmark$ & $1.24^{+0.01}_{-0.02}$ & $16.83$ \\
\rule{0pt}{12pt}
Model 3 & $\times$ & Rule out  & $4.2\times 10^{-33}$ \\
\rule{0pt}{12pt}
Model 4 & $\checkmark$ & $26.9^{+0.2}_{-0.2}$  & $2.19$ \\
\hline \hline 
\end{tabular}
\caption{Current cosmological constraints on \ac{PGW}+\ac{TIGW} in four distinct models, where symbol $\checkmark$ and symbol $\times$ respectively indicate whether the model can or cannot dominate \ac{PTA} observations. }
\label{ta:1}
\end{table}

\noindent
\textbf{Constraints on \acp{PGW}}: 
In Sec.~\ref{sec:4.0}, we examine how current cosmological observations constrain \acp{PGW} on small scales. Beyond direct and indirect measurements of the energy density spectrum of gravitational waves, large-amplitude \acp{PGW} can also induce higher-order density perturbations, potentially leading to the formation of \acp{PBH}. By combining constraints from \acp{PBH} observations and the \ac{SGWB} observations, we can jointly infer the physical properties of \acp{PGW} on small scales.

For a power spectrum of \ac{PGW} with amplitude $A_h$, the two-point correlation function of the second-order \acp{TIGW} $\langle h^{\lambda,(2)}_{\mathbf{k}} h^{\lambda^{\prime},(2)}_{\mathbf{k}^{\prime}}\rangle$, as well as the corresponding energy density spectrum, scale proportionally with $A^2_h$. It is important to note that the two-point correlation function $\langle h^{\lambda,(3)}_{\mathbf{k}} h^{\lambda^{\prime},(1)}_{\mathbf{k}^{\prime}}\rangle$, arising from the first-order and third-order gravitational waves, also contributes an energy density spectrum proportional to $A_h^2$ \cite{Chen:2022dah}. Both $\langle h^{\lambda,(2)}_{\mathbf{k}} h^{\lambda^{\prime},(2)}_{\mathbf{k}^{\prime}}\rangle$ and $\langle h^{\lambda,(3)}_{\mathbf{k}} h^{\lambda^{\prime},(1)}_{\mathbf{k}^{\prime}}\rangle$ represent second-order corrections to the total energy density spectrum. To fully characterize the second-order corrections, a systematic analysis of $\langle h^{\lambda,(3)}_{\mathbf{k}} h^{\lambda^{\prime},(1)}_{\mathbf{k}^{\prime}}\rangle$ is necessary. When the effects of $\langle h^{\lambda,(3)}_{\mathbf{k}} h^{\lambda^{\prime},(1)}_{\mathbf{k}^{\prime}}\rangle$ are included, the resulting energy density spectrum increases, implying a reduced amplitude $A_h$ as inferred from current cosmological data. Therefore, the results presented in this work still serve as an upper bound on the amplitude of \acp{PGW}.

As discussed in Sec.~\ref{sec:2.0}, the sound speed $c_s$ does not affect the equation of motion or the energy density spectrum of second-order \acp{TIGW}. However, for third-order \acp{TIGW}, the presence of second-order induced density perturbation $\rho^{(2)}$ introduces $c_s$ into the equation of motion of third-order \acp{TIGW}, thereby impacting the corresponding two-point correlation function and energy spectrum. In particular, sound speed effects arise only at the third order and beyond for \acp{TIGW}, which contrasts significantly with the \acp{SIGW}. A complete evaluation of the higher-order effects and the influence of various physical processes on \acp{TIGW} might be systematically explored in future work.
\begin{table}[h!]
\centering
\begin{tabular}{lccc}
\hline \hline \acp{TIGW} & Source  & $c_s$  & Gauge  \\
\hline \rule{0pt}{12pt}
Second-order  &  $h^{(1)}_{ij}$ & $\times$ & $\times$ \\
\rule{0pt}{12pt}
Third-order & $h^{(1)}_{ij}$ and $A^{(2)}$ & $\checkmark$ & $\checkmark$ \\
\hline \hline 
\end{tabular}
\caption{The differences between second-order and third-order \acp{TIGW}, where $A^{(2)}$ denotes all types of second-order cosmological perturbations. Symbols $\checkmark$ and $\times$ respectively denote dependence and independence on the sound speed parameter $c_s$ and gauge choice. }
\label{ta:2}
\end{table}

In Sec.~\ref{sec:2.1}, we examined the equation of motion for second-order \acp{TIGW}. In contrast to the \acp{SIGW}, we found that these second-order \acp{TIGW} exhibit gauge independence. However, this gauge invariance can not extend to the third-order \acp{TIGW}. In addition to the source term composed of three first-order tensor perturbations, third-order tensor modes can also arise from combinations of second-order perturbations and first-order tensor perturbations \cite{Chang:2023vjk}. Since second-order scalar perturbations induced by first-order tensor perturbation are gauge dependent, the source terms of third-order \acp{TIGW} vary under different gauge choices, giving rise to potential gauge dependence. In Table.~\ref{ta:2}, we highlight the differences between second-order \acp{TIGW} and third-order \acp{TIGW}. The gauge dependence of third-order and higher-order \acp{TIGW}, and whether they exhibit approximate gauge invariance similar to that of second-order \acp{SIGW}, remains an unresolved and important question.

\section*{Data availability}\label{sec:6.0}
The data that support the findings of this article are openly available in Ref. \cite{Nanograv:KDE}.

\vspace{0.3cm}
\begin{acknowledgements} 
This work has been funded by the National Nature Science Foundation of China under grant No. 12447127.  
\end{acknowledgements}

\bibliography{biblio}

%merlin.mbs apsrev4-1.bst 2010-07-25 4.21a (PWD, AO, DPC) hacked
%Control: key (0)
%Control: author (8) initials jnrlst
%Control: editor formatted (1) identically to author
%Control: production of article title (-1) disabled
%Control: page (0) single
%Control: year (1) truncated
%Control: production of eprint (0) enabled
\begin{thebibliography}{143}%
\makeatletter
\providecommand \@ifxundefined [1]{%
 \@ifx{#1\undefined}
}%
\providecommand \@ifnum [1]{%
 \ifnum #1\expandafter \@firstoftwo
 \else \expandafter \@secondoftwo
 \fi
}%
\providecommand \@ifx [1]{%
 \ifx #1\expandafter \@firstoftwo
 \else \expandafter \@secondoftwo
 \fi
}%
\providecommand \natexlab [1]{#1}%
\providecommand \enquote  [1]{``#1''}%
\providecommand \bibnamefont  [1]{#1}%
\providecommand \bibfnamefont [1]{#1}%
\providecommand \citenamefont [1]{#1}%
\providecommand \href@noop [0]{\@secondoftwo}%
\providecommand \href [0]{\begingroup \@sanitize@url \@href}%
\providecommand \@href[1]{\@@startlink{#1}\@@href}%
\providecommand \@@href[1]{\endgroup#1\@@endlink}%
\providecommand \@sanitize@url [0]{\catcode `\\12\catcode `\$12\catcode `\&12\catcode `\#12\catcode `\^12\catcode `\_12\catcode `\%12\relax}%
\providecommand \@@startlink[1]{}%
\providecommand \@@endlink[0]{}%
\providecommand \url  [0]{\begingroup\@sanitize@url \@url }%
\providecommand \@url [1]{\endgroup\@href {#1}{\urlprefix }}%
\providecommand \urlprefix  [0]{URL }%
\providecommand \Eprint [0]{\href }%
\providecommand \doibase [0]{http://dx.doi.org/}%
\providecommand \selectlanguage [0]{\@gobble}%
\providecommand \bibinfo  [0]{\@secondoftwo}%
\providecommand \bibfield  [0]{\@secondoftwo}%
\providecommand \translation [1]{[#1]}%
\providecommand \BibitemOpen [0]{}%
\providecommand \bibitemStop [0]{}%
\providecommand \bibitemNoStop [0]{.\EOS\space}%
\providecommand \EOS [0]{\spacefactor3000\relax}%
\providecommand \BibitemShut  [1]{\csname bibitem#1\endcsname}%
\let\auto@bib@innerbib\@empty
%</preamble>
\bibitem [{\citenamefont {Baumann}(2022)}]{Baumann:2022mni}%
  \BibitemOpen
  \bibfield  {author} {\bibinfo {author} {\bibfnamefont {D.}~\bibnamefont {Baumann}},\ }\href {\doibase 10.1017/9781108937092} {\emph {\bibinfo {title} {{Cosmology}}}}\ (\bibinfo  {publisher} {Cambridge University Press},\ \bibinfo {year} {2022})\BibitemShut {NoStop}%
\bibitem [{\citenamefont {Bardeen}(1980)}]{Bardeen:1980kt}%
  \BibitemOpen
  \bibfield  {author} {\bibinfo {author} {\bibfnamefont {J.~M.}\ \bibnamefont {Bardeen}},\ }\href {\doibase 10.1103/PhysRevD.22.1882} {\bibfield  {journal} {\bibinfo  {journal} {Phys. Rev. D}\ }\textbf {\bibinfo {volume} {22}},\ \bibinfo {pages} {1882} (\bibinfo {year} {1980})}\BibitemShut {NoStop}%
\bibitem [{\citenamefont {Kodama}\ and\ \citenamefont {Sasaki}(1984)}]{Kodama:1984ziu}%
  \BibitemOpen
  \bibfield  {author} {\bibinfo {author} {\bibfnamefont {H.}~\bibnamefont {Kodama}}\ and\ \bibinfo {author} {\bibfnamefont {M.}~\bibnamefont {Sasaki}},\ }\href {\doibase 10.1143/PTPS.78.1} {\bibfield  {journal} {\bibinfo  {journal} {Prog. Theor. Phys. Suppl.}\ }\textbf {\bibinfo {volume} {78}},\ \bibinfo {pages} {1} (\bibinfo {year} {1984})}\BibitemShut {NoStop}%
\bibitem [{\citenamefont {Malik}\ and\ \citenamefont {Wands}(2009)}]{Malik:2008im}%
  \BibitemOpen
  \bibfield  {author} {\bibinfo {author} {\bibfnamefont {K.~A.}\ \bibnamefont {Malik}}\ and\ \bibinfo {author} {\bibfnamefont {D.}~\bibnamefont {Wands}},\ }\href {\doibase 10.1016/j.physrep.2009.03.001} {\bibfield  {journal} {\bibinfo  {journal} {Phys. Rept.}\ }\textbf {\bibinfo {volume} {475}},\ \bibinfo {pages} {1} (\bibinfo {year} {2009})},\ \Eprint {http://arxiv.org/abs/0809.4944} {arXiv:0809.4944 [astro-ph]} \BibitemShut {NoStop}%
\bibitem [{\citenamefont {Lyth}\ and\ \citenamefont {Rodriguez}(2005)}]{Lyth:2005fi}%
  \BibitemOpen
  \bibfield  {author} {\bibinfo {author} {\bibfnamefont {D.~H.}\ \bibnamefont {Lyth}}\ and\ \bibinfo {author} {\bibfnamefont {Y.}~\bibnamefont {Rodriguez}},\ }\href {\doibase 10.1103/PhysRevLett.95.121302} {\bibfield  {journal} {\bibinfo  {journal} {Phys. Rev. Lett.}\ }\textbf {\bibinfo {volume} {95}},\ \bibinfo {pages} {121302} (\bibinfo {year} {2005})},\ \Eprint {http://arxiv.org/abs/astro-ph/0504045} {arXiv:astro-ph/0504045} \BibitemShut {NoStop}%
\bibitem [{\citenamefont {Weinberg}(2005)}]{Weinberg:2005vy}%
  \BibitemOpen
  \bibfield  {author} {\bibinfo {author} {\bibfnamefont {S.}~\bibnamefont {Weinberg}},\ }\href {\doibase 10.1103/PhysRevD.72.043514} {\bibfield  {journal} {\bibinfo  {journal} {Phys. Rev. D}\ }\textbf {\bibinfo {volume} {72}},\ \bibinfo {pages} {043514} (\bibinfo {year} {2005})},\ \Eprint {http://arxiv.org/abs/hep-th/0506236} {arXiv:hep-th/0506236} \BibitemShut {NoStop}%
\bibitem [{\citenamefont {Bassett}\ \emph {et~al.}(2006)\citenamefont {Bassett}, \citenamefont {Tsujikawa},\ and\ \citenamefont {Wands}}]{Bassett:2005xm}%
  \BibitemOpen
  \bibfield  {author} {\bibinfo {author} {\bibfnamefont {B.~A.}\ \bibnamefont {Bassett}}, \bibinfo {author} {\bibfnamefont {S.}~\bibnamefont {Tsujikawa}}, \ and\ \bibinfo {author} {\bibfnamefont {D.}~\bibnamefont {Wands}},\ }\href {\doibase 10.1103/RevModPhys.78.537} {\bibfield  {journal} {\bibinfo  {journal} {Rev. Mod. Phys.}\ }\textbf {\bibinfo {volume} {78}},\ \bibinfo {pages} {537} (\bibinfo {year} {2006})},\ \Eprint {http://arxiv.org/abs/astro-ph/0507632} {arXiv:astro-ph/0507632} \BibitemShut {NoStop}%
\bibitem [{\citenamefont {Arkani-Hamed}\ \emph {et~al.}(2004)\citenamefont {Arkani-Hamed}, \citenamefont {Creminelli}, \citenamefont {Mukohyama},\ and\ \citenamefont {Zaldarriaga}}]{Arkani-Hamed:2003juy}%
  \BibitemOpen
  \bibfield  {author} {\bibinfo {author} {\bibfnamefont {N.}~\bibnamefont {Arkani-Hamed}}, \bibinfo {author} {\bibfnamefont {P.}~\bibnamefont {Creminelli}}, \bibinfo {author} {\bibfnamefont {S.}~\bibnamefont {Mukohyama}}, \ and\ \bibinfo {author} {\bibfnamefont {M.}~\bibnamefont {Zaldarriaga}},\ }\href {\doibase 10.1088/1475-7516/2004/04/001} {\bibfield  {journal} {\bibinfo  {journal} {JCAP}\ }\textbf {\bibinfo {volume} {04}},\ \bibinfo {pages} {001} (\bibinfo {year} {2004})},\ \Eprint {http://arxiv.org/abs/hep-th/0312100} {arXiv:hep-th/0312100} \BibitemShut {NoStop}%
\bibitem [{\citenamefont {Kachru}\ \emph {et~al.}(2003)\citenamefont {Kachru}, \citenamefont {Kallosh}, \citenamefont {Linde}, \citenamefont {Maldacena}, \citenamefont {McAllister},\ and\ \citenamefont {Trivedi}}]{Kachru:2003sx}%
  \BibitemOpen
  \bibfield  {author} {\bibinfo {author} {\bibfnamefont {S.}~\bibnamefont {Kachru}}, \bibinfo {author} {\bibfnamefont {R.}~\bibnamefont {Kallosh}}, \bibinfo {author} {\bibfnamefont {A.~D.}\ \bibnamefont {Linde}}, \bibinfo {author} {\bibfnamefont {J.~M.}\ \bibnamefont {Maldacena}}, \bibinfo {author} {\bibfnamefont {L.~P.}\ \bibnamefont {McAllister}}, \ and\ \bibinfo {author} {\bibfnamefont {S.~P.}\ \bibnamefont {Trivedi}},\ }\href {\doibase 10.1088/1475-7516/2003/10/013} {\bibfield  {journal} {\bibinfo  {journal} {JCAP}\ }\textbf {\bibinfo {volume} {10}},\ \bibinfo {pages} {013} (\bibinfo {year} {2003})},\ \Eprint {http://arxiv.org/abs/hep-th/0308055} {arXiv:hep-th/0308055} \BibitemShut {NoStop}%
\bibitem [{\citenamefont {Armendariz-Picon}\ \emph {et~al.}(1999)\citenamefont {Armendariz-Picon}, \citenamefont {Damour},\ and\ \citenamefont {Mukhanov}}]{Armendariz-Picon:1999hyi}%
  \BibitemOpen
  \bibfield  {author} {\bibinfo {author} {\bibfnamefont {C.}~\bibnamefont {Armendariz-Picon}}, \bibinfo {author} {\bibfnamefont {T.}~\bibnamefont {Damour}}, \ and\ \bibinfo {author} {\bibfnamefont {V.~F.}\ \bibnamefont {Mukhanov}},\ }\href {\doibase 10.1016/S0370-2693(99)00603-6} {\bibfield  {journal} {\bibinfo  {journal} {Phys. Lett. B}\ }\textbf {\bibinfo {volume} {458}},\ \bibinfo {pages} {209} (\bibinfo {year} {1999})},\ \Eprint {http://arxiv.org/abs/hep-th/9904075} {arXiv:hep-th/9904075} \BibitemShut {NoStop}%
\bibitem [{\citenamefont {Peebles}\ and\ \citenamefont {Vilenkin}(1999)}]{Peebles:1998qn}%
  \BibitemOpen
  \bibfield  {author} {\bibinfo {author} {\bibfnamefont {P.~J.~E.}\ \bibnamefont {Peebles}}\ and\ \bibinfo {author} {\bibfnamefont {A.}~\bibnamefont {Vilenkin}},\ }\href {\doibase 10.1103/PhysRevD.59.063505} {\bibfield  {journal} {\bibinfo  {journal} {Phys. Rev. D}\ }\textbf {\bibinfo {volume} {59}},\ \bibinfo {pages} {063505} (\bibinfo {year} {1999})},\ \Eprint {http://arxiv.org/abs/astro-ph/9810509} {arXiv:astro-ph/9810509} \BibitemShut {NoStop}%
\bibitem [{\citenamefont {Kobayashi}\ \emph {et~al.}(2010)\citenamefont {Kobayashi}, \citenamefont {Yamaguchi},\ and\ \citenamefont {Yokoyama}}]{Kobayashi:2010cm}%
  \BibitemOpen
  \bibfield  {author} {\bibinfo {author} {\bibfnamefont {T.}~\bibnamefont {Kobayashi}}, \bibinfo {author} {\bibfnamefont {M.}~\bibnamefont {Yamaguchi}}, \ and\ \bibinfo {author} {\bibfnamefont {J.}~\bibnamefont {Yokoyama}},\ }\href {\doibase 10.1103/PhysRevLett.105.231302} {\bibfield  {journal} {\bibinfo  {journal} {Phys. Rev. Lett.}\ }\textbf {\bibinfo {volume} {105}},\ \bibinfo {pages} {231302} (\bibinfo {year} {2010})},\ \Eprint {http://arxiv.org/abs/1008.0603} {arXiv:1008.0603 [hep-th]} \BibitemShut {NoStop}%
\bibitem [{\citenamefont {Bezrukov}\ \emph {et~al.}(2011)\citenamefont {Bezrukov}, \citenamefont {Magnin}, \citenamefont {Shaposhnikov},\ and\ \citenamefont {Sibiryakov}}]{Bezrukov:2010jz}%
  \BibitemOpen
  \bibfield  {author} {\bibinfo {author} {\bibfnamefont {F.}~\bibnamefont {Bezrukov}}, \bibinfo {author} {\bibfnamefont {A.}~\bibnamefont {Magnin}}, \bibinfo {author} {\bibfnamefont {M.}~\bibnamefont {Shaposhnikov}}, \ and\ \bibinfo {author} {\bibfnamefont {S.}~\bibnamefont {Sibiryakov}},\ }\href {\doibase 10.1007/JHEP01(2011)016} {\bibfield  {journal} {\bibinfo  {journal} {JHEP}\ }\textbf {\bibinfo {volume} {01}},\ \bibinfo {pages} {016} (\bibinfo {year} {2011})},\ \Eprint {http://arxiv.org/abs/1008.5157} {arXiv:1008.5157 [hep-ph]} \BibitemShut {NoStop}%
\bibitem [{\citenamefont {Arkani-Hamed}\ \emph {et~al.}(2020)\citenamefont {Arkani-Hamed}, \citenamefont {Baumann}, \citenamefont {Lee},\ and\ \citenamefont {Pimentel}}]{Arkani-Hamed:2018kmz}%
  \BibitemOpen
  \bibfield  {author} {\bibinfo {author} {\bibfnamefont {N.}~\bibnamefont {Arkani-Hamed}}, \bibinfo {author} {\bibfnamefont {D.}~\bibnamefont {Baumann}}, \bibinfo {author} {\bibfnamefont {H.}~\bibnamefont {Lee}}, \ and\ \bibinfo {author} {\bibfnamefont {G.~L.}\ \bibnamefont {Pimentel}},\ }\href {\doibase 10.1007/JHEP04(2020)105} {\bibfield  {journal} {\bibinfo  {journal} {JHEP}\ }\textbf {\bibinfo {volume} {04}},\ \bibinfo {pages} {105} (\bibinfo {year} {2020})},\ \Eprint {http://arxiv.org/abs/1811.00024} {arXiv:1811.00024 [hep-th]} \BibitemShut {NoStop}%
\bibitem [{\citenamefont {Dom\`enech}(2021)}]{Domenech:2021ztg}%
  \BibitemOpen
  \bibfield  {author} {\bibinfo {author} {\bibfnamefont {G.}~\bibnamefont {Dom\`enech}},\ }\href {\doibase 10.3390/universe7110398} {\bibfield  {journal} {\bibinfo  {journal} {Universe}\ }\textbf {\bibinfo {volume} {7}},\ \bibinfo {pages} {398} (\bibinfo {year} {2021})},\ \Eprint {http://arxiv.org/abs/2109.01398} {arXiv:2109.01398 [gr-qc]} \BibitemShut {NoStop}%
\bibitem [{\citenamefont {Ananda}\ \emph {et~al.}(2007)\citenamefont {Ananda}, \citenamefont {Clarkson},\ and\ \citenamefont {Wands}}]{Ananda:2006af}%
  \BibitemOpen
  \bibfield  {author} {\bibinfo {author} {\bibfnamefont {K.~N.}\ \bibnamefont {Ananda}}, \bibinfo {author} {\bibfnamefont {C.}~\bibnamefont {Clarkson}}, \ and\ \bibinfo {author} {\bibfnamefont {D.}~\bibnamefont {Wands}},\ }\href {\doibase 10.1103/PhysRevD.75.123518} {\bibfield  {journal} {\bibinfo  {journal} {Phys. Rev. D}\ }\textbf {\bibinfo {volume} {75}},\ \bibinfo {pages} {123518} (\bibinfo {year} {2007})},\ \Eprint {http://arxiv.org/abs/gr-qc/0612013} {arXiv:gr-qc/0612013} \BibitemShut {NoStop}%
\bibitem [{\citenamefont {Baumann}\ \emph {et~al.}(2007)\citenamefont {Baumann}, \citenamefont {Steinhardt}, \citenamefont {Takahashi},\ and\ \citenamefont {Ichiki}}]{Baumann:2007zm}%
  \BibitemOpen
  \bibfield  {author} {\bibinfo {author} {\bibfnamefont {D.}~\bibnamefont {Baumann}}, \bibinfo {author} {\bibfnamefont {P.~J.}\ \bibnamefont {Steinhardt}}, \bibinfo {author} {\bibfnamefont {K.}~\bibnamefont {Takahashi}}, \ and\ \bibinfo {author} {\bibfnamefont {K.}~\bibnamefont {Ichiki}},\ }\href {\doibase 10.1103/PhysRevD.76.084019} {\bibfield  {journal} {\bibinfo  {journal} {Phys. Rev. D}\ }\textbf {\bibinfo {volume} {76}},\ \bibinfo {pages} {084019} (\bibinfo {year} {2007})},\ \Eprint {http://arxiv.org/abs/hep-th/0703290} {arXiv:hep-th/0703290} \BibitemShut {NoStop}%
\bibitem [{\citenamefont {Chang}\ \emph {et~al.}(2024{\natexlab{a}})\citenamefont {Chang}, \citenamefont {Kuang}, \citenamefont {Wu},\ and\ \citenamefont {Zhou}}]{Chang:2023vjk}%
  \BibitemOpen
  \bibfield  {author} {\bibinfo {author} {\bibfnamefont {Z.}~\bibnamefont {Chang}}, \bibinfo {author} {\bibfnamefont {Y.-T.}\ \bibnamefont {Kuang}}, \bibinfo {author} {\bibfnamefont {D.}~\bibnamefont {Wu}}, \ and\ \bibinfo {author} {\bibfnamefont {J.-Z.}\ \bibnamefont {Zhou}},\ }\href {\doibase 10.1088/1475-7516/2024/04/044} {\bibfield  {journal} {\bibinfo  {journal} {JCAP}\ }\textbf {\bibinfo {volume} {2024}},\ \bibinfo {pages} {044} (\bibinfo {year} {2024}{\natexlab{a}})},\ \Eprint {http://arxiv.org/abs/2312.14409} {arXiv:2312.14409 [astro-ph.CO]} \BibitemShut {NoStop}%
\bibitem [{\citenamefont {Bari}\ \emph {et~al.}(2022)\citenamefont {Bari}, \citenamefont {Ricciardone}, \citenamefont {Bartolo}, \citenamefont {Bertacca},\ and\ \citenamefont {Matarrese}}]{Bari:2021xvf}%
  \BibitemOpen
  \bibfield  {author} {\bibinfo {author} {\bibfnamefont {P.}~\bibnamefont {Bari}}, \bibinfo {author} {\bibfnamefont {A.}~\bibnamefont {Ricciardone}}, \bibinfo {author} {\bibfnamefont {N.}~\bibnamefont {Bartolo}}, \bibinfo {author} {\bibfnamefont {D.}~\bibnamefont {Bertacca}}, \ and\ \bibinfo {author} {\bibfnamefont {S.}~\bibnamefont {Matarrese}},\ }\href {\doibase 10.1103/PhysRevLett.129.091301} {\bibfield  {journal} {\bibinfo  {journal} {Phys. Rev. Lett.}\ }\textbf {\bibinfo {volume} {129}},\ \bibinfo {pages} {091301} (\bibinfo {year} {2022})},\ \Eprint {http://arxiv.org/abs/2111.06884} {arXiv:2111.06884 [astro-ph.CO]} \BibitemShut {NoStop}%
\bibitem [{\citenamefont {Bari}\ \emph {et~al.}(2023)\citenamefont {Bari}, \citenamefont {Bertacca}, \citenamefont {Bartolo}, \citenamefont {Ricciardone}, \citenamefont {Giardiello},\ and\ \citenamefont {Matarrese}}]{Bari:2022grh}%
  \BibitemOpen
  \bibfield  {author} {\bibinfo {author} {\bibfnamefont {P.}~\bibnamefont {Bari}}, \bibinfo {author} {\bibfnamefont {D.}~\bibnamefont {Bertacca}}, \bibinfo {author} {\bibfnamefont {N.}~\bibnamefont {Bartolo}}, \bibinfo {author} {\bibfnamefont {A.}~\bibnamefont {Ricciardone}}, \bibinfo {author} {\bibfnamefont {S.}~\bibnamefont {Giardiello}}, \ and\ \bibinfo {author} {\bibfnamefont {S.}~\bibnamefont {Matarrese}},\ }\href {\doibase 10.1088/1475-7516/2023/07/034} {\bibfield  {journal} {\bibinfo  {journal} {JCAP}\ }\textbf {\bibinfo {volume} {07}},\ \bibinfo {pages} {034} (\bibinfo {year} {2023})},\ \Eprint {http://arxiv.org/abs/2209.05329} {arXiv:2209.05329 [astro-ph.CO]} \BibitemShut {NoStop}%
\bibitem [{\citenamefont {Abdelaziz}\ \emph {et~al.}(2025)\citenamefont {Abdelaziz}, \citenamefont {Bari}, \citenamefont {Matarrese},\ and\ \citenamefont {Ricciardone}}]{Abdelaziz:2025qpn}%
  \BibitemOpen
  \bibfield  {author} {\bibinfo {author} {\bibfnamefont {M.}~\bibnamefont {Abdelaziz}}, \bibinfo {author} {\bibfnamefont {P.}~\bibnamefont {Bari}}, \bibinfo {author} {\bibfnamefont {S.}~\bibnamefont {Matarrese}}, \ and\ \bibinfo {author} {\bibfnamefont {A.}~\bibnamefont {Ricciardone}},\ }\href {\doibase 10.1103/bb22-pq2m} {\bibfield  {journal} {\bibinfo  {journal} {Phys. Rev. D}\ }\textbf {\bibinfo {volume} {112}},\ \bibinfo {pages} {023505} (\bibinfo {year} {2025})},\ \Eprint {http://arxiv.org/abs/2504.07063} {arXiv:2504.07063 [astro-ph.CO]} \BibitemShut {NoStop}%
\bibitem [{\citenamefont {Chang}\ \emph {et~al.}(2024{\natexlab{b}})\citenamefont {Chang}, \citenamefont {Kuang}, \citenamefont {Zhang},\ and\ \citenamefont {Zhou}}]{Chang:2022aqk}%
  \BibitemOpen
  \bibfield  {author} {\bibinfo {author} {\bibfnamefont {Z.}~\bibnamefont {Chang}}, \bibinfo {author} {\bibfnamefont {Y.-T.}\ \bibnamefont {Kuang}}, \bibinfo {author} {\bibfnamefont {X.}~\bibnamefont {Zhang}}, \ and\ \bibinfo {author} {\bibfnamefont {J.-Z.}\ \bibnamefont {Zhou}},\ }\href {\doibase 10.3390/universe10010039} {\bibfield  {journal} {\bibinfo  {journal} {Universe}\ }\textbf {\bibinfo {volume} {10}},\ \bibinfo {pages} {39} (\bibinfo {year} {2024}{\natexlab{b}})},\ \Eprint {http://arxiv.org/abs/2211.11948} {arXiv:2211.11948 [astro-ph.CO]} \BibitemShut {NoStop}%
\bibitem [{\citenamefont {Zhou}\ \emph {et~al.}(2023)\citenamefont {Zhou}, \citenamefont {Kuang}, \citenamefont {Chang}, \citenamefont {Zhang},\ and\ \citenamefont {Zhu}}]{Zhou:2023itl}%
  \BibitemOpen
  \bibfield  {author} {\bibinfo {author} {\bibfnamefont {J.-Z.}\ \bibnamefont {Zhou}}, \bibinfo {author} {\bibfnamefont {Y.-T.}\ \bibnamefont {Kuang}}, \bibinfo {author} {\bibfnamefont {Z.}~\bibnamefont {Chang}}, \bibinfo {author} {\bibfnamefont {X.}~\bibnamefont {Zhang}}, \ and\ \bibinfo {author} {\bibfnamefont {Q.-H.}\ \bibnamefont {Zhu}},\ }\href@noop {} {\  (\bibinfo {year} {2023})},\ \Eprint {http://arxiv.org/abs/2307.02067} {arXiv:2307.02067 [astro-ph.CO]} \BibitemShut {NoStop}%
\bibitem [{\citenamefont {Nakama}\ and\ \citenamefont {Suyama}(2016)}]{Nakama:2016enz}%
  \BibitemOpen
  \bibfield  {author} {\bibinfo {author} {\bibfnamefont {T.}~\bibnamefont {Nakama}}\ and\ \bibinfo {author} {\bibfnamefont {T.}~\bibnamefont {Suyama}},\ }\href {\doibase 10.1103/PhysRevD.94.043507} {\bibfield  {journal} {\bibinfo  {journal} {Phys. Rev. D}\ }\textbf {\bibinfo {volume} {94}},\ \bibinfo {pages} {043507} (\bibinfo {year} {2016})},\ \Eprint {http://arxiv.org/abs/1605.04482} {arXiv:1605.04482 [gr-qc]} \BibitemShut {NoStop}%
\bibitem [{\citenamefont {Nakama}\ and\ \citenamefont {Suyama}(2015)}]{Nakama:2015nea}%
  \BibitemOpen
  \bibfield  {author} {\bibinfo {author} {\bibfnamefont {T.}~\bibnamefont {Nakama}}\ and\ \bibinfo {author} {\bibfnamefont {T.}~\bibnamefont {Suyama}},\ }\href {\doibase 10.1103/PhysRevD.92.121304} {\bibfield  {journal} {\bibinfo  {journal} {Phys. Rev. D}\ }\textbf {\bibinfo {volume} {92}},\ \bibinfo {pages} {121304} (\bibinfo {year} {2015})},\ \Eprint {http://arxiv.org/abs/1506.05228} {arXiv:1506.05228 [gr-qc]} \BibitemShut {NoStop}%
\bibitem [{\citenamefont {De~Luca}\ \emph {et~al.}(2023)\citenamefont {De~Luca}, \citenamefont {Kehagias},\ and\ \citenamefont {Riotto}}]{DeLuca:2023tun}%
  \BibitemOpen
  \bibfield  {author} {\bibinfo {author} {\bibfnamefont {V.}~\bibnamefont {De~Luca}}, \bibinfo {author} {\bibfnamefont {A.}~\bibnamefont {Kehagias}}, \ and\ \bibinfo {author} {\bibfnamefont {A.}~\bibnamefont {Riotto}},\ }\href {\doibase 10.1103/PhysRevD.108.063531} {\bibfield  {journal} {\bibinfo  {journal} {Phys. Rev. D}\ }\textbf {\bibinfo {volume} {108}},\ \bibinfo {pages} {063531} (\bibinfo {year} {2023})},\ \Eprint {http://arxiv.org/abs/2307.13633} {arXiv:2307.13633 [astro-ph.CO]} \BibitemShut {NoStop}%
\bibitem [{\citenamefont {Mollerach}\ \emph {et~al.}(2004)\citenamefont {Mollerach}, \citenamefont {Harari},\ and\ \citenamefont {Matarrese}}]{Mollerach:2003nq}%
  \BibitemOpen
  \bibfield  {author} {\bibinfo {author} {\bibfnamefont {S.}~\bibnamefont {Mollerach}}, \bibinfo {author} {\bibfnamefont {D.}~\bibnamefont {Harari}}, \ and\ \bibinfo {author} {\bibfnamefont {S.}~\bibnamefont {Matarrese}},\ }\href {\doibase 10.1103/PhysRevD.69.063002} {\bibfield  {journal} {\bibinfo  {journal} {Phys. Rev. D}\ }\textbf {\bibinfo {volume} {69}},\ \bibinfo {pages} {063002} (\bibinfo {year} {2004})},\ \Eprint {http://arxiv.org/abs/astro-ph/0310711} {arXiv:astro-ph/0310711} \BibitemShut {NoStop}%
\bibitem [{\citenamefont {Cyr}\ \emph {et~al.}(2024)\citenamefont {Cyr}, \citenamefont {Kite}, \citenamefont {Chluba}, \citenamefont {Hill}, \citenamefont {Jeong}, \citenamefont {Acharya}, \citenamefont {Bolliet},\ and\ \citenamefont {Patil}}]{Cyr:2023pgw}%
  \BibitemOpen
  \bibfield  {author} {\bibinfo {author} {\bibfnamefont {B.}~\bibnamefont {Cyr}}, \bibinfo {author} {\bibfnamefont {T.}~\bibnamefont {Kite}}, \bibinfo {author} {\bibfnamefont {J.}~\bibnamefont {Chluba}}, \bibinfo {author} {\bibfnamefont {J.~C.}\ \bibnamefont {Hill}}, \bibinfo {author} {\bibfnamefont {D.}~\bibnamefont {Jeong}}, \bibinfo {author} {\bibfnamefont {S.~K.}\ \bibnamefont {Acharya}}, \bibinfo {author} {\bibfnamefont {B.}~\bibnamefont {Bolliet}}, \ and\ \bibinfo {author} {\bibfnamefont {S.~P.}\ \bibnamefont {Patil}},\ }\href {\doibase 10.1093/mnras/stad3861} {\bibfield  {journal} {\bibinfo  {journal} {Mon. Not. Roy. Astron. Soc.}\ }\textbf {\bibinfo {volume} {528}},\ \bibinfo {pages} {883} (\bibinfo {year} {2024})},\ \Eprint {http://arxiv.org/abs/2309.02366} {arXiv:2309.02366 [astro-ph.CO]} \BibitemShut {NoStop}%
\bibitem [{\citenamefont {Gurian}\ \emph {et~al.}(2021)\citenamefont {Gurian}, \citenamefont {Jeong}, \citenamefont {Hwang},\ and\ \citenamefont {Noh}}]{Gurian:2021rfv}%
  \BibitemOpen
  \bibfield  {author} {\bibinfo {author} {\bibfnamefont {J.}~\bibnamefont {Gurian}}, \bibinfo {author} {\bibfnamefont {D.}~\bibnamefont {Jeong}}, \bibinfo {author} {\bibfnamefont {J.-c.}\ \bibnamefont {Hwang}}, \ and\ \bibinfo {author} {\bibfnamefont {H.}~\bibnamefont {Noh}},\ }\href {\doibase 10.1103/PhysRevD.104.083534} {\bibfield  {journal} {\bibinfo  {journal} {Phys. Rev. D}\ }\textbf {\bibinfo {volume} {104}},\ \bibinfo {pages} {083534} (\bibinfo {year} {2021})},\ \Eprint {http://arxiv.org/abs/2104.03330} {arXiv:2104.03330 [astro-ph.CO]} \BibitemShut {NoStop}%
\bibitem [{\citenamefont {D'Amico}\ \emph {et~al.}(2008)\citenamefont {D'Amico}, \citenamefont {Bartolo}, \citenamefont {Matarrese},\ and\ \citenamefont {Riotto}}]{DAmico:2007ngk}%
  \BibitemOpen
  \bibfield  {author} {\bibinfo {author} {\bibfnamefont {G.}~\bibnamefont {D'Amico}}, \bibinfo {author} {\bibfnamefont {N.}~\bibnamefont {Bartolo}}, \bibinfo {author} {\bibfnamefont {S.}~\bibnamefont {Matarrese}}, \ and\ \bibinfo {author} {\bibfnamefont {A.}~\bibnamefont {Riotto}},\ }\href {\doibase 10.1088/1475-7516/2008/01/005} {\bibfield  {journal} {\bibinfo  {journal} {JCAP}\ }\textbf {\bibinfo {volume} {01}},\ \bibinfo {pages} {005} (\bibinfo {year} {2008})},\ \Eprint {http://arxiv.org/abs/0707.2894} {arXiv:0707.2894 [astro-ph]} \BibitemShut {NoStop}%
\bibitem [{\citenamefont {Yamauchi}\ \emph {et~al.}(2012)\citenamefont {Yamauchi}, \citenamefont {Namikawa},\ and\ \citenamefont {Taruya}}]{Yamauchi:2012bc}%
  \BibitemOpen
  \bibfield  {author} {\bibinfo {author} {\bibfnamefont {D.}~\bibnamefont {Yamauchi}}, \bibinfo {author} {\bibfnamefont {T.}~\bibnamefont {Namikawa}}, \ and\ \bibinfo {author} {\bibfnamefont {A.}~\bibnamefont {Taruya}},\ }\href {\doibase 10.1088/1475-7516/2012/10/030} {\bibfield  {journal} {\bibinfo  {journal} {JCAP}\ }\textbf {\bibinfo {volume} {10}},\ \bibinfo {pages} {030} (\bibinfo {year} {2012})},\ \Eprint {http://arxiv.org/abs/1205.2139} {arXiv:1205.2139 [astro-ph.CO]} \BibitemShut {NoStop}%
\bibitem [{\citenamefont {Saga}\ \emph {et~al.}(2015{\natexlab{a}})\citenamefont {Saga}, \citenamefont {Yamauchi},\ and\ \citenamefont {Ichiki}}]{Saga:2015apa}%
  \BibitemOpen
  \bibfield  {author} {\bibinfo {author} {\bibfnamefont {S.}~\bibnamefont {Saga}}, \bibinfo {author} {\bibfnamefont {D.}~\bibnamefont {Yamauchi}}, \ and\ \bibinfo {author} {\bibfnamefont {K.}~\bibnamefont {Ichiki}},\ }\href {\doibase 10.1103/PhysRevD.92.063533} {\bibfield  {journal} {\bibinfo  {journal} {Phys. Rev. D}\ }\textbf {\bibinfo {volume} {92}},\ \bibinfo {pages} {063533} (\bibinfo {year} {2015}{\natexlab{a}})},\ \Eprint {http://arxiv.org/abs/1505.02774} {arXiv:1505.02774 [astro-ph.CO]} \BibitemShut {NoStop}%
\bibitem [{\citenamefont {Aghanim}\ \emph {et~al.}(2020)\citenamefont {Aghanim} \emph {et~al.}}]{Planck:2018vyg}%
  \BibitemOpen
  \bibfield  {author} {\bibinfo {author} {\bibfnamefont {N.}~\bibnamefont {Aghanim}} \emph {et~al.} (\bibinfo {collaboration} {Planck}),\ }\href {\doibase 10.1051/0004-6361/201833910} {\bibfield  {journal} {\bibinfo  {journal} {Astron. Astrophys.}\ }\textbf {\bibinfo {volume} {641}},\ \bibinfo {pages} {A6} (\bibinfo {year} {2020})},\ \bibinfo {note} {[Erratum: Astron.Astrophys. 652, C4 (2021)]},\ \Eprint {http://arxiv.org/abs/1807.06209} {arXiv:1807.06209 [astro-ph.CO]} \BibitemShut {NoStop}%
\bibitem [{\citenamefont {Afzal}\ \emph {et~al.}(2023)\citenamefont {Afzal} \emph {et~al.}}]{NANOGrav:2023hvm}%
  \BibitemOpen
  \bibfield  {author} {\bibinfo {author} {\bibfnamefont {A.}~\bibnamefont {Afzal}} \emph {et~al.} (\bibinfo {collaboration} {NANOGrav}),\ }\href {\doibase 10.3847/2041-8213/acdc91} {\bibfield  {journal} {\bibinfo  {journal} {Astrophys. J. Lett.}\ }\textbf {\bibinfo {volume} {951}},\ \bibinfo {pages} {L11} (\bibinfo {year} {2023})},\ \bibinfo {note} {[Erratum: Astrophys.J.Lett. 971, L27 (2024), Erratum: Astrophys.J. 971, L27 (2024)]},\ \Eprint {http://arxiv.org/abs/2306.16219} {arXiv:2306.16219 [astro-ph.HE]} \BibitemShut {NoStop}%
\bibitem [{\citenamefont {Agazie}\ \emph {et~al.}(2023)\citenamefont {Agazie} \emph {et~al.}}]{NANOGrav:2023gor}%
  \BibitemOpen
  \bibfield  {author} {\bibinfo {author} {\bibfnamefont {G.}~\bibnamefont {Agazie}} \emph {et~al.} (\bibinfo {collaboration} {NANOGrav}),\ }\href {\doibase 10.3847/2041-8213/acdac6} {\bibfield  {journal} {\bibinfo  {journal} {Astrophys. J. Lett.}\ }\textbf {\bibinfo {volume} {951}},\ \bibinfo {pages} {L8} (\bibinfo {year} {2023})},\ \Eprint {http://arxiv.org/abs/2306.16213} {arXiv:2306.16213 [astro-ph.HE]} \BibitemShut {NoStop}%
\bibitem [{\citenamefont {Antoniadis}\ \emph {et~al.}(2023)\citenamefont {Antoniadis} \emph {et~al.}}]{EPTA:2023fyk}%
  \BibitemOpen
  \bibfield  {author} {\bibinfo {author} {\bibfnamefont {J.}~\bibnamefont {Antoniadis}} \emph {et~al.} (\bibinfo {collaboration} {EPTA, InPTA:}),\ }\href {\doibase 10.1051/0004-6361/202346844} {\bibfield  {journal} {\bibinfo  {journal} {Astron. Astrophys.}\ }\textbf {\bibinfo {volume} {678}},\ \bibinfo {pages} {A50} (\bibinfo {year} {2023})},\ \Eprint {http://arxiv.org/abs/2306.16214} {arXiv:2306.16214 [astro-ph.HE]} \BibitemShut {NoStop}%
\bibitem [{\citenamefont {Reardon}\ \emph {et~al.}(2023)\citenamefont {Reardon} \emph {et~al.}}]{Reardon:2023gzh}%
  \BibitemOpen
  \bibfield  {author} {\bibinfo {author} {\bibfnamefont {D.~J.}\ \bibnamefont {Reardon}} \emph {et~al.},\ }\href {\doibase 10.3847/2041-8213/acdd02} {\bibfield  {journal} {\bibinfo  {journal} {Astrophys. J. Lett.}\ }\textbf {\bibinfo {volume} {951}},\ \bibinfo {pages} {L6} (\bibinfo {year} {2023})},\ \Eprint {http://arxiv.org/abs/2306.16215} {arXiv:2306.16215 [astro-ph.HE]} \BibitemShut {NoStop}%
\bibitem [{\citenamefont {Xu}\ \emph {et~al.}(2023)\citenamefont {Xu} \emph {et~al.}}]{Xu:2023wog}%
  \BibitemOpen
  \bibfield  {author} {\bibinfo {author} {\bibfnamefont {H.}~\bibnamefont {Xu}} \emph {et~al.},\ }\href {\doibase 10.1088/1674-4527/acdfa5} {\bibfield  {journal} {\bibinfo  {journal} {Res. Astron. Astrophys.}\ }\textbf {\bibinfo {volume} {23}},\ \bibinfo {pages} {075024} (\bibinfo {year} {2023})},\ \Eprint {http://arxiv.org/abs/2306.16216} {arXiv:2306.16216 [astro-ph.HE]} \BibitemShut {NoStop}%
\bibitem [{\citenamefont {Assadullahi}\ and\ \citenamefont {Wands}(2009)}]{Assadullahi:2009nf}%
  \BibitemOpen
  \bibfield  {author} {\bibinfo {author} {\bibfnamefont {H.}~\bibnamefont {Assadullahi}}\ and\ \bibinfo {author} {\bibfnamefont {D.}~\bibnamefont {Wands}},\ }\href {\doibase 10.1103/PhysRevD.79.083511} {\bibfield  {journal} {\bibinfo  {journal} {Phys. Rev. D}\ }\textbf {\bibinfo {volume} {79}},\ \bibinfo {pages} {083511} (\bibinfo {year} {2009})},\ \Eprint {http://arxiv.org/abs/0901.0989} {arXiv:0901.0989 [astro-ph.CO]} \BibitemShut {NoStop}%
\bibitem [{\citenamefont {Alabidi}\ \emph {et~al.}(2013)\citenamefont {Alabidi}, \citenamefont {Kohri}, \citenamefont {Sasaki},\ and\ \citenamefont {Sendouda}}]{Alabidi:2013lya}%
  \BibitemOpen
  \bibfield  {author} {\bibinfo {author} {\bibfnamefont {L.}~\bibnamefont {Alabidi}}, \bibinfo {author} {\bibfnamefont {K.}~\bibnamefont {Kohri}}, \bibinfo {author} {\bibfnamefont {M.}~\bibnamefont {Sasaki}}, \ and\ \bibinfo {author} {\bibfnamefont {Y.}~\bibnamefont {Sendouda}},\ }\href {\doibase 10.1088/1475-7516/2013/05/033} {\bibfield  {journal} {\bibinfo  {journal} {JCAP}\ }\textbf {\bibinfo {volume} {05}},\ \bibinfo {pages} {033} (\bibinfo {year} {2013})},\ \Eprint {http://arxiv.org/abs/1303.4519} {arXiv:1303.4519 [astro-ph.CO]} \BibitemShut {NoStop}%
\bibitem [{\citenamefont {Dom\`enech}\ \emph {et~al.}(2021)\citenamefont {Dom\`enech}, \citenamefont {Lin},\ and\ \citenamefont {Sasaki}}]{Domenech:2020ssp}%
  \BibitemOpen
  \bibfield  {author} {\bibinfo {author} {\bibfnamefont {G.}~\bibnamefont {Dom\`enech}}, \bibinfo {author} {\bibfnamefont {C.}~\bibnamefont {Lin}}, \ and\ \bibinfo {author} {\bibfnamefont {M.}~\bibnamefont {Sasaki}},\ }\href {\doibase 10.1088/1475-7516/2021/11/E01} {\bibfield  {journal} {\bibinfo  {journal} {JCAP}\ }\textbf {\bibinfo {volume} {04}},\ \bibinfo {pages} {062} (\bibinfo {year} {2021})},\ \bibinfo {note} {[Erratum: JCAP 11, E01 (2021)]},\ \Eprint {http://arxiv.org/abs/2012.08151} {arXiv:2012.08151 [gr-qc]} \BibitemShut {NoStop}%
\bibitem [{\citenamefont {Dom\`enech}\ \emph {et~al.}(2020)\citenamefont {Dom\`enech}, \citenamefont {Pi},\ and\ \citenamefont {Sasaki}}]{Domenech:2020kqm}%
  \BibitemOpen
  \bibfield  {author} {\bibinfo {author} {\bibfnamefont {G.}~\bibnamefont {Dom\`enech}}, \bibinfo {author} {\bibfnamefont {S.}~\bibnamefont {Pi}}, \ and\ \bibinfo {author} {\bibfnamefont {M.}~\bibnamefont {Sasaki}},\ }\href {\doibase 10.1088/1475-7516/2020/08/017} {\bibfield  {journal} {\bibinfo  {journal} {JCAP}\ }\textbf {\bibinfo {volume} {08}},\ \bibinfo {pages} {017} (\bibinfo {year} {2020})},\ \Eprint {http://arxiv.org/abs/2005.12314} {arXiv:2005.12314 [gr-qc]} \BibitemShut {NoStop}%
\bibitem [{\citenamefont {Inomata}\ \emph {et~al.}(2019{\natexlab{a}})\citenamefont {Inomata}, \citenamefont {Kohri}, \citenamefont {Nakama},\ and\ \citenamefont {Terada}}]{Inomata:2019zqy}%
  \BibitemOpen
  \bibfield  {author} {\bibinfo {author} {\bibfnamefont {K.}~\bibnamefont {Inomata}}, \bibinfo {author} {\bibfnamefont {K.}~\bibnamefont {Kohri}}, \bibinfo {author} {\bibfnamefont {T.}~\bibnamefont {Nakama}}, \ and\ \bibinfo {author} {\bibfnamefont {T.}~\bibnamefont {Terada}},\ }\href {\doibase 10.1088/1475-7516/2019/10/071} {\bibfield  {journal} {\bibinfo  {journal} {JCAP}\ }\textbf {\bibinfo {volume} {10}},\ \bibinfo {pages} {071} (\bibinfo {year} {2019}{\natexlab{a}})},\ \bibinfo {note} {[Erratum: JCAP 08, E01 (2023)]},\ \Eprint {http://arxiv.org/abs/1904.12878} {arXiv:1904.12878 [astro-ph.CO]} \BibitemShut {NoStop}%
\bibitem [{\citenamefont {Wang}\ \emph {et~al.}(2019)\citenamefont {Wang}, \citenamefont {Terada},\ and\ \citenamefont {Kohri}}]{Wang:2019kaf}%
  \BibitemOpen
  \bibfield  {author} {\bibinfo {author} {\bibfnamefont {S.}~\bibnamefont {Wang}}, \bibinfo {author} {\bibfnamefont {T.}~\bibnamefont {Terada}}, \ and\ \bibinfo {author} {\bibfnamefont {K.}~\bibnamefont {Kohri}},\ }\href {\doibase 10.1103/PhysRevD.99.103531} {\bibfield  {journal} {\bibinfo  {journal} {Phys. Rev. D}\ }\textbf {\bibinfo {volume} {99}},\ \bibinfo {pages} {103531} (\bibinfo {year} {2019})},\ \bibinfo {note} {[Erratum: Phys.Rev.D 101, 069901 (2020)]},\ \Eprint {http://arxiv.org/abs/1903.05924} {arXiv:1903.05924 [astro-ph.CO]} \BibitemShut {NoStop}%
\bibitem [{\citenamefont {Liu}\ \emph {et~al.}(2023)\citenamefont {Liu}, \citenamefont {Chen},\ and\ \citenamefont {Huang}}]{Liu:2023pau}%
  \BibitemOpen
  \bibfield  {author} {\bibinfo {author} {\bibfnamefont {L.}~\bibnamefont {Liu}}, \bibinfo {author} {\bibfnamefont {Z.-C.}\ \bibnamefont {Chen}}, \ and\ \bibinfo {author} {\bibfnamefont {Q.-G.}\ \bibnamefont {Huang}},\ }\href {\doibase 10.1088/1475-7516/2023/11/071} {\bibfield  {journal} {\bibinfo  {journal} {JCAP}\ }\textbf {\bibinfo {volume} {11}},\ \bibinfo {pages} {071} (\bibinfo {year} {2023})},\ \Eprint {http://arxiv.org/abs/2307.14911} {arXiv:2307.14911 [astro-ph.CO]} \BibitemShut {NoStop}%
\bibitem [{\citenamefont {Dom\`enech}\ \emph {et~al.}(2024)\citenamefont {Dom\`enech}, \citenamefont {Pi}, \citenamefont {Wang},\ and\ \citenamefont {Wang}}]{Domenech:2024rks}%
  \BibitemOpen
  \bibfield  {author} {\bibinfo {author} {\bibfnamefont {G.}~\bibnamefont {Dom\`enech}}, \bibinfo {author} {\bibfnamefont {S.}~\bibnamefont {Pi}}, \bibinfo {author} {\bibfnamefont {A.}~\bibnamefont {Wang}}, \ and\ \bibinfo {author} {\bibfnamefont {J.}~\bibnamefont {Wang}},\ }\href {\doibase 10.1088/1475-7516/2024/08/054} {\bibfield  {journal} {\bibinfo  {journal} {JCAP}\ }\textbf {\bibinfo {volume} {08}},\ \bibinfo {pages} {054} (\bibinfo {year} {2024})},\ \Eprint {http://arxiv.org/abs/2402.18965} {arXiv:2402.18965 [astro-ph.CO]} \BibitemShut {NoStop}%
\bibitem [{\citenamefont {Harigaya}\ \emph {et~al.}(2023)\citenamefont {Harigaya}, \citenamefont {Inomata},\ and\ \citenamefont {Terada}}]{Harigaya:2023pmw}%
  \BibitemOpen
  \bibfield  {author} {\bibinfo {author} {\bibfnamefont {K.}~\bibnamefont {Harigaya}}, \bibinfo {author} {\bibfnamefont {K.}~\bibnamefont {Inomata}}, \ and\ \bibinfo {author} {\bibfnamefont {T.}~\bibnamefont {Terada}},\ }\href {\doibase 10.1103/PhysRevD.108.123538} {\bibfield  {journal} {\bibinfo  {journal} {Phys. Rev. D}\ }\textbf {\bibinfo {volume} {108}},\ \bibinfo {pages} {123538} (\bibinfo {year} {2023})},\ \Eprint {http://arxiv.org/abs/2309.00228} {arXiv:2309.00228 [astro-ph.CO]} \BibitemShut {NoStop}%
\bibitem [{\citenamefont {Zhu}\ \emph {et~al.}(2024)\citenamefont {Zhu}, \citenamefont {Zhao}, \citenamefont {Wang},\ and\ \citenamefont {Zhang}}]{Zhu:2023gmx}%
  \BibitemOpen
  \bibfield  {author} {\bibinfo {author} {\bibfnamefont {Q.-H.}\ \bibnamefont {Zhu}}, \bibinfo {author} {\bibfnamefont {Z.-C.}\ \bibnamefont {Zhao}}, \bibinfo {author} {\bibfnamefont {S.}~\bibnamefont {Wang}}, \ and\ \bibinfo {author} {\bibfnamefont {X.}~\bibnamefont {Zhang}},\ }\href {\doibase 10.1088/1674-1137/ad79d5} {\bibfield  {journal} {\bibinfo  {journal} {Chin. Phys. C}\ }\textbf {\bibinfo {volume} {48}},\ \bibinfo {pages} {125105} (\bibinfo {year} {2024})},\ \Eprint {http://arxiv.org/abs/2307.13574} {arXiv:2307.13574 [astro-ph.CO]} \BibitemShut {NoStop}%
\bibitem [{\citenamefont {Pearce}\ \emph {et~al.}(2024)\citenamefont {Pearce}, \citenamefont {Pearce}, \citenamefont {White},\ and\ \citenamefont {Balazs}}]{Pearce:2023kxp}%
  \BibitemOpen
  \bibfield  {author} {\bibinfo {author} {\bibfnamefont {M.}~\bibnamefont {Pearce}}, \bibinfo {author} {\bibfnamefont {L.}~\bibnamefont {Pearce}}, \bibinfo {author} {\bibfnamefont {G.}~\bibnamefont {White}}, \ and\ \bibinfo {author} {\bibfnamefont {C.}~\bibnamefont {Balazs}},\ }\href {\doibase 10.1088/1475-7516/2024/06/021} {\bibfield  {journal} {\bibinfo  {journal} {JCAP}\ }\textbf {\bibinfo {volume} {06}},\ \bibinfo {pages} {021} (\bibinfo {year} {2024})},\ \Eprint {http://arxiv.org/abs/2311.12340} {arXiv:2311.12340 [astro-ph.CO]} \BibitemShut {NoStop}%
\bibitem [{\citenamefont {Inomata}\ \emph {et~al.}(2019{\natexlab{b}})\citenamefont {Inomata}, \citenamefont {Kohri}, \citenamefont {Nakama},\ and\ \citenamefont {Terada}}]{Inomata:2019ivs}%
  \BibitemOpen
  \bibfield  {author} {\bibinfo {author} {\bibfnamefont {K.}~\bibnamefont {Inomata}}, \bibinfo {author} {\bibfnamefont {K.}~\bibnamefont {Kohri}}, \bibinfo {author} {\bibfnamefont {T.}~\bibnamefont {Nakama}}, \ and\ \bibinfo {author} {\bibfnamefont {T.}~\bibnamefont {Terada}},\ }\href {\doibase 10.1103/PhysRevD.108.049901} {\bibfield  {journal} {\bibinfo  {journal} {Phys. Rev. D}\ }\textbf {\bibinfo {volume} {100}},\ \bibinfo {pages} {043532} (\bibinfo {year} {2019}{\natexlab{b}})},\ \bibinfo {note} {[Erratum: Phys.Rev.D 108, 049901 (2023)]},\ \Eprint {http://arxiv.org/abs/1904.12879} {arXiv:1904.12879 [astro-ph.CO]} \BibitemShut {NoStop}%
\bibitem [{\citenamefont {Zhang}\ \emph {et~al.}(2022)\citenamefont {Zhang}, \citenamefont {Zhou},\ and\ \citenamefont {Chang}}]{Zhang:2022dgx}%
  \BibitemOpen
  \bibfield  {author} {\bibinfo {author} {\bibfnamefont {X.}~\bibnamefont {Zhang}}, \bibinfo {author} {\bibfnamefont {J.-Z.}\ \bibnamefont {Zhou}}, \ and\ \bibinfo {author} {\bibfnamefont {Z.}~\bibnamefont {Chang}},\ }\href {\doibase 10.1140/epjc/s10052-022-10742-x} {\bibfield  {journal} {\bibinfo  {journal} {Eur. Phys. J. C}\ }\textbf {\bibinfo {volume} {82}},\ \bibinfo {pages} {781} (\bibinfo {year} {2022})},\ \Eprint {http://arxiv.org/abs/2208.12948} {arXiv:2208.12948 [astro-ph.CO]} \BibitemShut {NoStop}%
\bibitem [{\citenamefont {Mangilli}\ \emph {et~al.}(2008)\citenamefont {Mangilli}, \citenamefont {Bartolo}, \citenamefont {Matarrese},\ and\ \citenamefont {Riotto}}]{Mangilli:2008bw}%
  \BibitemOpen
  \bibfield  {author} {\bibinfo {author} {\bibfnamefont {A.}~\bibnamefont {Mangilli}}, \bibinfo {author} {\bibfnamefont {N.}~\bibnamefont {Bartolo}}, \bibinfo {author} {\bibfnamefont {S.}~\bibnamefont {Matarrese}}, \ and\ \bibinfo {author} {\bibfnamefont {A.}~\bibnamefont {Riotto}},\ }\href {\doibase 10.1103/PhysRevD.78.083517} {\bibfield  {journal} {\bibinfo  {journal} {Phys. Rev. D}\ }\textbf {\bibinfo {volume} {78}},\ \bibinfo {pages} {083517} (\bibinfo {year} {2008})},\ \Eprint {http://arxiv.org/abs/0805.3234} {arXiv:0805.3234 [astro-ph]} \BibitemShut {NoStop}%
\bibitem [{\citenamefont {Yu}\ and\ \citenamefont {Wang}(2025)}]{Yu:2024xmz}%
  \BibitemOpen
  \bibfield  {author} {\bibinfo {author} {\bibfnamefont {Y.-H.}\ \bibnamefont {Yu}}\ and\ \bibinfo {author} {\bibfnamefont {S.}~\bibnamefont {Wang}},\ }\href {\doibase 10.1007/s11433-024-2499-9} {\bibfield  {journal} {\bibinfo  {journal} {Sci. China Phys. Mech. Astron.}\ }\textbf {\bibinfo {volume} {68}},\ \bibinfo {pages} {210412} (\bibinfo {year} {2025})},\ \Eprint {http://arxiv.org/abs/2405.02960} {arXiv:2405.02960 [astro-ph.CO]} \BibitemShut {NoStop}%
\bibitem [{\citenamefont {Loverde}\ and\ \citenamefont {Weiner}(2023)}]{Loverde:2022wih}%
  \BibitemOpen
  \bibfield  {author} {\bibinfo {author} {\bibfnamefont {M.}~\bibnamefont {Loverde}}\ and\ \bibinfo {author} {\bibfnamefont {Z.~J.}\ \bibnamefont {Weiner}},\ }\href {\doibase 10.1088/1475-7516/2023/02/064} {\bibfield  {journal} {\bibinfo  {journal} {JCAP}\ }\textbf {\bibinfo {volume} {02}},\ \bibinfo {pages} {064} (\bibinfo {year} {2023})},\ \Eprint {http://arxiv.org/abs/2208.11714} {arXiv:2208.11714 [astro-ph.CO]} \BibitemShut {NoStop}%
\bibitem [{\citenamefont {Saga}\ \emph {et~al.}(2015{\natexlab{b}})\citenamefont {Saga}, \citenamefont {Ichiki},\ and\ \citenamefont {Sugiyama}}]{Saga:2014jca}%
  \BibitemOpen
  \bibfield  {author} {\bibinfo {author} {\bibfnamefont {S.}~\bibnamefont {Saga}}, \bibinfo {author} {\bibfnamefont {K.}~\bibnamefont {Ichiki}}, \ and\ \bibinfo {author} {\bibfnamefont {N.}~\bibnamefont {Sugiyama}},\ }\href {\doibase 10.1103/PhysRevD.91.024030} {\bibfield  {journal} {\bibinfo  {journal} {Phys. Rev. D}\ }\textbf {\bibinfo {volume} {91}},\ \bibinfo {pages} {024030} (\bibinfo {year} {2015}{\natexlab{b}})},\ \Eprint {http://arxiv.org/abs/1412.1081} {arXiv:1412.1081 [astro-ph.CO]} \BibitemShut {NoStop}%
\bibitem [{\citenamefont {Sui}\ \emph {et~al.}(2024)\citenamefont {Sui}, \citenamefont {Liu}, \citenamefont {Yang},\ and\ \citenamefont {Cai}}]{Sui:2024nip}%
  \BibitemOpen
  \bibfield  {author} {\bibinfo {author} {\bibfnamefont {X.-B.}\ \bibnamefont {Sui}}, \bibinfo {author} {\bibfnamefont {J.}~\bibnamefont {Liu}}, \bibinfo {author} {\bibfnamefont {X.-Y.}\ \bibnamefont {Yang}}, \ and\ \bibinfo {author} {\bibfnamefont {R.-G.}\ \bibnamefont {Cai}},\ }\href {\doibase 10.1103/PhysRevD.110.103541} {\bibfield  {journal} {\bibinfo  {journal} {Phys. Rev. D}\ }\textbf {\bibinfo {volume} {110}},\ \bibinfo {pages} {103541} (\bibinfo {year} {2024})},\ \Eprint {http://arxiv.org/abs/2407.04220} {arXiv:2407.04220 [astro-ph.CO]} \BibitemShut {NoStop}%
\bibitem [{\citenamefont {Madge}\ \emph {et~al.}(2023)\citenamefont {Madge}, \citenamefont {Morgante}, \citenamefont {Puchades-Ib\'a\~nez}, \citenamefont {Ramberg}, \citenamefont {Ratzinger}, \citenamefont {Schenk},\ and\ \citenamefont {Schwaller}}]{Madge:2023dxc}%
  \BibitemOpen
  \bibfield  {author} {\bibinfo {author} {\bibfnamefont {E.}~\bibnamefont {Madge}}, \bibinfo {author} {\bibfnamefont {E.}~\bibnamefont {Morgante}}, \bibinfo {author} {\bibfnamefont {C.}~\bibnamefont {Puchades-Ib\'a\~nez}}, \bibinfo {author} {\bibfnamefont {N.}~\bibnamefont {Ramberg}}, \bibinfo {author} {\bibfnamefont {W.}~\bibnamefont {Ratzinger}}, \bibinfo {author} {\bibfnamefont {S.}~\bibnamefont {Schenk}}, \ and\ \bibinfo {author} {\bibfnamefont {P.}~\bibnamefont {Schwaller}},\ }\href {\doibase 10.1007/JHEP10(2023)171} {\bibfield  {journal} {\bibinfo  {journal} {JHEP}\ }\textbf {\bibinfo {volume} {10}},\ \bibinfo {pages} {171} (\bibinfo {year} {2023})},\ \Eprint {http://arxiv.org/abs/2306.14856} {arXiv:2306.14856 [hep-ph]} \BibitemShut {NoStop}%
\bibitem [{\citenamefont {Dom{\`e}nech}\ and\ \citenamefont {Pi}(2022)}]{Domenech:2020ers}%
  \BibitemOpen
  \bibfield  {author} {\bibinfo {author} {\bibfnamefont {G.}~\bibnamefont {Dom{\`e}nech}}\ and\ \bibinfo {author} {\bibfnamefont {S.}~\bibnamefont {Pi}},\ }\href {\doibase 10.1007/s11433-021-1839-6} {\bibfield  {journal} {\bibinfo  {journal} {Sci. China Phys. Mech. Astron.}\ }\textbf {\bibinfo {volume} {65}},\ \bibinfo {pages} {230411} (\bibinfo {year} {2022})},\ \Eprint {http://arxiv.org/abs/2010.03976} {arXiv:2010.03976 [astro-ph.CO]} \BibitemShut {NoStop}%
\bibitem [{\citenamefont {Hwang}\ \emph {et~al.}(2017)\citenamefont {Hwang}, \citenamefont {Jeong},\ and\ \citenamefont {Noh}}]{Hwang:2017oxa}%
  \BibitemOpen
  \bibfield  {author} {\bibinfo {author} {\bibfnamefont {J.-C.}\ \bibnamefont {Hwang}}, \bibinfo {author} {\bibfnamefont {D.}~\bibnamefont {Jeong}}, \ and\ \bibinfo {author} {\bibfnamefont {H.}~\bibnamefont {Noh}},\ }\href {\doibase 10.3847/1538-4357/aa74be} {\bibfield  {journal} {\bibinfo  {journal} {Astrophys. J.}\ }\textbf {\bibinfo {volume} {842}},\ \bibinfo {pages} {46} (\bibinfo {year} {2017})},\ \Eprint {http://arxiv.org/abs/1704.03500} {arXiv:1704.03500 [astro-ph.CO]} \BibitemShut {NoStop}%
\bibitem [{\citenamefont {De~Luca}\ \emph {et~al.}(2020)\citenamefont {De~Luca}, \citenamefont {Franciolini}, \citenamefont {Kehagias},\ and\ \citenamefont {Riotto}}]{DeLuca:2019ufz}%
  \BibitemOpen
  \bibfield  {author} {\bibinfo {author} {\bibfnamefont {V.}~\bibnamefont {De~Luca}}, \bibinfo {author} {\bibfnamefont {G.}~\bibnamefont {Franciolini}}, \bibinfo {author} {\bibfnamefont {A.}~\bibnamefont {Kehagias}}, \ and\ \bibinfo {author} {\bibfnamefont {A.}~\bibnamefont {Riotto}},\ }\href {\doibase 10.1088/1475-7516/2020/03/014} {\bibfield  {journal} {\bibinfo  {journal} {JCAP}\ }\textbf {\bibinfo {volume} {03}},\ \bibinfo {pages} {014} (\bibinfo {year} {2020})},\ \Eprint {http://arxiv.org/abs/1911.09689} {arXiv:1911.09689 [gr-qc]} \BibitemShut {NoStop}%
\bibitem [{\citenamefont {Lu}\ \emph {et~al.}(2020)\citenamefont {Lu}, \citenamefont {Ali}, \citenamefont {Gong}, \citenamefont {Lin},\ and\ \citenamefont {Zhang}}]{Lu:2020diy}%
  \BibitemOpen
  \bibfield  {author} {\bibinfo {author} {\bibfnamefont {Y.}~\bibnamefont {Lu}}, \bibinfo {author} {\bibfnamefont {A.}~\bibnamefont {Ali}}, \bibinfo {author} {\bibfnamefont {Y.}~\bibnamefont {Gong}}, \bibinfo {author} {\bibfnamefont {J.}~\bibnamefont {Lin}}, \ and\ \bibinfo {author} {\bibfnamefont {F.}~\bibnamefont {Zhang}},\ }\href {\doibase 10.1103/PhysRevD.102.083503(2020)} {\bibfield  {journal} {\bibinfo  {journal} {Phys. Rev. D}\ }\textbf {\bibinfo {volume} {102}},\ \bibinfo {pages} {083503} (\bibinfo {year} {2020})},\ \Eprint {http://arxiv.org/abs/2006.03450} {arXiv:2006.03450 [gr-qc]} \BibitemShut {NoStop}%
\bibitem [{\citenamefont {Tomikawa}\ and\ \citenamefont {Kobayashi}(2020)}]{Tomikawa:2019tvi}%
  \BibitemOpen
  \bibfield  {author} {\bibinfo {author} {\bibfnamefont {K.}~\bibnamefont {Tomikawa}}\ and\ \bibinfo {author} {\bibfnamefont {T.}~\bibnamefont {Kobayashi}},\ }\href {\doibase 10.1103/PhysRevD.101.083529} {\bibfield  {journal} {\bibinfo  {journal} {Phys. Rev. D}\ }\textbf {\bibinfo {volume} {101}},\ \bibinfo {pages} {083529} (\bibinfo {year} {2020})},\ \Eprint {http://arxiv.org/abs/1910.01880} {arXiv:1910.01880 [gr-qc]} \BibitemShut {NoStop}%
\bibitem [{\citenamefont {Dom{\`e}nech}\ and\ \citenamefont {Sasaki}(2021)}]{Domenech:2020xin}%
  \BibitemOpen
  \bibfield  {author} {\bibinfo {author} {\bibfnamefont {G.}~\bibnamefont {Dom{\`e}nech}}\ and\ \bibinfo {author} {\bibfnamefont {M.}~\bibnamefont {Sasaki}},\ }\href {\doibase 10.1103/PhysRevD.103.063531} {\bibfield  {journal} {\bibinfo  {journal} {Phys. Rev. D}\ }\textbf {\bibinfo {volume} {103}},\ \bibinfo {pages} {063531} (\bibinfo {year} {2021})},\ \Eprint {http://arxiv.org/abs/2012.14016} {arXiv:2012.14016 [gr-qc]} \BibitemShut {NoStop}%
\bibitem [{\citenamefont {Inomata}\ and\ \citenamefont {Terada}(2020)}]{Inomata:2019yww}%
  \BibitemOpen
  \bibfield  {author} {\bibinfo {author} {\bibfnamefont {K.}~\bibnamefont {Inomata}}\ and\ \bibinfo {author} {\bibfnamefont {T.}~\bibnamefont {Terada}},\ }\href {\doibase 10.1103/PhysRevD.101.023523} {\bibfield  {journal} {\bibinfo  {journal} {Phys. Rev. D}\ }\textbf {\bibinfo {volume} {101}},\ \bibinfo {pages} {023523} (\bibinfo {year} {2020})},\ \Eprint {http://arxiv.org/abs/1912.00785} {arXiv:1912.00785 [gr-qc]} \BibitemShut {NoStop}%
\bibitem [{\citenamefont {Yuan}\ \emph {et~al.}(2020)\citenamefont {Yuan}, \citenamefont {Chen},\ and\ \citenamefont {Huang}}]{Yuan:2019fwv}%
  \BibitemOpen
  \bibfield  {author} {\bibinfo {author} {\bibfnamefont {C.}~\bibnamefont {Yuan}}, \bibinfo {author} {\bibfnamefont {Z.-C.}\ \bibnamefont {Chen}}, \ and\ \bibinfo {author} {\bibfnamefont {Q.-G.}\ \bibnamefont {Huang}},\ }\href {\doibase 10.1103/PhysRevD.101.063018} {\bibfield  {journal} {\bibinfo  {journal} {Phys. Rev. D}\ }\textbf {\bibinfo {volume} {101}},\ \bibinfo {pages} {6} (\bibinfo {year} {2020})},\ \Eprint {http://arxiv.org/abs/1912.00885} {arXiv:1912.00885 [astro-ph.CO]} \BibitemShut {NoStop}%
\bibitem [{\citenamefont {Yuan}\ \emph {et~al.}(2025)\citenamefont {Yuan}, \citenamefont {Chen},\ and\ \citenamefont {Liu}}]{Yuan:2024qfz}%
  \BibitemOpen
  \bibfield  {author} {\bibinfo {author} {\bibfnamefont {C.}~\bibnamefont {Yuan}}, \bibinfo {author} {\bibfnamefont {Z.-C.}\ \bibnamefont {Chen}}, \ and\ \bibinfo {author} {\bibfnamefont {L.}~\bibnamefont {Liu}},\ }\href {\doibase 10.1103/PhysRevD.111.103528} {\bibfield  {journal} {\bibinfo  {journal} {Phys. Rev. D}\ }\textbf {\bibinfo {volume} {111}},\ \bibinfo {pages} {103528} (\bibinfo {year} {2025})},\ \Eprint {http://arxiv.org/abs/2410.18996} {arXiv:2410.18996 [gr-qc]} \BibitemShut {NoStop}%
\bibitem [{\citenamefont {Kugarajh}(2025)}]{Kugarajh:2025pjl}%
  \BibitemOpen
  \bibfield  {author} {\bibinfo {author} {\bibfnamefont {A.~A.}\ \bibnamefont {Kugarajh}},\ }\href {\doibase 10.1088/1361-6382/ade2b3} {\  (\bibinfo {year} {2025}),\ 10.1088/1361-6382/ade2b3},\ \Eprint {http://arxiv.org/abs/2503.00083} {arXiv:2503.00083 [gr-qc]} \BibitemShut {NoStop}%
\bibitem [{\citenamefont {Dom{\`e}nech}(2020)}]{Domenech:2019quo}%
  \BibitemOpen
  \bibfield  {author} {\bibinfo {author} {\bibfnamefont {G.}~\bibnamefont {Dom{\`e}nech}},\ }\href {\doibase 10.1142/S0218271820500285} {\bibfield  {journal} {\bibinfo  {journal} {Int. J. Mod. Phys. D}\ }\textbf {\bibinfo {volume} {29}},\ \bibinfo {pages} {2050028} (\bibinfo {year} {2020})},\ \Eprint {http://arxiv.org/abs/1912.05583} {arXiv:1912.05583 [gr-qc]} \BibitemShut {NoStop}%
\bibitem [{\citenamefont {Gong}(2022)}]{Gong:2019mui}%
  \BibitemOpen
  \bibfield  {author} {\bibinfo {author} {\bibfnamefont {J.-O.}\ \bibnamefont {Gong}},\ }\href {\doibase 10.3847/1538-4357/ac3a6c} {\bibfield  {journal} {\bibinfo  {journal} {Astrophys. J.}\ }\textbf {\bibinfo {volume} {925}},\ \bibinfo {pages} {102} (\bibinfo {year} {2022})},\ \Eprint {http://arxiv.org/abs/1909.12708} {arXiv:1909.12708 [gr-qc]} \BibitemShut {NoStop}%
\bibitem [{\citenamefont {Balaji}\ \emph {et~al.}(2023)\citenamefont {Balaji}, \citenamefont {Dom{\`e}nech},\ and\ \citenamefont {Franciolini}}]{Balaji:2023ehk}%
  \BibitemOpen
  \bibfield  {author} {\bibinfo {author} {\bibfnamefont {S.}~\bibnamefont {Balaji}}, \bibinfo {author} {\bibfnamefont {G.}~\bibnamefont {Dom{\`e}nech}}, \ and\ \bibinfo {author} {\bibfnamefont {G.}~\bibnamefont {Franciolini}},\ }\href {\doibase 10.1088/1475-7516/2023/10/041} {\bibfield  {journal} {\bibinfo  {journal} {JCAP}\ }\textbf {\bibinfo {volume} {10}},\ \bibinfo {pages} {041} (\bibinfo {year} {2023})},\ \Eprint {http://arxiv.org/abs/2307.08552} {arXiv:2307.08552 [gr-qc]} \BibitemShut {NoStop}%
\bibitem [{\citenamefont {Jin}\ \emph {et~al.}(2023)\citenamefont {Jin}, \citenamefont {Chen}, \citenamefont {Yi}, \citenamefont {You}, \citenamefont {Liu},\ and\ \citenamefont {Wu}}]{Jin:2023wri}%
  \BibitemOpen
  \bibfield  {author} {\bibinfo {author} {\bibfnamefont {J.-H.}\ \bibnamefont {Jin}}, \bibinfo {author} {\bibfnamefont {Z.-C.}\ \bibnamefont {Chen}}, \bibinfo {author} {\bibfnamefont {Z.}~\bibnamefont {Yi}}, \bibinfo {author} {\bibfnamefont {Z.-Q.}\ \bibnamefont {You}}, \bibinfo {author} {\bibfnamefont {L.}~\bibnamefont {Liu}}, \ and\ \bibinfo {author} {\bibfnamefont {Y.}~\bibnamefont {Wu}},\ }\href {\doibase 10.1088/1475-7516/2023/09/016} {\bibfield  {journal} {\bibinfo  {journal} {JCAP}\ }\textbf {\bibinfo {volume} {09}},\ \bibinfo {pages} {016} (\bibinfo {year} {2023})},\ \Eprint {http://arxiv.org/abs/2307.08687} {arXiv:2307.08687 [astro-ph.CO]} \BibitemShut {NoStop}%
\bibitem [{\citenamefont {Liu}\ \emph {et~al.}(2024)\citenamefont {Liu}, \citenamefont {Wu},\ and\ \citenamefont {Chen}}]{Liu:2023hpw}%
  \BibitemOpen
  \bibfield  {author} {\bibinfo {author} {\bibfnamefont {L.}~\bibnamefont {Liu}}, \bibinfo {author} {\bibfnamefont {Y.}~\bibnamefont {Wu}}, \ and\ \bibinfo {author} {\bibfnamefont {Z.-C.}\ \bibnamefont {Chen}},\ }\href {\doibase 10.1088/1475-7516/2024/04/011} {\bibfield  {journal} {\bibinfo  {journal} {JCAP}\ }\textbf {\bibinfo {volume} {04}},\ \bibinfo {pages} {011} (\bibinfo {year} {2024})},\ \Eprint {http://arxiv.org/abs/2310.16500} {arXiv:2310.16500 [astro-ph.CO]} \BibitemShut {NoStop}%
\bibitem [{\citenamefont {Li}\ \emph {et~al.}(2024)\citenamefont {Li}, \citenamefont {Guo},\ and\ \citenamefont {Zu}}]{Li:2023uhu}%
  \BibitemOpen
  \bibfield  {author} {\bibinfo {author} {\bibfnamefont {J.}~\bibnamefont {Li}}, \bibinfo {author} {\bibfnamefont {G.-H.}\ \bibnamefont {Guo}}, \ and\ \bibinfo {author} {\bibfnamefont {Y.}~\bibnamefont {Zu}},\ }\href {\doibase 10.1140/epjc/s10052-024-13448-4} {\bibfield  {journal} {\bibinfo  {journal} {Eur. Phys. J. C}\ }\textbf {\bibinfo {volume} {84}},\ \bibinfo {pages} {1083} (\bibinfo {year} {2024})},\ \Eprint {http://arxiv.org/abs/2312.04589} {arXiv:2312.04589 [gr-qc]} \BibitemShut {NoStop}%
\bibitem [{\citenamefont {Chen}\ \emph {et~al.}(2024)\citenamefont {Chen}, \citenamefont {Li}, \citenamefont {Liu},\ and\ \citenamefont {Yi}}]{Chen:2024fir}%
  \BibitemOpen
  \bibfield  {author} {\bibinfo {author} {\bibfnamefont {Z.-C.}\ \bibnamefont {Chen}}, \bibinfo {author} {\bibfnamefont {J.}~\bibnamefont {Li}}, \bibinfo {author} {\bibfnamefont {L.}~\bibnamefont {Liu}}, \ and\ \bibinfo {author} {\bibfnamefont {Z.}~\bibnamefont {Yi}},\ }\href {\doibase 10.1103/PhysRevD.109.L101302} {\bibfield  {journal} {\bibinfo  {journal} {Phys. Rev. D}\ }\textbf {\bibinfo {volume} {109}},\ \bibinfo {pages} {L101302} (\bibinfo {year} {2024})},\ \Eprint {http://arxiv.org/abs/2401.09818} {arXiv:2401.09818 [gr-qc]} \BibitemShut {NoStop}%
\bibitem [{\citenamefont {Wang}\ \emph {et~al.}(2025{\natexlab{a}})\citenamefont {Wang}, \citenamefont {Chen},\ and\ \citenamefont {Xie}}]{Wang:2025aon}%
  \BibitemOpen
  \bibfield  {author} {\bibinfo {author} {\bibfnamefont {L.-S.}\ \bibnamefont {Wang}}, \bibinfo {author} {\bibfnamefont {L.-Y.}\ \bibnamefont {Chen}}, \ and\ \bibinfo {author} {\bibfnamefont {Q.-T.}\ \bibnamefont {Xie}},\ }\href@noop {} {\  (\bibinfo {year} {2025}{\natexlab{a}})},\ \Eprint {http://arxiv.org/abs/2503.01460} {arXiv:2503.01460 [astro-ph.CO]} \BibitemShut {NoStop}%
\bibitem [{\citenamefont {Zhou}\ \emph {et~al.}(2024)\citenamefont {Zhou}, \citenamefont {Kuang}, \citenamefont {Wu}, \citenamefont {Chen}, \citenamefont {L\"u},\ and\ \citenamefont {Chang}}]{Zhou:2024doz}%
  \BibitemOpen
  \bibfield  {author} {\bibinfo {author} {\bibfnamefont {J.-Z.}\ \bibnamefont {Zhou}}, \bibinfo {author} {\bibfnamefont {Y.-T.}\ \bibnamefont {Kuang}}, \bibinfo {author} {\bibfnamefont {D.}~\bibnamefont {Wu}}, \bibinfo {author} {\bibfnamefont {F.-Y.}\ \bibnamefont {Chen}}, \bibinfo {author} {\bibfnamefont {H.}~\bibnamefont {L\"u}}, \ and\ \bibinfo {author} {\bibfnamefont {Z.}~\bibnamefont {Chang}},\ }\href {\doibase 10.1088/1475-7516/2024/12/021} {\bibfield  {journal} {\bibinfo  {journal} {JCAP}\ }\textbf {\bibinfo {volume} {12}},\ \bibinfo {pages} {021} (\bibinfo {year} {2024})},\ \Eprint {http://arxiv.org/abs/2409.07702} {arXiv:2409.07702 [gr-qc]} \BibitemShut {NoStop}%
\bibitem [{\citenamefont {Maggiore}(2000)}]{Maggiore:1999vm}%
  \BibitemOpen
  \bibfield  {author} {\bibinfo {author} {\bibfnamefont {M.}~\bibnamefont {Maggiore}},\ }\href {\doibase 10.1016/S0370-1573(99)00102-7} {\bibfield  {journal} {\bibinfo  {journal} {Phys. Rept.}\ }\textbf {\bibinfo {volume} {331}},\ \bibinfo {pages} {283} (\bibinfo {year} {2000})},\ \Eprint {http://arxiv.org/abs/gr-qc/9909001} {arXiv:gr-qc/9909001} \BibitemShut {NoStop}%
\bibitem [{\citenamefont {Yuan}\ and\ \citenamefont {Huang}(2021)}]{Yuan:2021qgz}%
  \BibitemOpen
  \bibfield  {author} {\bibinfo {author} {\bibfnamefont {C.}~\bibnamefont {Yuan}}\ and\ \bibinfo {author} {\bibfnamefont {Q.-G.}\ \bibnamefont {Huang}},\ }\href {\doibase 10.1016/j.isci.2021.102860} {\bibfield  {journal} {\bibinfo  {journal} {iScience}\ }\textbf {\bibinfo {volume} {24}},\ \bibinfo {pages} {102860} (\bibinfo {year} {2021})},\ \Eprint {http://arxiv.org/abs/2103.04739} {arXiv:2103.04739 [astro-ph.GA]} \BibitemShut {NoStop}%
\bibitem [{\citenamefont {Gorji}\ and\ \citenamefont {Sasaki}(2023)}]{Gorji:2023ziy}%
  \BibitemOpen
  \bibfield  {author} {\bibinfo {author} {\bibfnamefont {M.~A.}\ \bibnamefont {Gorji}}\ and\ \bibinfo {author} {\bibfnamefont {M.}~\bibnamefont {Sasaki}},\ }\href {\doibase 10.1016/j.physletb.2023.138236} {\bibfield  {journal} {\bibinfo  {journal} {Phys. Lett. B}\ }\textbf {\bibinfo {volume} {846}},\ \bibinfo {pages} {138236} (\bibinfo {year} {2023})},\ \Eprint {http://arxiv.org/abs/2302.14080} {arXiv:2302.14080 [gr-qc]} \BibitemShut {NoStop}%
\bibitem [{\citenamefont {Gorji}\ \emph {et~al.}(2023)\citenamefont {Gorji}, \citenamefont {Sasaki},\ and\ \citenamefont {Suyama}}]{Gorji:2023sil}%
  \BibitemOpen
  \bibfield  {author} {\bibinfo {author} {\bibfnamefont {M.~A.}\ \bibnamefont {Gorji}}, \bibinfo {author} {\bibfnamefont {M.}~\bibnamefont {Sasaki}}, \ and\ \bibinfo {author} {\bibfnamefont {T.}~\bibnamefont {Suyama}},\ }\href {\doibase 10.1016/j.physletb.2023.138214} {\bibfield  {journal} {\bibinfo  {journal} {Phys. Lett. B}\ }\textbf {\bibinfo {volume} {846}},\ \bibinfo {pages} {138214} (\bibinfo {year} {2023})},\ \Eprint {http://arxiv.org/abs/2307.13109} {arXiv:2307.13109 [astro-ph.CO]} \BibitemShut {NoStop}%
\bibitem [{\citenamefont {Biagetti}\ \emph {et~al.}(2013)\citenamefont {Biagetti}, \citenamefont {Fasiello},\ and\ \citenamefont {Riotto}}]{Biagetti:2013kwa}%
  \BibitemOpen
  \bibfield  {author} {\bibinfo {author} {\bibfnamefont {M.}~\bibnamefont {Biagetti}}, \bibinfo {author} {\bibfnamefont {M.}~\bibnamefont {Fasiello}}, \ and\ \bibinfo {author} {\bibfnamefont {A.}~\bibnamefont {Riotto}},\ }\href {\doibase 10.1103/PhysRevD.88.103518} {\bibfield  {journal} {\bibinfo  {journal} {Phys. Rev. D}\ }\textbf {\bibinfo {volume} {88}},\ \bibinfo {pages} {103518} (\bibinfo {year} {2013})},\ \Eprint {http://arxiv.org/abs/1305.7241} {arXiv:1305.7241 [astro-ph.CO]} \BibitemShut {NoStop}%
\bibitem [{\citenamefont {Fu}\ \emph {et~al.}(2024)\citenamefont {Fu}, \citenamefont {Liu}, \citenamefont {Yang}, \citenamefont {Yu},\ and\ \citenamefont {Zhang}}]{Fu:2023aab}%
  \BibitemOpen
  \bibfield  {author} {\bibinfo {author} {\bibfnamefont {C.}~\bibnamefont {Fu}}, \bibinfo {author} {\bibfnamefont {J.}~\bibnamefont {Liu}}, \bibinfo {author} {\bibfnamefont {X.-Y.}\ \bibnamefont {Yang}}, \bibinfo {author} {\bibfnamefont {W.-W.}\ \bibnamefont {Yu}}, \ and\ \bibinfo {author} {\bibfnamefont {Y.}~\bibnamefont {Zhang}},\ }\href {\doibase 10.1103/PhysRevD.109.063526} {\bibfield  {journal} {\bibinfo  {journal} {Phys. Rev. D}\ }\textbf {\bibinfo {volume} {109}},\ \bibinfo {pages} {063526} (\bibinfo {year} {2024})},\ \Eprint {http://arxiv.org/abs/2308.15329} {arXiv:2308.15329 [astro-ph.CO]} \BibitemShut {NoStop}%
\bibitem [{\citenamefont {Cai}\ \emph {et~al.}(2022)\citenamefont {Cai}, \citenamefont {Fu},\ and\ \citenamefont {Yu}}]{Cai:2021uup}%
  \BibitemOpen
  \bibfield  {author} {\bibinfo {author} {\bibfnamefont {R.-G.}\ \bibnamefont {Cai}}, \bibinfo {author} {\bibfnamefont {C.}~\bibnamefont {Fu}}, \ and\ \bibinfo {author} {\bibfnamefont {W.-W.}\ \bibnamefont {Yu}},\ }\href {\doibase 10.1103/PhysRevD.105.103520} {\bibfield  {journal} {\bibinfo  {journal} {Phys. Rev. D}\ }\textbf {\bibinfo {volume} {105}},\ \bibinfo {pages} {103520} (\bibinfo {year} {2022})},\ \Eprint {http://arxiv.org/abs/2112.04794} {arXiv:2112.04794 [astro-ph.CO]} \BibitemShut {NoStop}%
\bibitem [{\citenamefont {Jiang}\ \emph {et~al.}(2024)\citenamefont {Jiang}, \citenamefont {Cai}, \citenamefont {Wang},\ and\ \citenamefont {Piao}}]{Jiang:2024woi}%
  \BibitemOpen
  \bibfield  {author} {\bibinfo {author} {\bibfnamefont {Z.-W.}\ \bibnamefont {Jiang}}, \bibinfo {author} {\bibfnamefont {Y.}~\bibnamefont {Cai}}, \bibinfo {author} {\bibfnamefont {F.}~\bibnamefont {Wang}}, \ and\ \bibinfo {author} {\bibfnamefont {Y.-S.}\ \bibnamefont {Piao}},\ }\href {\doibase 10.1007/JHEP09(2024)067} {\bibfield  {journal} {\bibinfo  {journal} {JHEP}\ }\textbf {\bibinfo {volume} {09}},\ \bibinfo {pages} {067} (\bibinfo {year} {2024})},\ \Eprint {http://arxiv.org/abs/2406.16549} {arXiv:2406.16549 [astro-ph.CO]} \BibitemShut {NoStop}%
\bibitem [{\citenamefont {Cai}\ \emph {et~al.}(2021)\citenamefont {Cai}, \citenamefont {Lin}, \citenamefont {Wang},\ and\ \citenamefont {Yan}}]{Cai:2020ovp}%
  \BibitemOpen
  \bibfield  {author} {\bibinfo {author} {\bibfnamefont {Y.-F.}\ \bibnamefont {Cai}}, \bibinfo {author} {\bibfnamefont {C.}~\bibnamefont {Lin}}, \bibinfo {author} {\bibfnamefont {B.}~\bibnamefont {Wang}}, \ and\ \bibinfo {author} {\bibfnamefont {S.-F.}\ \bibnamefont {Yan}},\ }\href {\doibase 10.1103/PhysRevLett.126.071303} {\bibfield  {journal} {\bibinfo  {journal} {Phys. Rev. Lett.}\ }\textbf {\bibinfo {volume} {126}},\ \bibinfo {pages} {071303} (\bibinfo {year} {2021})},\ \Eprint {http://arxiv.org/abs/2009.09833} {arXiv:2009.09833 [gr-qc]} \BibitemShut {NoStop}%
\bibitem [{\citenamefont {Addazi}\ \emph {et~al.}(2024)\citenamefont {Addazi}, \citenamefont {Aldabergenov},\ and\ \citenamefont {Cai}}]{Addazi:2024gew}%
  \BibitemOpen
  \bibfield  {author} {\bibinfo {author} {\bibfnamefont {A.}~\bibnamefont {Addazi}}, \bibinfo {author} {\bibfnamefont {Y.}~\bibnamefont {Aldabergenov}}, \ and\ \bibinfo {author} {\bibfnamefont {Y.}~\bibnamefont {Cai}},\ }\href {\doibase 10.1103/PhysRevD.110.123530} {\bibfield  {journal} {\bibinfo  {journal} {Phys. Rev. D}\ }\textbf {\bibinfo {volume} {110}},\ \bibinfo {pages} {123530} (\bibinfo {year} {2024})},\ \Eprint {http://arxiv.org/abs/2408.05091} {arXiv:2408.05091 [gr-qc]} \BibitemShut {NoStop}%
\bibitem [{\citenamefont {Guzzetti}\ \emph {et~al.}(2016)\citenamefont {Guzzetti}, \citenamefont {Bartolo}, \citenamefont {Liguori},\ and\ \citenamefont {Matarrese}}]{Guzzetti:2016mkm}%
  \BibitemOpen
  \bibfield  {author} {\bibinfo {author} {\bibfnamefont {M.~C.}\ \bibnamefont {Guzzetti}}, \bibinfo {author} {\bibfnamefont {N.}~\bibnamefont {Bartolo}}, \bibinfo {author} {\bibfnamefont {M.}~\bibnamefont {Liguori}}, \ and\ \bibinfo {author} {\bibfnamefont {S.}~\bibnamefont {Matarrese}},\ }\href {\doibase 10.1393/ncr/i2016-10127-1} {\bibfield  {journal} {\bibinfo  {journal} {Riv. Nuovo Cim.}\ }\textbf {\bibinfo {volume} {39}},\ \bibinfo {pages} {399} (\bibinfo {year} {2016})},\ \Eprint {http://arxiv.org/abs/1605.01615} {arXiv:1605.01615 [astro-ph.CO]} \BibitemShut {NoStop}%
\bibitem [{\citenamefont {Li}\ \emph {et~al.}(2020)\citenamefont {Li}, \citenamefont {Rao},\ and\ \citenamefont {Zhao}}]{Li:2020xjt}%
  \BibitemOpen
  \bibfield  {author} {\bibinfo {author} {\bibfnamefont {M.}~\bibnamefont {Li}}, \bibinfo {author} {\bibfnamefont {H.}~\bibnamefont {Rao}}, \ and\ \bibinfo {author} {\bibfnamefont {D.}~\bibnamefont {Zhao}},\ }\href {\doibase 10.1088/1475-7516/2020/11/023} {\bibfield  {journal} {\bibinfo  {journal} {JCAP}\ }\textbf {\bibinfo {volume} {11}},\ \bibinfo {pages} {023} (\bibinfo {year} {2020})},\ \Eprint {http://arxiv.org/abs/2007.08038} {arXiv:2007.08038 [gr-qc]} \BibitemShut {NoStop}%
\bibitem [{\citenamefont {Li}\ \emph {et~al.}(2021)\citenamefont {Li}, \citenamefont {Rao},\ and\ \citenamefont {Tong}}]{Li:2021wij}%
  \BibitemOpen
  \bibfield  {author} {\bibinfo {author} {\bibfnamefont {M.}~\bibnamefont {Li}}, \bibinfo {author} {\bibfnamefont {H.}~\bibnamefont {Rao}}, \ and\ \bibinfo {author} {\bibfnamefont {Y.}~\bibnamefont {Tong}},\ }\href {\doibase 10.1103/PhysRevD.104.084077} {\bibfield  {journal} {\bibinfo  {journal} {Phys. Rev. D}\ }\textbf {\bibinfo {volume} {104}},\ \bibinfo {pages} {084077} (\bibinfo {year} {2021})},\ \Eprint {http://arxiv.org/abs/2104.05917} {arXiv:2104.05917 [gr-qc]} \BibitemShut {NoStop}%
\bibitem [{\citenamefont {Zhou}\ \emph {et~al.}(2025{\natexlab{a}})\citenamefont {Zhou}, \citenamefont {Kuang}, \citenamefont {Chang},\ and\ \citenamefont {L\"u}}]{Zhou:2024yke}%
  \BibitemOpen
  \bibfield  {author} {\bibinfo {author} {\bibfnamefont {J.-Z.}\ \bibnamefont {Zhou}}, \bibinfo {author} {\bibfnamefont {Y.-T.}\ \bibnamefont {Kuang}}, \bibinfo {author} {\bibfnamefont {Z.}~\bibnamefont {Chang}}, \ and\ \bibinfo {author} {\bibfnamefont {H.}~\bibnamefont {L\"u}},\ }\href {\doibase 10.3847/1538-4357/ada6b4} {\bibfield  {journal} {\bibinfo  {journal} {Astrophys. J.}\ }\textbf {\bibinfo {volume} {979}},\ \bibinfo {pages} {178} (\bibinfo {year} {2025}{\natexlab{a}})},\ \Eprint {http://arxiv.org/abs/2410.10111} {arXiv:2410.10111 [astro-ph.CO]} \BibitemShut {NoStop}%
\bibitem [{\citenamefont {Cang}\ \emph {et~al.}(2023)\citenamefont {Cang}, \citenamefont {Ma},\ and\ \citenamefont {Gao}}]{Cang:2022jyc}%
  \BibitemOpen
  \bibfield  {author} {\bibinfo {author} {\bibfnamefont {J.}~\bibnamefont {Cang}}, \bibinfo {author} {\bibfnamefont {Y.-Z.}\ \bibnamefont {Ma}}, \ and\ \bibinfo {author} {\bibfnamefont {Y.}~\bibnamefont {Gao}},\ }\href {\doibase 10.3847/1538-4357/acc949} {\bibfield  {journal} {\bibinfo  {journal} {Astrophys. J.}\ }\textbf {\bibinfo {volume} {949}},\ \bibinfo {pages} {64} (\bibinfo {year} {2023})},\ \Eprint {http://arxiv.org/abs/2210.03476} {arXiv:2210.03476 [astro-ph.CO]} \BibitemShut {NoStop}%
\bibitem [{\citenamefont {Ben-Dayan}\ \emph {et~al.}(2019)\citenamefont {Ben-Dayan}, \citenamefont {Keating}, \citenamefont {Leon},\ and\ \citenamefont {Wolfson}}]{Ben-Dayan:2019gll}%
  \BibitemOpen
  \bibfield  {author} {\bibinfo {author} {\bibfnamefont {I.}~\bibnamefont {Ben-Dayan}}, \bibinfo {author} {\bibfnamefont {B.}~\bibnamefont {Keating}}, \bibinfo {author} {\bibfnamefont {D.}~\bibnamefont {Leon}}, \ and\ \bibinfo {author} {\bibfnamefont {I.}~\bibnamefont {Wolfson}},\ }\href {\doibase 10.1088/1475-7516/2019/06/007} {\bibfield  {journal} {\bibinfo  {journal} {JCAP}\ }\textbf {\bibinfo {volume} {06}},\ \bibinfo {pages} {007} (\bibinfo {year} {2019})},\ \Eprint {http://arxiv.org/abs/1903.11843} {arXiv:1903.11843 [astro-ph.CO]} \BibitemShut {NoStop}%
\bibitem [{\citenamefont {Wang}\ \emph {et~al.}(2024{\natexlab{a}})\citenamefont {Wang}, \citenamefont {Zhao},\ and\ \citenamefont {Zhu}}]{Wang:2023sij}%
  \BibitemOpen
  \bibfield  {author} {\bibinfo {author} {\bibfnamefont {S.}~\bibnamefont {Wang}}, \bibinfo {author} {\bibfnamefont {Z.-C.}\ \bibnamefont {Zhao}}, \ and\ \bibinfo {author} {\bibfnamefont {Q.-H.}\ \bibnamefont {Zhu}},\ }\href {\doibase 10.1103/PhysRevResearch.6.013207} {\bibfield  {journal} {\bibinfo  {journal} {Phys. Rev. Res.}\ }\textbf {\bibinfo {volume} {6}},\ \bibinfo {pages} {013207} (\bibinfo {year} {2024}{\natexlab{a}})},\ \Eprint {http://arxiv.org/abs/2307.03095} {arXiv:2307.03095 [astro-ph.CO]} \BibitemShut {NoStop}%
\bibitem [{\citenamefont {Auclair}\ \emph {et~al.}(2023)\citenamefont {Auclair} \emph {et~al.}}]{LISACosmologyWorkingGroup:2022jok}%
  \BibitemOpen
  \bibfield  {author} {\bibinfo {author} {\bibfnamefont {P.}~\bibnamefont {Auclair}} \emph {et~al.} (\bibinfo {collaboration} {LISA Cosmology Working Group}),\ }\href {\doibase 10.1007/s41114-023-00045-2} {\bibfield  {journal} {\bibinfo  {journal} {Living Rev. Rel.}\ }\textbf {\bibinfo {volume} {26}},\ \bibinfo {pages} {5} (\bibinfo {year} {2023})},\ \Eprint {http://arxiv.org/abs/2204.05434} {arXiv:2204.05434 [astro-ph.CO]} \BibitemShut {NoStop}%
\bibitem [{\citenamefont {Iacconi}\ \emph {et~al.}(2024)\citenamefont {Iacconi}, \citenamefont {Bacchi}, \citenamefont {Guimar\~aes},\ and\ \citenamefont {Falciano}}]{Iacconi:2024hmg}%
  \BibitemOpen
  \bibfield  {author} {\bibinfo {author} {\bibfnamefont {L.}~\bibnamefont {Iacconi}}, \bibinfo {author} {\bibfnamefont {M.}~\bibnamefont {Bacchi}}, \bibinfo {author} {\bibfnamefont {L.~F.}\ \bibnamefont {Guimar\~aes}}, \ and\ \bibinfo {author} {\bibfnamefont {F.~T.}\ \bibnamefont {Falciano}},\ }\href@noop {} {\  (\bibinfo {year} {2024})},\ \Eprint {http://arxiv.org/abs/2412.02544} {arXiv:2412.02544 [astro-ph.CO]} \BibitemShut {NoStop}%
\bibitem [{\citenamefont {Flauger}\ \emph {et~al.}(2021)\citenamefont {Flauger}, \citenamefont {Karnesis}, \citenamefont {Nardini}, \citenamefont {Pieroni}, \citenamefont {Ricciardone},\ and\ \citenamefont {Torrado}}]{Flauger:2020qyi}%
  \BibitemOpen
  \bibfield  {author} {\bibinfo {author} {\bibfnamefont {R.}~\bibnamefont {Flauger}}, \bibinfo {author} {\bibfnamefont {N.}~\bibnamefont {Karnesis}}, \bibinfo {author} {\bibfnamefont {G.}~\bibnamefont {Nardini}}, \bibinfo {author} {\bibfnamefont {M.}~\bibnamefont {Pieroni}}, \bibinfo {author} {\bibfnamefont {A.}~\bibnamefont {Ricciardone}}, \ and\ \bibinfo {author} {\bibfnamefont {J.}~\bibnamefont {Torrado}},\ }\href {\doibase 10.1088/1475-7516/2021/01/059} {\bibfield  {journal} {\bibinfo  {journal} {JCAP}\ }\textbf {\bibinfo {volume} {01}},\ \bibinfo {pages} {059} (\bibinfo {year} {2021})},\ \Eprint {http://arxiv.org/abs/2009.11845} {arXiv:2009.11845 [astro-ph.CO]} \BibitemShut {NoStop}%
\bibitem [{\citenamefont {Wu}\ \emph {et~al.}(2021)\citenamefont {Wu} \emph {et~al.}}]{TaijiScientific:2021qgx}%
  \BibitemOpen
  \bibfield  {author} {\bibinfo {author} {\bibfnamefont {Y.-L.}\ \bibnamefont {Wu}} \emph {et~al.} (\bibinfo {collaboration} {Taiji Scientific}),\ }\href {\doibase 10.1038/s42005-021-00529-z} {\bibfield  {journal} {\bibinfo  {journal} {Commun. Phys.}\ }\textbf {\bibinfo {volume} {4}},\ \bibinfo {pages} {34} (\bibinfo {year} {2021})}\BibitemShut {NoStop}%
\bibitem [{\citenamefont {Chluba}\ and\ \citenamefont {Grin}(2013)}]{Chluba:2013dna}%
  \BibitemOpen
  \bibfield  {author} {\bibinfo {author} {\bibfnamefont {J.}~\bibnamefont {Chluba}}\ and\ \bibinfo {author} {\bibfnamefont {D.}~\bibnamefont {Grin}},\ }\href {\doibase 10.1093/mnras/stt1129} {\bibfield  {journal} {\bibinfo  {journal} {Mon. Not. Roy. Astron. Soc.}\ }\textbf {\bibinfo {volume} {434}},\ \bibinfo {pages} {1619} (\bibinfo {year} {2013})},\ \Eprint {http://arxiv.org/abs/1304.4596} {arXiv:1304.4596 [astro-ph.CO]} \BibitemShut {NoStop}%
\bibitem [{\citenamefont {Bringmann}\ \emph {et~al.}(2025)\citenamefont {Bringmann}, \citenamefont {Croon},\ and\ \citenamefont {Sevillano~Mu{\~n}oz}}]{Bringmann:2025cht}%
  \BibitemOpen
  \bibfield  {author} {\bibinfo {author} {\bibfnamefont {T.}~\bibnamefont {Bringmann}}, \bibinfo {author} {\bibfnamefont {D.}~\bibnamefont {Croon}}, \ and\ \bibinfo {author} {\bibfnamefont {S.}~\bibnamefont {Sevillano~Mu{\~n}oz}},\ }\href@noop {} {\  (\bibinfo {year} {2025})},\ \Eprint {http://arxiv.org/abs/2506.20704} {arXiv:2506.20704 [astro-ph.CO]} \BibitemShut {NoStop}%
\bibitem [{\citenamefont {Zhou}\ \emph {et~al.}(2025{\natexlab{b}})\citenamefont {Zhou}, \citenamefont {Li},\ and\ \citenamefont {Wu}}]{Zhou:2025djn}%
  \BibitemOpen
  \bibfield  {author} {\bibinfo {author} {\bibfnamefont {J.-Z.}\ \bibnamefont {Zhou}}, \bibinfo {author} {\bibfnamefont {Z.-C.}\ \bibnamefont {Li}}, \ and\ \bibinfo {author} {\bibfnamefont {D.}~\bibnamefont {Wu}},\ }\href@noop {} {\  (\bibinfo {year} {2025}{\natexlab{b}})},\ \Eprint {http://arxiv.org/abs/2505.22614} {arXiv:2505.22614 [astro-ph.CO]} \BibitemShut {NoStop}%
\bibitem [{\citenamefont {Tagliazucchi}\ \emph {et~al.}(2025)\citenamefont {Tagliazucchi}, \citenamefont {Braglia}, \citenamefont {Finelli},\ and\ \citenamefont {Pieroni}}]{Tagliazucchi:2023dai}%
  \BibitemOpen
  \bibfield  {author} {\bibinfo {author} {\bibfnamefont {M.}~\bibnamefont {Tagliazucchi}}, \bibinfo {author} {\bibfnamefont {M.}~\bibnamefont {Braglia}}, \bibinfo {author} {\bibfnamefont {F.}~\bibnamefont {Finelli}}, \ and\ \bibinfo {author} {\bibfnamefont {M.}~\bibnamefont {Pieroni}},\ }\href {\doibase 10.1103/PhysRevD.111.L021305} {\bibfield  {journal} {\bibinfo  {journal} {Phys. Rev. D}\ }\textbf {\bibinfo {volume} {111}},\ \bibinfo {pages} {L021305} (\bibinfo {year} {2025})},\ \Eprint {http://arxiv.org/abs/2310.08527} {arXiv:2310.08527 [astro-ph.CO]} \BibitemShut {NoStop}%
\bibitem [{\citenamefont {Chluba}\ \emph {et~al.}(2012{\natexlab{a}})\citenamefont {Chluba}, \citenamefont {Erickcek},\ and\ \citenamefont {Ben-Dayan}}]{Chluba:2012we}%
  \BibitemOpen
  \bibfield  {author} {\bibinfo {author} {\bibfnamefont {J.}~\bibnamefont {Chluba}}, \bibinfo {author} {\bibfnamefont {A.~L.}\ \bibnamefont {Erickcek}}, \ and\ \bibinfo {author} {\bibfnamefont {I.}~\bibnamefont {Ben-Dayan}},\ }\href {\doibase 10.1088/0004-637X/758/2/76} {\bibfield  {journal} {\bibinfo  {journal} {Astrophys. J.}\ }\textbf {\bibinfo {volume} {758}},\ \bibinfo {pages} {76} (\bibinfo {year} {2012}{\natexlab{a}})},\ \Eprint {http://arxiv.org/abs/1203.2681} {arXiv:1203.2681 [astro-ph.CO]} \BibitemShut {NoStop}%
\bibitem [{\citenamefont {Chluba}\ \emph {et~al.}(2012{\natexlab{b}})\citenamefont {Chluba}, \citenamefont {Khatri},\ and\ \citenamefont {Sunyaev}}]{Chluba:2012gq}%
  \BibitemOpen
  \bibfield  {author} {\bibinfo {author} {\bibfnamefont {J.}~\bibnamefont {Chluba}}, \bibinfo {author} {\bibfnamefont {R.}~\bibnamefont {Khatri}}, \ and\ \bibinfo {author} {\bibfnamefont {R.~A.}\ \bibnamefont {Sunyaev}},\ }\href {\doibase 10.1111/j.1365-2966.2012.21474.x} {\bibfield  {journal} {\bibinfo  {journal} {Mon. Not. Roy. Astron. Soc.}\ }\textbf {\bibinfo {volume} {425}},\ \bibinfo {pages} {1129} (\bibinfo {year} {2012}{\natexlab{b}})},\ \Eprint {http://arxiv.org/abs/1202.0057} {arXiv:1202.0057 [astro-ph.CO]} \BibitemShut {NoStop}%
\bibitem [{\citenamefont {Sharma}\ \emph {et~al.}(2024)\citenamefont {Sharma}, \citenamefont {Lesgourgues},\ and\ \citenamefont {Byrnes}}]{Sharma:2024img}%
  \BibitemOpen
  \bibfield  {author} {\bibinfo {author} {\bibfnamefont {D.}~\bibnamefont {Sharma}}, \bibinfo {author} {\bibfnamefont {J.}~\bibnamefont {Lesgourgues}}, \ and\ \bibinfo {author} {\bibfnamefont {C.~T.}\ \bibnamefont {Byrnes}},\ }\href {\doibase 10.1088/1475-7516/2024/07/090} {\bibfield  {journal} {\bibinfo  {journal} {JCAP}\ }\textbf {\bibinfo {volume} {07}},\ \bibinfo {pages} {090} (\bibinfo {year} {2024})},\ \Eprint {http://arxiv.org/abs/2404.18474} {arXiv:2404.18474 [astro-ph.CO]} \BibitemShut {NoStop}%
\bibitem [{\citenamefont {Byrnes}\ \emph {et~al.}(2024)\citenamefont {Byrnes}, \citenamefont {Lesgourgues},\ and\ \citenamefont {Sharma}}]{Byrnes:2024vjt}%
  \BibitemOpen
  \bibfield  {author} {\bibinfo {author} {\bibfnamefont {C.~T.}\ \bibnamefont {Byrnes}}, \bibinfo {author} {\bibfnamefont {J.}~\bibnamefont {Lesgourgues}}, \ and\ \bibinfo {author} {\bibfnamefont {D.}~\bibnamefont {Sharma}},\ }\href {\doibase 10.1088/1475-7516/2024/09/012} {\bibfield  {journal} {\bibinfo  {journal} {JCAP}\ }\textbf {\bibinfo {volume} {09}},\ \bibinfo {pages} {012} (\bibinfo {year} {2024})},\ \Eprint {http://arxiv.org/abs/2404.18475} {arXiv:2404.18475 [astro-ph.CO]} \BibitemShut {NoStop}%
\bibitem [{\citenamefont {Clarke}\ \emph {et~al.}(2020)\citenamefont {Clarke}, \citenamefont {Copeland},\ and\ \citenamefont {Moss}}]{Clarke:2020bil}%
  \BibitemOpen
  \bibfield  {author} {\bibinfo {author} {\bibfnamefont {T.~J.}\ \bibnamefont {Clarke}}, \bibinfo {author} {\bibfnamefont {E.~J.}\ \bibnamefont {Copeland}}, \ and\ \bibinfo {author} {\bibfnamefont {A.}~\bibnamefont {Moss}},\ }\href {\doibase 10.1088/1475-7516/2020/10/002} {\bibfield  {journal} {\bibinfo  {journal} {JCAP}\ }\textbf {\bibinfo {volume} {10}},\ \bibinfo {pages} {002} (\bibinfo {year} {2020})},\ \Eprint {http://arxiv.org/abs/2004.11396} {arXiv:2004.11396 [astro-ph.CO]} \BibitemShut {NoStop}%
\bibitem [{\citenamefont {Lamb}\ \emph {et~al.}(2023)\citenamefont {Lamb}, \citenamefont {Taylor},\ and\ \citenamefont {van Haasteren}}]{Lamb:2023jls}%
  \BibitemOpen
  \bibfield  {author} {\bibinfo {author} {\bibfnamefont {W.~G.}\ \bibnamefont {Lamb}}, \bibinfo {author} {\bibfnamefont {S.~R.}\ \bibnamefont {Taylor}}, \ and\ \bibinfo {author} {\bibfnamefont {R.}~\bibnamefont {van Haasteren}},\ }\href {\doibase 10.1103/PhysRevD.108.103019} {\bibfield  {journal} {\bibinfo  {journal} {Phys. Rev. D}\ }\textbf {\bibinfo {volume} {108}},\ \bibinfo {pages} {103019} (\bibinfo {year} {2023})},\ \Eprint {http://arxiv.org/abs/2303.15442} {arXiv:2303.15442 [astro-ph.HE]} \BibitemShut {NoStop}%
\bibitem [{\citenamefont {Moore}\ and\ \citenamefont {Vecchio}(2021)}]{Moore:2021ibq}%
  \BibitemOpen
  \bibfield  {author} {\bibinfo {author} {\bibfnamefont {C.~J.}\ \bibnamefont {Moore}}\ and\ \bibinfo {author} {\bibfnamefont {A.}~\bibnamefont {Vecchio}},\ }\href {\doibase 10.1038/s41550-021-01489-8} {\bibfield  {journal} {\bibinfo  {journal} {Nature Astron.}\ }\textbf {\bibinfo {volume} {5}},\ \bibinfo {pages} {1268} (\bibinfo {year} {2021})},\ \Eprint {http://arxiv.org/abs/2104.15130} {arXiv:2104.15130 [astro-ph.CO]} \BibitemShut {NoStop}%
\bibitem [{\citenamefont {Mitridate}\ \emph {et~al.}(2023)\citenamefont {Mitridate}, \citenamefont {Wright}, \citenamefont {von Eckardstein}, \citenamefont {Schr\"oder}, \citenamefont {Nay}, \citenamefont {Olum}, \citenamefont {Schmitz},\ and\ \citenamefont {Trickle}}]{Mitridate:2023oar}%
  \BibitemOpen
  \bibfield  {author} {\bibinfo {author} {\bibfnamefont {A.}~\bibnamefont {Mitridate}}, \bibinfo {author} {\bibfnamefont {D.}~\bibnamefont {Wright}}, \bibinfo {author} {\bibfnamefont {R.}~\bibnamefont {von Eckardstein}}, \bibinfo {author} {\bibfnamefont {T.}~\bibnamefont {Schr\"oder}}, \bibinfo {author} {\bibfnamefont {J.}~\bibnamefont {Nay}}, \bibinfo {author} {\bibfnamefont {K.}~\bibnamefont {Olum}}, \bibinfo {author} {\bibfnamefont {K.}~\bibnamefont {Schmitz}}, \ and\ \bibinfo {author} {\bibfnamefont {T.}~\bibnamefont {Trickle}},\ }\href@noop {} {\  (\bibinfo {year} {2023})},\ \Eprint {http://arxiv.org/abs/2306.16377} {arXiv:2306.16377 [hep-ph]} \BibitemShut {NoStop}%
\bibitem [{\citenamefont {Collaboration}(2023)}]{Nanograv:KDE}%
  \BibitemOpen
  \bibfield  {author} {\bibinfo {author} {\bibfnamefont {T.~N.}\ \bibnamefont {Collaboration}},\ }\href {\doibase 10.5281/zenodo.10344086} {\enquote {\bibinfo {title} {Kde representations of the gravitational wave background free spectra present in the nanograv 15-year dataset},}\ } (\bibinfo {year} {2023})\BibitemShut {NoStop}%
\bibitem [{\citenamefont {Ashton}\ \emph {et~al.}(2019)\citenamefont {Ashton} \emph {et~al.}}]{bilby_paper}%
  \BibitemOpen
  \bibfield  {author} {\bibinfo {author} {\bibfnamefont {G.}~\bibnamefont {Ashton}} \emph {et~al.},\ }\href {\doibase 10.3847/1538-4365/ab06fc} {\bibfield  {journal} {\bibinfo  {journal} {Astrophys. J. Suppl.}\ }\textbf {\bibinfo {volume} {241}},\ \bibinfo {pages} {27} (\bibinfo {year} {2019})},\ \Eprint {http://arxiv.org/abs/1811.02042} {arXiv:1811.02042 [astro-ph.IM]} \BibitemShut {NoStop}%
\bibitem [{\citenamefont {Speagle}(2020)}]{Speagle:2019ivv}%
  \BibitemOpen
  \bibfield  {author} {\bibinfo {author} {\bibfnamefont {J.~S.}\ \bibnamefont {Speagle}},\ }\href {\doibase 10.1093/mnras/staa278} {\bibfield  {journal} {\bibinfo  {journal} {Mon. Not. Roy. Astron. Soc.}\ }\textbf {\bibinfo {volume} {493}},\ \bibinfo {pages} {3132} (\bibinfo {year} {2020})},\ \Eprint {http://arxiv.org/abs/1904.02180} {arXiv:1904.02180 [astro-ph.IM]} \BibitemShut {NoStop}%
\bibitem [{\citenamefont {Koposov}\ \emph {et~al.}(2024)\citenamefont {Koposov}, \citenamefont {Speagle}, \citenamefont {Barbary}, \citenamefont {Ashton}, \citenamefont {Bennett}, \citenamefont {Buchner}, \citenamefont {Scheffler}, \citenamefont {Cook}, \citenamefont {Talbot}, \citenamefont {Guillochon}, \citenamefont {Cubillos}, \citenamefont {Ramos}, \citenamefont {Dartiailh}, \citenamefont {Ilya}, \citenamefont {Tollerud}, \citenamefont {Lang}, \citenamefont {Johnson}, \citenamefont {jtmendel}, \citenamefont {Higson}, \citenamefont {Vandal}, \citenamefont {Daylan}, \citenamefont {Angus}, \citenamefont {patelR}, \citenamefont {Cargile}, \citenamefont {Sheehan}, \citenamefont {Pitkin}, \citenamefont {Kirk}, \citenamefont {Leja}, \citenamefont {joezuntz},\ and\ \citenamefont {Goldstein}}]{dynesty_software}%
  \BibitemOpen
  \bibfield  {author} {\bibinfo {author} {\bibfnamefont {S.}~\bibnamefont {Koposov}}, \bibinfo {author} {\bibfnamefont {J.}~\bibnamefont {Speagle}}, \bibinfo {author} {\bibfnamefont {K.}~\bibnamefont {Barbary}}, \bibinfo {author} {\bibfnamefont {G.}~\bibnamefont {Ashton}}, \bibinfo {author} {\bibfnamefont {E.}~\bibnamefont {Bennett}}, \bibinfo {author} {\bibfnamefont {J.}~\bibnamefont {Buchner}}, \bibinfo {author} {\bibfnamefont {C.}~\bibnamefont {Scheffler}}, \bibinfo {author} {\bibfnamefont {B.}~\bibnamefont {Cook}}, \bibinfo {author} {\bibfnamefont {C.}~\bibnamefont {Talbot}}, \bibinfo {author} {\bibfnamefont {J.}~\bibnamefont {Guillochon}}, \bibinfo {author} {\bibfnamefont {P.}~\bibnamefont {Cubillos}}, \bibinfo {author} {\bibfnamefont {A.~A.}\ \bibnamefont {Ramos}}, \bibinfo {author} {\bibfnamefont {M.}~\bibnamefont {Dartiailh}}, \bibinfo {author} {\bibnamefont {Ilya}}, \bibinfo {author} {\bibfnamefont {E.}~\bibnamefont {Tollerud}}, \bibinfo {author} {\bibfnamefont {D.}~\bibnamefont {Lang}}, \bibinfo {author} {\bibfnamefont {B.}~\bibnamefont {Johnson}}, \bibinfo {author} {\bibnamefont {jtmendel}}, \bibinfo {author} {\bibfnamefont {E.}~\bibnamefont {Higson}}, \bibinfo {author} {\bibfnamefont {T.}~\bibnamefont {Vandal}}, \bibinfo {author} {\bibfnamefont {T.}~\bibnamefont {Daylan}}, \bibinfo {author} {\bibfnamefont {R.}~\bibnamefont {Angus}}, \bibinfo {author} {\bibnamefont {patelR}}, \bibinfo {author} {\bibfnamefont {P.}~\bibnamefont {Cargile}}, \bibinfo {author} {\bibfnamefont {P.}~\bibnamefont {Sheehan}}, \bibinfo {author} {\bibfnamefont {M.}~\bibnamefont {Pitkin}}, \bibinfo {author} {\bibfnamefont {M.}~\bibnamefont {Kirk}}, \bibinfo {author} {\bibfnamefont {J.}~\bibnamefont {Leja}}, \bibinfo {author} {\bibnamefont {joezuntz}}, \ and\ \bibinfo {author} {\bibfnamefont {D.}~\bibnamefont {Goldstein}},\ }\href {\doibase 10.5281/zenodo.12537467} {\enquote {\bibinfo {title} {joshspeagle/dynesty: v2.1.4},}\ } (\bibinfo {year} {2024})\BibitemShut {NoStop}%
\bibitem [{\citenamefont {Carr}(1975)}]{Carr:1975qj}%
  \BibitemOpen
  \bibfield  {author} {\bibinfo {author} {\bibfnamefont {B.~J.}\ \bibnamefont {Carr}},\ }\href {\doibase 10.1086/153853} {\bibfield  {journal} {\bibinfo  {journal} {Astrophys. J.}\ }\textbf {\bibinfo {volume} {201}},\ \bibinfo {pages} {1} (\bibinfo {year} {1975})}\BibitemShut {NoStop}%
\bibitem [{\citenamefont {Carr}\ \emph {et~al.}(2021)\citenamefont {Carr}, \citenamefont {Kohri}, \citenamefont {Sendouda},\ and\ \citenamefont {Yokoyama}}]{Carr:2020gox}%
  \BibitemOpen
  \bibfield  {author} {\bibinfo {author} {\bibfnamefont {B.}~\bibnamefont {Carr}}, \bibinfo {author} {\bibfnamefont {K.}~\bibnamefont {Kohri}}, \bibinfo {author} {\bibfnamefont {Y.}~\bibnamefont {Sendouda}}, \ and\ \bibinfo {author} {\bibfnamefont {J.}~\bibnamefont {Yokoyama}},\ }\href {\doibase 10.1088/1361-6633/ac1e31} {\bibfield  {journal} {\bibinfo  {journal} {Rept. Prog. Phys.}\ }\textbf {\bibinfo {volume} {84}},\ \bibinfo {pages} {116902} (\bibinfo {year} {2021})},\ \Eprint {http://arxiv.org/abs/2002.12778} {arXiv:2002.12778 [astro-ph.CO]} \BibitemShut {NoStop}%
\bibitem [{\citenamefont {Carr}\ and\ \citenamefont {Kuhnel}(2020)}]{Carr:2020xqk}%
  \BibitemOpen
  \bibfield  {author} {\bibinfo {author} {\bibfnamefont {B.}~\bibnamefont {Carr}}\ and\ \bibinfo {author} {\bibfnamefont {F.}~\bibnamefont {Kuhnel}},\ }\href {\doibase 10.1146/annurev-nucl-050520-125911} {\bibfield  {journal} {\bibinfo  {journal} {Ann. Rev. Nucl. Part. Sci.}\ }\textbf {\bibinfo {volume} {70}},\ \bibinfo {pages} {355} (\bibinfo {year} {2020})},\ \Eprint {http://arxiv.org/abs/2006.02838} {arXiv:2006.02838 [astro-ph.CO]} \BibitemShut {NoStop}%
\bibitem [{\citenamefont {Carr}\ \emph {et~al.}(2016)\citenamefont {Carr}, \citenamefont {Kuhnel},\ and\ \citenamefont {Sandstad}}]{Carr:2016drx}%
  \BibitemOpen
  \bibfield  {author} {\bibinfo {author} {\bibfnamefont {B.}~\bibnamefont {Carr}}, \bibinfo {author} {\bibfnamefont {F.}~\bibnamefont {Kuhnel}}, \ and\ \bibinfo {author} {\bibfnamefont {M.}~\bibnamefont {Sandstad}},\ }\href {\doibase 10.1103/PhysRevD.94.083504} {\bibfield  {journal} {\bibinfo  {journal} {Phys. Rev. D}\ }\textbf {\bibinfo {volume} {94}},\ \bibinfo {pages} {083504} (\bibinfo {year} {2016})},\ \Eprint {http://arxiv.org/abs/1607.06077} {arXiv:1607.06077 [astro-ph.CO]} \BibitemShut {NoStop}%
\bibitem [{\citenamefont {Carr}\ \emph {et~al.}(2010)\citenamefont {Carr}, \citenamefont {Kohri}, \citenamefont {Sendouda},\ and\ \citenamefont {Yokoyama}}]{Carr:2009jm}%
  \BibitemOpen
  \bibfield  {author} {\bibinfo {author} {\bibfnamefont {B.~J.}\ \bibnamefont {Carr}}, \bibinfo {author} {\bibfnamefont {K.}~\bibnamefont {Kohri}}, \bibinfo {author} {\bibfnamefont {Y.}~\bibnamefont {Sendouda}}, \ and\ \bibinfo {author} {\bibfnamefont {J.}~\bibnamefont {Yokoyama}},\ }\href {\doibase 10.1103/PhysRevD.81.104019} {\bibfield  {journal} {\bibinfo  {journal} {Phys. Rev. D}\ }\textbf {\bibinfo {volume} {81}},\ \bibinfo {pages} {104019} (\bibinfo {year} {2010})},\ \Eprint {http://arxiv.org/abs/0912.5297} {arXiv:0912.5297 [astro-ph.CO]} \BibitemShut {NoStop}%
\bibitem [{\citenamefont {Musco}(2019)}]{Musco:2018rwt}%
  \BibitemOpen
  \bibfield  {author} {\bibinfo {author} {\bibfnamefont {I.}~\bibnamefont {Musco}},\ }\href {\doibase 10.1103/PhysRevD.100.123524} {\bibfield  {journal} {\bibinfo  {journal} {Phys. Rev. D}\ }\textbf {\bibinfo {volume} {100}},\ \bibinfo {pages} {123524} (\bibinfo {year} {2019})},\ \Eprint {http://arxiv.org/abs/1809.02127} {arXiv:1809.02127 [gr-qc]} \BibitemShut {NoStop}%
\bibitem [{\citenamefont {Germani}\ and\ \citenamefont {Musco}(2019)}]{Germani:2018jgr}%
  \BibitemOpen
  \bibfield  {author} {\bibinfo {author} {\bibfnamefont {C.}~\bibnamefont {Germani}}\ and\ \bibinfo {author} {\bibfnamefont {I.}~\bibnamefont {Musco}},\ }\href {\doibase 10.1103/PhysRevLett.122.141302} {\bibfield  {journal} {\bibinfo  {journal} {Phys. Rev. Lett.}\ }\textbf {\bibinfo {volume} {122}},\ \bibinfo {pages} {141302} (\bibinfo {year} {2019})},\ \Eprint {http://arxiv.org/abs/1805.04087} {arXiv:1805.04087 [astro-ph.CO]} \BibitemShut {NoStop}%
\bibitem [{\citenamefont {Ferrante}\ \emph {et~al.}(2023)\citenamefont {Ferrante}, \citenamefont {Franciolini}, \citenamefont {Iovino},\ and\ \citenamefont {Urbano}}]{Ferrante:2022mui}%
  \BibitemOpen
  \bibfield  {author} {\bibinfo {author} {\bibfnamefont {G.}~\bibnamefont {Ferrante}}, \bibinfo {author} {\bibfnamefont {G.}~\bibnamefont {Franciolini}}, \bibinfo {author} {\bibfnamefont {A.}~\bibnamefont {Iovino}, \bibfnamefont {Junior.}}, \ and\ \bibinfo {author} {\bibfnamefont {A.}~\bibnamefont {Urbano}},\ }\href {\doibase 10.1103/PhysRevD.107.043520} {\bibfield  {journal} {\bibinfo  {journal} {Phys. Rev. D}\ }\textbf {\bibinfo {volume} {107}},\ \bibinfo {pages} {043520} (\bibinfo {year} {2023})},\ \Eprint {http://arxiv.org/abs/2211.01728} {arXiv:2211.01728 [astro-ph.CO]} \BibitemShut {NoStop}%
\bibitem [{\citenamefont {Shimada}\ \emph {et~al.}(2025)\citenamefont {Shimada}, \citenamefont {Escriv{\'a}}, \citenamefont {Saito}, \citenamefont {Uehara},\ and\ \citenamefont {Yoo}}]{Shimada:2024eec}%
  \BibitemOpen
  \bibfield  {author} {\bibinfo {author} {\bibfnamefont {M.}~\bibnamefont {Shimada}}, \bibinfo {author} {\bibfnamefont {A.}~\bibnamefont {Escriv{\'a}}}, \bibinfo {author} {\bibfnamefont {D.}~\bibnamefont {Saito}}, \bibinfo {author} {\bibfnamefont {K.}~\bibnamefont {Uehara}}, \ and\ \bibinfo {author} {\bibfnamefont {C.-M.}\ \bibnamefont {Yoo}},\ }\href {\doibase 10.1088/1475-7516/2025/02/018} {\bibfield  {journal} {\bibinfo  {journal} {JCAP}\ }\textbf {\bibinfo {volume} {02}},\ \bibinfo {pages} {018} (\bibinfo {year} {2025})},\ \Eprint {http://arxiv.org/abs/2411.07648} {arXiv:2411.07648 [gr-qc]} \BibitemShut {NoStop}%
\bibitem [{\citenamefont {Iovino}\ \emph {et~al.}(2024)\citenamefont {Iovino}, \citenamefont {Perna}, \citenamefont {Riotto},\ and\ \citenamefont {Veerm\"ae}}]{Iovino:2024tyg}%
  \BibitemOpen
  \bibfield  {author} {\bibinfo {author} {\bibfnamefont {A.~J.}\ \bibnamefont {Iovino}}, \bibinfo {author} {\bibfnamefont {G.}~\bibnamefont {Perna}}, \bibinfo {author} {\bibfnamefont {A.}~\bibnamefont {Riotto}}, \ and\ \bibinfo {author} {\bibfnamefont {H.}~\bibnamefont {Veerm\"ae}},\ }\href {\doibase 10.1088/1475-7516/2024/10/050} {\bibfield  {journal} {\bibinfo  {journal} {JCAP}\ }\textbf {\bibinfo {volume} {10}},\ \bibinfo {pages} {050} (\bibinfo {year} {2024})},\ \Eprint {http://arxiv.org/abs/2406.20089} {arXiv:2406.20089 [astro-ph.CO]} \BibitemShut {NoStop}%
\bibitem [{\citenamefont {Musco}\ \emph {et~al.}(2021)\citenamefont {Musco}, \citenamefont {De~Luca}, \citenamefont {Franciolini},\ and\ \citenamefont {Riotto}}]{Musco:2020jjb}%
  \BibitemOpen
  \bibfield  {author} {\bibinfo {author} {\bibfnamefont {I.}~\bibnamefont {Musco}}, \bibinfo {author} {\bibfnamefont {V.}~\bibnamefont {De~Luca}}, \bibinfo {author} {\bibfnamefont {G.}~\bibnamefont {Franciolini}}, \ and\ \bibinfo {author} {\bibfnamefont {A.}~\bibnamefont {Riotto}},\ }\href {\doibase 10.1103/PhysRevD.103.063538} {\bibfield  {journal} {\bibinfo  {journal} {Phys. Rev. D}\ }\textbf {\bibinfo {volume} {103}},\ \bibinfo {pages} {063538} (\bibinfo {year} {2021})},\ \Eprint {http://arxiv.org/abs/2011.03014} {arXiv:2011.03014 [astro-ph.CO]} \BibitemShut {NoStop}%
\bibitem [{\citenamefont {Young}\ \emph {et~al.}(2014)\citenamefont {Young}, \citenamefont {Byrnes},\ and\ \citenamefont {Sasaki}}]{Young:2014ana}%
  \BibitemOpen
  \bibfield  {author} {\bibinfo {author} {\bibfnamefont {S.}~\bibnamefont {Young}}, \bibinfo {author} {\bibfnamefont {C.~T.}\ \bibnamefont {Byrnes}}, \ and\ \bibinfo {author} {\bibfnamefont {M.}~\bibnamefont {Sasaki}},\ }\href {\doibase 10.1088/1475-7516/2014/07/045} {\bibfield  {journal} {\bibinfo  {journal} {JCAP}\ }\textbf {\bibinfo {volume} {07}},\ \bibinfo {pages} {045} (\bibinfo {year} {2014})},\ \Eprint {http://arxiv.org/abs/1405.7023} {arXiv:1405.7023 [gr-qc]} \BibitemShut {NoStop}%
\bibitem [{\citenamefont {Green}\ \emph {et~al.}(2004)\citenamefont {Green}, \citenamefont {Liddle}, \citenamefont {Malik},\ and\ \citenamefont {Sasaki}}]{Green:2004wb}%
  \BibitemOpen
  \bibfield  {author} {\bibinfo {author} {\bibfnamefont {A.~M.}\ \bibnamefont {Green}}, \bibinfo {author} {\bibfnamefont {A.~R.}\ \bibnamefont {Liddle}}, \bibinfo {author} {\bibfnamefont {K.~A.}\ \bibnamefont {Malik}}, \ and\ \bibinfo {author} {\bibfnamefont {M.}~\bibnamefont {Sasaki}},\ }\href {\doibase 10.1103/PhysRevD.70.041502} {\bibfield  {journal} {\bibinfo  {journal} {Phys. Rev. D}\ }\textbf {\bibinfo {volume} {70}},\ \bibinfo {pages} {041502} (\bibinfo {year} {2004})},\ \Eprint {http://arxiv.org/abs/astro-ph/0403181} {arXiv:astro-ph/0403181} \BibitemShut {NoStop}%
\bibitem [{\citenamefont {De~Luca}\ \emph {et~al.}(2019)\citenamefont {De~Luca}, \citenamefont {Franciolini}, \citenamefont {Kehagias}, \citenamefont {Peloso}, \citenamefont {Riotto},\ and\ \citenamefont {{\"U}nal}}]{DeLuca:2019qsy}%
  \BibitemOpen
  \bibfield  {author} {\bibinfo {author} {\bibfnamefont {V.}~\bibnamefont {De~Luca}}, \bibinfo {author} {\bibfnamefont {G.}~\bibnamefont {Franciolini}}, \bibinfo {author} {\bibfnamefont {A.}~\bibnamefont {Kehagias}}, \bibinfo {author} {\bibfnamefont {M.}~\bibnamefont {Peloso}}, \bibinfo {author} {\bibfnamefont {A.}~\bibnamefont {Riotto}}, \ and\ \bibinfo {author} {\bibfnamefont {C.}~\bibnamefont {{\"U}nal}},\ }\href {\doibase 10.1088/1475-7516/2019/07/048} {\bibfield  {journal} {\bibinfo  {journal} {JCAP}\ }\textbf {\bibinfo {volume} {07}},\ \bibinfo {pages} {048} (\bibinfo {year} {2019})},\ \Eprint {http://arxiv.org/abs/1904.00970} {arXiv:1904.00970 [astro-ph.CO]} \BibitemShut {NoStop}%
\bibitem [{\citenamefont {Ballesteros}\ and\ \citenamefont {Taoso}(2018)}]{Ballesteros:2017fsr}%
  \BibitemOpen
  \bibfield  {author} {\bibinfo {author} {\bibfnamefont {G.}~\bibnamefont {Ballesteros}}\ and\ \bibinfo {author} {\bibfnamefont {M.}~\bibnamefont {Taoso}},\ }\href {\doibase 10.1103/PhysRevD.97.023501} {\bibfield  {journal} {\bibinfo  {journal} {Phys. Rev. D}\ }\textbf {\bibinfo {volume} {97}},\ \bibinfo {pages} {023501} (\bibinfo {year} {2018})},\ \Eprint {http://arxiv.org/abs/1709.05565} {arXiv:1709.05565 [hep-ph]} \BibitemShut {NoStop}%
\bibitem [{\citenamefont {Kannike}\ \emph {et~al.}(2017)\citenamefont {Kannike}, \citenamefont {Marzola}, \citenamefont {Raidal},\ and\ \citenamefont {Veerm{\"a}e}}]{Kannike:2017bxn}%
  \BibitemOpen
  \bibfield  {author} {\bibinfo {author} {\bibfnamefont {K.}~\bibnamefont {Kannike}}, \bibinfo {author} {\bibfnamefont {L.}~\bibnamefont {Marzola}}, \bibinfo {author} {\bibfnamefont {M.}~\bibnamefont {Raidal}}, \ and\ \bibinfo {author} {\bibfnamefont {H.}~\bibnamefont {Veerm{\"a}e}},\ }\href {\doibase 10.1088/1475-7516/2017/09/020} {\bibfield  {journal} {\bibinfo  {journal} {JCAP}\ }\textbf {\bibinfo {volume} {09}},\ \bibinfo {pages} {020} (\bibinfo {year} {2017})},\ \Eprint {http://arxiv.org/abs/1705.06225} {arXiv:1705.06225 [astro-ph.CO]} \BibitemShut {NoStop}%
\bibitem [{\citenamefont {Ezquiaga}\ \emph {et~al.}(2018)\citenamefont {Ezquiaga}, \citenamefont {Garcia-Bellido},\ and\ \citenamefont {Ruiz~Morales}}]{Ezquiaga:2017fvi}%
  \BibitemOpen
  \bibfield  {author} {\bibinfo {author} {\bibfnamefont {J.~M.}\ \bibnamefont {Ezquiaga}}, \bibinfo {author} {\bibfnamefont {J.}~\bibnamefont {Garcia-Bellido}}, \ and\ \bibinfo {author} {\bibfnamefont {E.}~\bibnamefont {Ruiz~Morales}},\ }\href {\doibase 10.1016/j.physletb.2017.11.039} {\bibfield  {journal} {\bibinfo  {journal} {Phys. Lett. B}\ }\textbf {\bibinfo {volume} {776}},\ \bibinfo {pages} {345} (\bibinfo {year} {2018})},\ \Eprint {http://arxiv.org/abs/1705.04861} {arXiv:1705.04861 [astro-ph.CO]} \BibitemShut {NoStop}%
\bibitem [{\citenamefont {Garcia-Bellido}\ and\ \citenamefont {Ruiz~Morales}(2017)}]{Garcia-Bellido:2017mdw}%
  \BibitemOpen
  \bibfield  {author} {\bibinfo {author} {\bibfnamefont {J.}~\bibnamefont {Garcia-Bellido}}\ and\ \bibinfo {author} {\bibfnamefont {E.}~\bibnamefont {Ruiz~Morales}},\ }\href {\doibase 10.1016/j.dark.2017.09.007} {\bibfield  {journal} {\bibinfo  {journal} {Phys. Dark Univ.}\ }\textbf {\bibinfo {volume} {18}},\ \bibinfo {pages} {47} (\bibinfo {year} {2017})},\ \Eprint {http://arxiv.org/abs/1702.03901} {arXiv:1702.03901 [astro-ph.CO]} \BibitemShut {NoStop}%
\bibitem [{\citenamefont {Firouzjahi}\ and\ \citenamefont {Riotto}(2024)}]{Firouzjahi:2023ahg}%
  \BibitemOpen
  \bibfield  {author} {\bibinfo {author} {\bibfnamefont {H.}~\bibnamefont {Firouzjahi}}\ and\ \bibinfo {author} {\bibfnamefont {A.}~\bibnamefont {Riotto}},\ }\href {\doibase 10.1088/1475-7516/2024/02/021} {\bibfield  {journal} {\bibinfo  {journal} {JCAP}\ }\textbf {\bibinfo {volume} {02}},\ \bibinfo {pages} {021} (\bibinfo {year} {2024})},\ \Eprint {http://arxiv.org/abs/2304.07801} {arXiv:2304.07801 [astro-ph.CO]} \BibitemShut {NoStop}%
\bibitem [{\citenamefont {Ferraz}\ and\ \citenamefont {Rosa}(2025)}]{Ferraz:2024bvd}%
  \BibitemOpen
  \bibfield  {author} {\bibinfo {author} {\bibfnamefont {P.~B.}\ \bibnamefont {Ferraz}}\ and\ \bibinfo {author} {\bibfnamefont {J.~G.}\ \bibnamefont {Rosa}},\ }\href {\doibase 10.1088/1475-7516/2025/03/040} {\bibfield  {journal} {\bibinfo  {journal} {JCAP}\ }\textbf {\bibinfo {volume} {03}},\ \bibinfo {pages} {040} (\bibinfo {year} {2025})},\ \Eprint {http://arxiv.org/abs/2410.10996} {arXiv:2410.10996 [hep-ph]} \BibitemShut {NoStop}%
\bibitem [{\citenamefont {Kristiano}\ and\ \citenamefont {Yokoyama}(2024)}]{Kristiano:2022maq}%
  \BibitemOpen
  \bibfield  {author} {\bibinfo {author} {\bibfnamefont {J.}~\bibnamefont {Kristiano}}\ and\ \bibinfo {author} {\bibfnamefont {J.}~\bibnamefont {Yokoyama}},\ }\href {\doibase 10.1103/PhysRevLett.132.221003} {\bibfield  {journal} {\bibinfo  {journal} {Phys. Rev. Lett.}\ }\textbf {\bibinfo {volume} {132}},\ \bibinfo {pages} {221003} (\bibinfo {year} {2024})},\ \Eprint {http://arxiv.org/abs/2211.03395} {arXiv:2211.03395 [hep-th]} \BibitemShut {NoStop}%
\bibitem [{\citenamefont {Franciolini}\ \emph {et~al.}(2023)\citenamefont {Franciolini}, \citenamefont {Iovino}, \citenamefont {Vaskonen},\ and\ \citenamefont {Veermae}}]{Franciolini:2023pbf}%
  \BibitemOpen
  \bibfield  {author} {\bibinfo {author} {\bibfnamefont {G.}~\bibnamefont {Franciolini}}, \bibinfo {author} {\bibfnamefont {A.}~\bibnamefont {Iovino}, \bibfnamefont {Junior.}}, \bibinfo {author} {\bibfnamefont {V.}~\bibnamefont {Vaskonen}}, \ and\ \bibinfo {author} {\bibfnamefont {H.}~\bibnamefont {Veermae}},\ }\href {\doibase 10.1103/PhysRevLett.131.201401} {\bibfield  {journal} {\bibinfo  {journal} {Phys. Rev. Lett.}\ }\textbf {\bibinfo {volume} {131}},\ \bibinfo {pages} {201401} (\bibinfo {year} {2023})},\ \Eprint {http://arxiv.org/abs/2306.17149} {arXiv:2306.17149 [astro-ph.CO]} \BibitemShut {NoStop}%
\bibitem [{\citenamefont {Wang}\ \emph {et~al.}(2024{\natexlab{b}})\citenamefont {Wang}, \citenamefont {Zhao}, \citenamefont {Li},\ and\ \citenamefont {Zhu}}]{Wang:2023ost}%
  \BibitemOpen
  \bibfield  {author} {\bibinfo {author} {\bibfnamefont {S.}~\bibnamefont {Wang}}, \bibinfo {author} {\bibfnamefont {Z.-C.}\ \bibnamefont {Zhao}}, \bibinfo {author} {\bibfnamefont {J.-P.}\ \bibnamefont {Li}}, \ and\ \bibinfo {author} {\bibfnamefont {Q.-H.}\ \bibnamefont {Zhu}},\ }\href {\doibase 10.1103/PhysRevResearch.6.L012060} {\bibfield  {journal} {\bibinfo  {journal} {Phys. Rev. Res.}\ }\textbf {\bibinfo {volume} {6}},\ \bibinfo {pages} {L012060} (\bibinfo {year} {2024}{\natexlab{b}})},\ \Eprint {http://arxiv.org/abs/2307.00572} {arXiv:2307.00572 [astro-ph.CO]} \BibitemShut {NoStop}%
\bibitem [{\citenamefont {Wu}\ \emph {et~al.}(2025)\citenamefont {Wu}, \citenamefont {Li}, \citenamefont {Wu}, \citenamefont {Chen},\ and\ \citenamefont {Zhou}}]{Wu:2025gwt}%
  \BibitemOpen
  \bibfield  {author} {\bibinfo {author} {\bibfnamefont {D.}~\bibnamefont {Wu}}, \bibinfo {author} {\bibfnamefont {Z.-C.}\ \bibnamefont {Li}}, \bibinfo {author} {\bibfnamefont {P.-Y.}\ \bibnamefont {Wu}}, \bibinfo {author} {\bibfnamefont {F.-Y.}\ \bibnamefont {Chen}}, \ and\ \bibinfo {author} {\bibfnamefont {J.-Z.}\ \bibnamefont {Zhou}},\ }\href@noop {} {\  (\bibinfo {year} {2025})},\ \Eprint {http://arxiv.org/abs/2507.07836} {arXiv:2507.07836 [astro-ph.CO]} \BibitemShut {NoStop}%
\bibitem [{\citenamefont {Choudhury}\ \emph {et~al.}(2023)\citenamefont {Choudhury}, \citenamefont {Panda},\ and\ \citenamefont {Sami}}]{Choudhury:2023jlt}%
  \BibitemOpen
  \bibfield  {author} {\bibinfo {author} {\bibfnamefont {S.}~\bibnamefont {Choudhury}}, \bibinfo {author} {\bibfnamefont {S.}~\bibnamefont {Panda}}, \ and\ \bibinfo {author} {\bibfnamefont {M.}~\bibnamefont {Sami}},\ }\href {\doibase 10.1016/j.physletb.2023.138123} {\bibfield  {journal} {\bibinfo  {journal} {Phys. Lett. B}\ }\textbf {\bibinfo {volume} {845}},\ \bibinfo {pages} {138123} (\bibinfo {year} {2023})},\ \Eprint {http://arxiv.org/abs/2302.05655} {arXiv:2302.05655 [astro-ph.CO]} \BibitemShut {NoStop}%
\bibitem [{\citenamefont {Barker}\ \emph {et~al.}(2025)\citenamefont {Barker}, \citenamefont {Gladwyn},\ and\ \citenamefont {Zell}}]{Barker:2024mpz}%
  \BibitemOpen
  \bibfield  {author} {\bibinfo {author} {\bibfnamefont {W.}~\bibnamefont {Barker}}, \bibinfo {author} {\bibfnamefont {B.}~\bibnamefont {Gladwyn}}, \ and\ \bibinfo {author} {\bibfnamefont {S.}~\bibnamefont {Zell}},\ }\href {\doibase 10.1103/4hrv-zfch} {\bibfield  {journal} {\bibinfo  {journal} {Phys. Rev. D}\ }\textbf {\bibinfo {volume} {111}},\ \bibinfo {pages} {123033} (\bibinfo {year} {2025})},\ \Eprint {http://arxiv.org/abs/2410.11948} {arXiv:2410.11948 [astro-ph.CO]} \BibitemShut {NoStop}%
\bibitem [{\citenamefont {Su}\ \emph {et~al.}(2025)\citenamefont {Su}, \citenamefont {Li},\ and\ \citenamefont {Feng}}]{Su:2025mam}%
  \BibitemOpen
  \bibfield  {author} {\bibinfo {author} {\bibfnamefont {B.-Y.}\ \bibnamefont {Su}}, \bibinfo {author} {\bibfnamefont {N.}~\bibnamefont {Li}}, \ and\ \bibinfo {author} {\bibfnamefont {L.}~\bibnamefont {Feng}},\ }\href {\doibase 10.1140/epjc/s10052-025-13921-8} {\bibfield  {journal} {\bibinfo  {journal} {Eur. Phys. J. C}\ }\textbf {\bibinfo {volume} {85}},\ \bibinfo {pages} {197} (\bibinfo {year} {2025})},\ \Eprint {http://arxiv.org/abs/2502.14323} {arXiv:2502.14323 [astro-ph.CO]} \BibitemShut {NoStop}%
\bibitem [{\citenamefont {Wang}\ \emph {et~al.}(2025{\natexlab{b}})\citenamefont {Wang}, \citenamefont {Ma},\ and\ \citenamefont {Cai}}]{Wang:2024nmd}%
  \BibitemOpen
  \bibfield  {author} {\bibinfo {author} {\bibfnamefont {X.}~\bibnamefont {Wang}}, \bibinfo {author} {\bibfnamefont {X.-H.}\ \bibnamefont {Ma}}, \ and\ \bibinfo {author} {\bibfnamefont {Y.-F.}\ \bibnamefont {Cai}},\ }\href {\doibase 10.1142/S0218271825500270} {\bibfield  {journal} {\bibinfo  {journal} {Int. J. Mod. Phys. D}\ }\textbf {\bibinfo {volume} {34}},\ \bibinfo {pages} {2550027} (\bibinfo {year} {2025}{\natexlab{b}})},\ \Eprint {http://arxiv.org/abs/2412.19631} {arXiv:2412.19631 [astro-ph.CO]} \BibitemShut {NoStop}%
\bibitem [{\citenamefont {Inomata}(2021)}]{Inomata:2020cck}%
  \BibitemOpen
  \bibfield  {author} {\bibinfo {author} {\bibfnamefont {K.}~\bibnamefont {Inomata}},\ }\href {\doibase 10.1088/1475-7516/2021/03/013} {\bibfield  {journal} {\bibinfo  {journal} {JCAP}\ }\textbf {\bibinfo {volume} {03}},\ \bibinfo {pages} {013} (\bibinfo {year} {2021})},\ \Eprint {http://arxiv.org/abs/2008.12300} {arXiv:2008.12300 [gr-qc]} \BibitemShut {NoStop}%
\bibitem [{\citenamefont {Chen}\ \emph {et~al.}(2023)\citenamefont {Chen}, \citenamefont {Ota}, \citenamefont {Zhu},\ and\ \citenamefont {Zhu}}]{Chen:2022dah}%
  \BibitemOpen
  \bibfield  {author} {\bibinfo {author} {\bibfnamefont {C.}~\bibnamefont {Chen}}, \bibinfo {author} {\bibfnamefont {A.}~\bibnamefont {Ota}}, \bibinfo {author} {\bibfnamefont {H.-Y.}\ \bibnamefont {Zhu}}, \ and\ \bibinfo {author} {\bibfnamefont {Y.}~\bibnamefont {Zhu}},\ }\href {\doibase 10.1103/PhysRevD.107.083518} {\bibfield  {journal} {\bibinfo  {journal} {Phys. Rev. D}\ }\textbf {\bibinfo {volume} {107}},\ \bibinfo {pages} {083518} (\bibinfo {year} {2023})},\ \Eprint {http://arxiv.org/abs/2210.17176} {arXiv:2210.17176 [astro-ph.CO]} \BibitemShut {NoStop}%
\end{thebibliography}%

\end{document}